\documentclass[acmsmall,screen]{acmart}

\newcommand{\secref}[1]{Sec.~\ref{sec:#1}}
\newcommand{\tabref}[1]{Table~\ref{tab:#1}}
\newcommand{\figref}[1]{Fig.~\ref{fig:#1}}
\newcommand{\lineref}[1]{Line~\ref{line:#1}}

\newcommand{\maybelabelledregion}{\ensuremath{\tilde{r}}}

\newcommand{\lproj}[2]{\ensuremath{#1 \downarrow #2}}
\newcommand{\lprojtt}[2]{\ensuremath{\lproj{\mathtt{#1}}{\tick{\mathtt{#2}}}}}

\newcommand{\drawpcgnode}[4]{
    \node[pcg] (#1) at #4 {$\code{#2}$: $#3$};
}

\newcommand{\withproj}[6][-0.3]{
    \begin{pgfonlayer}{background}
    \drawpcgnode{#2}{#3}{#4}{#6}
    \end{pgfonlayer}
    \node[lproj] (#2proj) at ($(#2.east)+(#1,-0.4)$) {$\lprojtt{#3}{#5}$};
}

\newcommand{\drawderef}[2]{
    \node[hyperedge, fit=(#1proj)(#1)] (#1hyperedge) {};
    \draw[expand] (#1hyperedge) -- (#2);
}
\newcommand{\pcgplace}{\ensuremath{P}}
\newcommand{\lifetimeproj}{\ensuremath{\mathit{lp}}}
\newcommand{\rp}{\lifetimeproj}
\newcommand{\RP}{\ensuremath{\mathit{RP}}}

\newcommand{\tick}[1]{\ensuremath{\texttt{\textquotesingle}#1}}
\newcommand{\ticktt}[1]{\ensuremath{\texttt{\textquotesingle#1}}}

\newcommand{\remote}[1]{\ensuremath{\mathit{origin}(#1)}}

\newcommand{\borrowedgecolour}{brown}
\newcommand{\aliasedgecolour}{blue}
\newcommand{\outlivesedgecolour}{DarkGreen}

\newcommand{\borrowType}{\ensuremath{b}}
\newcommand{\unpackType}{\ensuremath{u}}
\newcommand{\borrowFlowType}{\ensuremath{f}}
\newcommand{\abstractionType}{\ensuremath{a}}
\newcommand{\aliasType}{\ensuremath{l}}
\newcommand{\derefType}{\ensuremath{d}}

\newcommand{\placecap}[2]{\ensuremath{#1 \colon #2}}
\newcommand{\placecaptt}[2]{\ensuremath{\texttt{#1} \colon #2}}

\newcommand{\opunpack}{\spoperator{\mathtt{unpack}}}
\newcommand{\opderef}{\spoperator{\mathtt{deref}}} % We call deref the same as unpack, so use the same letter
\newcommand{\oppack}{\spoperator{\mathtt{pack}}}
\newcommand{\opexpire}{\spoperator{\mathtt{expire}}}

\newcommand{\ppoint}{\ensuremath{l}}

\newcommand{\capability}{c}

\newcommand{\capexclusive}{\ensuremath{\mathbf{E}}}
\newcommand{\capread}{\ensuremath{\mathbf{R}}}
\newcommand{\capwrite}{\ensuremath{\mathbf{W}}}

\newcommand{\capnone}{\ensuremath{\emptyset}}

\newcommand{\etc}{{{etc.\@}}}

\newcommand{\eg}{{{e.g.\@}}}
\newcommand{\resp}{{{resp.\@}}}
\newcommand{\ie}{{{i.e.\@}}}
\newcommand{\cf}{{{cf.\@}}} % None of the other abbreviations are \it, so why should this one be
 \newcommand{\nocolour}{} % uncomment this to kill the edit colours/macros

\newcommand{\soutifcolour}[1]{\ifdefined\nocolour{}\else{\texorpdfstring{\sout{#1}}{#1}}\fi}

\newcommand{\as}[1]{\ifdefined\nocolour{#1}\else{\color{orange!70!black}{#1}}\fi}

\newcommand{\asfootnote}[1]{\ifdefined\nocolour{}\else{\as{\footnote{\as{ALEX: #1}}}}\fi}
\DeclareRobustCommand{\asout}[1]{\as{{\soutifcolour{#1}}}}

\newcommand{\zg}[1]{\ifdefined\nocolour{#1}\else{\color{blue}{#1}}\fi}

\DeclareRobustCommand{\zgout}[1]{\zg{{\soutifcolour{#1}}}}

\definecolor{abcolour}{rgb}{0.59, 0.48, 0.71}
\newcommand{\ab}[1]{\ifdefined\nocolour{#1}\else{\color{abcolour}{#1}}\fi}

\newcommand{\peter}[1]{\ifdefined\nocolour{#1}\else{\color{green!70!black}{#1}}\fi}

\newcommand{\jg}[1]{\ifdefined\nocolour{#1}\else{\color{orange}{#1}}\fi}
\newcommand{\jgnote}[1]{\ifdefined\nocolour{}\else{\color{orange}{JASPER: #1}}\fi}
\newcommand{\jgfootnote}[1]{\ifdefined\nocolour{}\else{\jg{\footnote{\jg{JASPER: #1}}}}\fi}
\DeclareRobustCommand{\jgout}[1]{\as{{\soutifcolour{#1}}}}

\newcommand{\markus}[1]{\ifdefined\nocolour{#1}\else{\color{pink!70!black}{#1}}\fi}

\newcommand{\markusfootnote}[1]{\ifdefined\nocolour{}\else{\markus{\footnote{\markus{MARKUS: #1}}}}\fi}

\newcommand{\intersect}{\cap}

\newcommand{\code}[1]{\texttt{#1}}

\def\NodeIDs{\nodeIDs}

\newcommand{\nodeIDs}{\ensuremath{\mathcal{N}}}
\newcommand{\nodeID}{\ensuremath{\alpha}}
\newcommand{\placeNode}{\operator{\textit{placeNode}}}
\newcommand{\nodes}{\operator{\textit{nodes}}}
\newcommand{\edges}{\operator{\textit{edges}}}
\newcommand{\places}{\operator{\textit{places}}}
\newcommand{\lifetimeprojs}{\operator{\textit{lfProjs}}}
\newcommand{\branchChoices}{\operator{\textit{branches}}}
\newcommand{\capabilities}{\operator{\textit{caps}}}

\newcommand{\future}{\texttt{FUT}}
\newcommand{\leaves}{\operator{\textit{leaves}}}
\newcommand{\roots}{\operator{\textit{roots}}}
\newcommand{\holes}{\operator{\textit{holes}}}
\newcommand{\comp}[2]{\ensuremath{#1}{\left[#2\right]}}

\newcommand{\nodeSet}{\ensuremath{S}}
\newcommand{\edgeType}{\ensuremath{t}}
\newcommand{\at}{\ensuremath{\;\mathtt{at}\;}}

\makeatletter
\def\node#1{\@ifnextchar\bgroup{\nodeTwo{#1}}{\nodeTwo{#1}{}}}
\newcommand{\nodeTwo}[2]{\roundBrackets{#1}_{#2}} % to do: make this pretty
\def\LPnode#1{\@ifnextchar\bgroup{\LPnodeTwo{#1}}{\LPnodeTwo{#1}{}}}
\newcommand{\LPnodeTwo}[2]{\angleBrackets{#1}_{#2}} % to do: make this pretty

\def\triple#1#2#3{\@ifnextchar\bgroup{\tripleFour{#1}{#2}{#3}}{\tripleThree{#1}{#2}{#3}}}
\newcommand{\tripleThree}[3]{\ensuremath{#1\;\stackrel{\footnotesize{#2}}{\Longrightarrow}\;#3}}
\newcommand{\tripleFour}[4]{\ensuremath{#1\;\stackrel{\footnotesize{#2{:}\;#3}}{\Longrightarrow}\;#4}}
\def\ghosttriple#1#2#3{\@ifnextchar\bgroup{\ghosttripleFour{#1}{#2}{#3}}{\ghosttripleThree{#1}{#2}{#3}}}
\newcommand{\ghosttripleThree}[3]{\ensuremath{#1\;\stackrel{\Longrightarrow}{\scriptstyle{#2}}\;#3}}
\newcommand{\ghosttripleFour}[4]{\ensuremath{#1\;\stackrel{\Longrightarrow}{\scriptstyle{#2{:}\;#3}}\;#4}}

\makeatother
\newcommand{\pcgop}{\texttt{op}}

\makeatletter
\def\edge#1#2{\@ifnextchar\bgroup{\edgeThree{#1}{#2}}{\edgeThree{}{#1}{#2}}}
\def\edgeThree#1#2#3{\ensuremath{#2\stackrel{\scriptscriptstyle{#1}}{\longrightarrow}#3}}

\def\operator#1{\@ifnextchar\bgroup {\operatorarg{\ensuremath{#1}}}{\ensuremath{#1}}}
\def\operatorarg#1#2{{#1}{\ensuremath{(#2)}}}
\def\spoperator#1{\@ifnextchar\bgroup{\spoperatorarg{\ensuremath{#1}}}{\ensuremath{#1}}}
\def\spoperatorarg#1#2{\ensuremath{#1\;#2}}

\def\fixedoperator#1{\@ifnextchar\bgroup {\fixedoperatorarg{#1}}{\ensuremath{#1}}}
\def\fixedoperatorarg#1#2{\fixedoperatorparse{#1}#2~}
\def\fixedoperatorparse#1#2,#3~{\ensuremath{{#2}{.}{#1}{(#3)}}}
\makeatother

\newskip \point \point =1pt
\setbox132=\hbox{\leavevmode\raise0\point\hbox{${\langle}\kern-2.5\point{\langle}$}}
\setbox133=\hbox{\raise0\point\hbox{${ \rangle}\kern-2.5\point{ \rangle}$}}
\def \angleBrackets#1{\copy132{#1}\copy133}
\setbox134=\hbox{\leavevmode\raise0\point\hbox{${(}\kern-2.5\point{(}$}}
\setbox135=\hbox{\raise0\point\hbox{${ )}\kern-2.5\point{ )}$}}
\def \roundBrackets#1{\copy134{#1}\copy135}
\setbox136=\hbox{\leavevmode\raise0\point\hbox{${\lfloor}\kern-3.0\point{\lfloor}$}}
\setbox137=\hbox{\raise0\point\hbox{${ \rfloor}\kern-3.0\point{ \rfloor}$}}

\setbox138=\hbox{\leavevmode\raise0\point\hbox{${[}\kern-1.5\point{[}$}}
\setbox139=\hbox{\raise0\point\hbox{${]}\kern-1.5\point{]}$}}

\renewcommand{\Vec}[1]{\ensuremath{\overrightharpoonup{#1}}}

\usepackage{listings}
\usepackage{subcaption}
\usepackage{tikz}
\usepackage{multirow}
\usepackage[allcommands, old-arrows]{overarrows}
\usepackage{dashbox}
\usepackage{algorithmicx}
\usetikzlibrary{arrows,shapes,positioning,calc,fit,decorations.pathmorphing}
\pgfdeclarelayer{background}
\pgfsetlayers{background,main}
\tikzset{
  octagon/.style= {
    shape=chamfered rectangle,
    draw,
    minimum width=.8in,
    chamfered rectangle angle=45,
    chamfered rectangle sep=3pt,
    text height=0.25ex,
    text depth=0.0ex,
  },
  remote/.style={
	draw,
	dashed,
	rectangle,
	minimum width=1cm,
	minimum height=0.5cm,
	align=center,
	font=\footnotesize
  },
  pcg/.style={
	draw,
	rectangle,
	minimum width=1cm,
	minimum height=0.5cm,
	align=center,
    fill=white,
	font=\footnotesize
  },
  lproj/.style={
    draw=blue,
    text=blue,
    octagon,
    minimum width=1.0cm,
    minimum height=0.5cm,
    align=center,
    fill=white,
	font=\footnotesize
  },
    hyperedge/.style={
		draw,
		dashed,
		rounded corners,
		inner sep=0.10 cm
	},
	abstract/.style={->, very thick, draw=black, decorate, decoration={snake, segment length=3mm}},
	coupled/.style={->, very thick, draw=black, decorate, decoration={snake, segment length=3mm}},
    expand/.style={->,thick, draw=black},
    outlives/.style={->, thick, draw=\outlivesedgecolour},
    borrow/.style={->, thick, draw=\borrowedgecolour},
    alias/.style={->, thick, dashed, draw=\aliasedgecolour},
}

\definecolor{DarkGreen}{RGB}{0,90,0}
\definecolor{DarkBlue}{RGB}{30,30,150}
\definecolor{DarkRed}{RGB}{160,20,10}
\lstdefinestyle{colouredRust}
{
	basicstyle=,
	identifierstyle=,
	keywordstyle=[1]\bfseries\color{DarkBlue},  % [1] reserved keywords
	keywordstyle=[2]\color{DarkGreen}, % [2] macro calls
	keywordstyle=[3]\color{DarkRed},   % [3] Prusti stuff
	keywordstyle=[4]\bfseries\color{DarkBlue},  % [4] primitive types
	keywordstyle=[5]\color{DarkGreen}, % [5] traits, types, variants etc
	keywordstyle=[6]\color{DarkGreen}, % [6] primitive values
	commentstyle=\color[gray]{0.4},
	stringstyle=\color{DarkRed},
	literate={!matches!}{!{\color{DarkGreen}matches!}}1,
	columns=spaceflexible,
	keepspaces=true,
	showspaces=false,
	showtabs=false,
	showstringspaces=true
}

\lstdefinestyle{boxed}{
	style=colouredRust,
	numbers=left,
	firstnumber=auto,
	numberblanklines=true,
	frame=trbL,
	numberstyle=\tiny,
	frame=leftline,
	numbersep=7pt,
	framesep=5pt,
	framerule=10pt,
	xleftmargin=15pt,
	backgroundcolor=\color[gray]{0.97},
	rulecolor=\color[gray]{0.90}
}

\lstdefinelanguage{Rust}
{
	sensitive,
	morekeywords=[1]{
		unsafe,async,await,move,
		use,pub,crate,super,self,Self,mod,
		struct,enum,fn,const,static,let,mut,ref,type,impl,dyn,trait,where,as,
		break,continue,if,else,while,for,loop,match,return,yield,in
	},
	morekeywords=[2]{
		assert!, assert_eq!, debug_assert!, debug_assert_eq!, debug_assert_ne!,
		panic!, unimplemented!, unreachable!,
		print!, println!, vec!, matches!
	},
	morekeywords=[3]{
		pure, trusted, ensures, requires, result, invariant, extern_spec,
		body_invariant!, predicate!,
		forall, old, exists,
		after_expiry, before_expiry, on_expiry, assert_on_expiry
	},
	morekeywords=[4]{
		bool,u8,u16,u32,u64,u128,i8,i16,i32,i64,i128,char,str,usize
	},
	morekeywords=[5]{
		Box, Option, Some, None, Ord, Equal, Less, Greater, Result, Ok, Err,
		T, Tree, Node, Empty, Timestamp, TryFromIntError, Utc, Height,
		MAX
	},
	morekeywords=[5]{
		true, false
	},
	morecomment=[l]{//},
	morecomment=[s]{/*}{*/},
	morecomment=[l]{///},
	morecomment=[s]{/*!}{*/},
	morecomment=[l]{//!},
	string=[b]{"},
	alsoletter={!},
}
\lstnewenvironment{rust}[1][1]{%
    \lstset{
    	language=Rust,
    	style=boxed,
    	basicstyle=\footnotesize\ttfamily,
    	escapechar=~,
		firstnumber=#1,
    	tabsize=2,
    }
}{}

\usepackage{algorithm}
\usepackage[noend]{algpseudocode}

\usepackage{fancyvrb}
\usepackage[normalem]{ulem}
\VerbatimFootnotes

\setcopyright{cc}
\setcctype{by}
\acmDOI{10.1145/3763122}
\acmYear{2025}
\acmJournal{PACMPL}
\acmVolume{9}
\acmNumber{OOPSLA2}
\acmArticle{344}
\acmMonth{10}
\received{2025-03-25}
\received[accepted]{2025-08-12}



\begin{document}

\title{Place Capability Graphs: A General-Purpose Model of Rust’s Ownership and Borrowing Guarantees}

\author{Zachary Grannan}
\orcid{0000-0002-7042-7013}
\affiliation{%
  \institution{University of British Columbia}
  \city{Vancouver}
  \country{Canada}
}
\email{zgrannan@cs.ubc.ca}

\author{Aurel Bílý}
\orcid{0000-0002-9284-9161}
\affiliation{%
  \institution{ETH Zurich}
  \city{Zurich}
  \country{Switzerland}
}
\email{aurel.bily@inf.ethz.ch}

\author{Jonáš Fiala}
\orcid{0009-0001-2121-7044}
\affiliation{%
  \institution{ETH Zurich}
  \city{Zurich}
  \country{Switzerland}
}
\email{jonas.fiala@inf.ethz.ch}

\author{Jasper Geer}
\orcid{0009-0006-6839-2305}
\affiliation{%
  \institution{University of British Columbia}
  \city{Vancouver}
  \country{Canada}
}
\email{jgeer35@cs.ubc.ca}

\author{Markus de Medeiros}
\orcid{0009-0005-3285-5032}
\affiliation{%
  \institution{New York University}
  \city{New York City}
  \country{USA}
}
\email{mjd9606@nyu.edu}

\author{Peter Müller}
\orcid{0000-0001-7001-2566}
\affiliation{%
  \institution{ETH Zurich}
  \city{Zurich}
  \country{Switzerland}
}
\email{peter.mueller@inf.ethz.ch}

\author{Alexander J. Summers}
\orcid{0000-0001-5554-9381}
\affiliation{%
  \institution{University of British Columbia}
  \city{Vancouver}
  \country{Canada}
}
\email{alex.summers@ubc.ca}

\begin{abstract}
  Rust's novel type system has proved an attractive target for verification and program analysis tools, due to the rich guarantees it provides for controlling aliasing and mutability. However, fully understanding, extracting and exploiting these guarantees is subtle and challenging: existing models for Rust's type checking either support a smaller idealised language disconnected from real-world Rust code, or come with severe limitations in terms of precise modelling of Rust borrows, composite types storing them, function signatures and loops.

  In this paper, we present \emph{Place Capability Graphs}: a novel model of Rust's type-checking results, which lifts these limitations, and which can be directly calculated from the Rust compiler's own programmatic representations and analyses. We demonstrate that our model supports over 97\% of Rust functions in the most popular public crates, and show its suitability as a general-purpose basis for verification and program analysis tools by developing promising new prototype versions of the existing Flowistry and Prusti tools.
\end{abstract}

\maketitle
\renewcommand{\shortauthors}{Z. Grannan, A. Bílý, J. Fiala, J. Geer, M. de Medeiros, P. Müller, A. J. Summers}

\section{Introduction}\label{sec:intro}
Rust is a popular programming language, combining competitive performance with a rich and expressive type system that eliminates many common programming errors and unintended side-effects. Rust's type system applies a notion of \emph{ownership} to all allocated memory, by default preventing access through any variable other than its owner, and defining its deallocation time (when its owner goes out of scope). Aliasing is restricted by a notion of \emph{borrowed references} which may temporarily access memory that they alias but do not own. As the name suggests, the ability to access borrowed memory is taken temporarily, and then restored to its lender. Rust allows multiple aliases to the same memory to be usable at once only if none can mutate its content.

The restrictive aliasing discipline of Rust's type system makes the language attractive as a static analysis and verification target, given the challenges of reasoning about aliasing in general (the infamous \emph{frame problem}). A wealth of formal techniques have grown up for Rust, including \as{formalisations}\asout{mechanisations in proof assistants} \cite{rustbelt,Oxide,rusthornbelt,aeneas}, (deductive) verification tools targeting specification-annotated Rust code \cite{prusti,creusot,verus-extended,Flux}, and \jgout{automated}static analyses \cite{Kani,Flowistry,Rudra}. These projects leverage the rich guarantees provided by Rust's type checking in the ways they internally model program memory, track aliasing and value information, and \jgout{(where appropriate)}interpret user-provided specifications of desired program properties.

\as{This paper does not define Rust's type-checking \emph{algorithms}, but rather defines a detailed model which substantially elaborates their \emph{results} in a way which reifies Rust's rich guarantees, and is exploitable for diverse program analysis and verification techniques and tools.}
Comprehensively modelling Rust's type checking is technically challenging for a variety of reasons. The properties implied at different program points are \emph{flow-sensitive}; for example, whether a variable currently owns memory is not visible from its declared type or its initialisation, as Rust allows \emph{move assignments} to transfer ownership between \emph{places} (Rust's terminology for expressions denoting locations). Similarly, whether a place has been borrowed-from, and for how long, are flow-sensitive properties. Rust's type checking includes \emph{borrow checking}, which determines whether consistent choices can be made for the extents of each borrow; the points at which borrows end are invisible in source code, but important for code understanding and analysis: this is when side-effects previously performed via the borrowed references become observable via the borrowed-from places.\asfootnote{This sentence might be a candidate for trimming.}

Capturing Rust's notions of borrowing for general types and programs yields several sources of complexity: (1) instances of composite types (such as structs) may store borrows. These types are declared with \emph{nested lifetimes}, \markus{which indicates to Rust's type checker} the possibly unbounded number of borrows stored within them; notions of borrowing are generalised to the sets of borrows stored in these types. (2) Rust allows for \emph{reborrowing}: transitively creating chains of borrows to parts of memory previously borrowed. These borrows can be created and expired differently in different paths through the program, so precisely tracking their effects requires \emph{path-sensitive} information. (3) Programs with loops can create (re)borrows unboundedly many times. To capture Rust's type checking in such cases, loops must be summarised in a way which matches the compiler's rules. (4) \markus{A model of Rust's borrowing must respect the modularity of Rust function calls, including the subtle ways in which Rust allows programmers to abstract over constraints between (nested) lifetimes in function signatures. }
We explain the full details of these (and other) challenges in \secref{motivation}.

\begin{table}[t]
\caption{Most-closely-related work, compared on the criteria motivating our model and paper.}\label{tab:relatedtable}

\setlength{\tabcolsep}{4pt}
%\newcolumntype{d}[1]{D{.}{.}{#1}}
%\newcommand{\h}[1]{\multicolumn{1}{c}{#1}}
%\newcommand{\hb}[1]{\multicolumn{1}{c|}{#1}} % used for adding extra lines; not really needed in this example.
\begin{center}
\vspace{-6pt}
\scalebox{0.75}{
\begin{tabular}{l|ccccc|cccc}
&\multicolumn{5}{|c}{Our Key Criteria (\cf{} page \pageref{ourkeycriteria})}&\multicolumn{4}{c}{Other Borrowing Challenges (\cf{} page 1)}\\
\midrule
Selected&(1) Function-&(2) Path-  &(3) Connect&(4) Most  &(5) Use for&Struct    &Loop       &Function   &Defines\\
Existing&Modular    &Sensitive  &from + to &Real-world  &Diverse   &Lifetime   &Reborrow   &Reborrow   &Borrow\\
Work    &Model      &Info.      &Rust repr.&Rust Code   &Analyses  &Params     &Summary    &Summary    &Checking\\
\midrule
Oxide \cite{Oxide}&\checkmark&  &${}^1$       &           &          &           &\checkmark &\checkmark &\checkmark \\
Aeneas \cite{Ho'24}&\checkmark& &${}^1$&           &\checkmark&           &${}^2$     &\checkmark &\checkmark \\
RustBelt \cite{rustbelt}&\checkmark&\checkmark& &        &          &\checkmark &           &\checkmark &\checkmark \\
\kern-4pt\begin{tabular}{l}Stacked / Tree\\ Borrows \cite{stacked-borrows,tree-borrows}\end{tabular}& &         &\checkmark&\checkmark  &          &\checkmark &           &            & \\
PCGs    &\checkmark &\checkmark &\checkmark&\checkmark&\checkmark&\checkmark&\checkmark&\checkmark& \\
\end{tabular}
}
\end{center}

\scalebox{0.75}{
\noindent
${}^1$ compilation \emph{from} Rust only; results for custom program representations only.
${}^2$ partial heuristic support: see \cite{Ho'24} for details.
}
\end{table}

In this paper, we present a novel model of the results of Rust's type checking, designed to address all of these technical challenges and to enable downstream tools to exploit the properties implied by Rust's design. Our model does \emph{not} define Rust type checking itself, but captures in detail the \emph{results} of \markus{successfully} type checking a Rust program as defined by the standard Rust compiler. It seamlessly combines solutions to two technical objectives: (1) where computable, all information a downstream analysis might require is \emph{precise}, and (2) where abstractions are necessary in general either for computability or modularity reasons (for loops, function calls and types containing sets of borrows)\asfootnote{shortened but didn't drop: the aim here is to differentiate where we aim for precision or abstraction.}, these abstractions are consistent with those employed by the compiler. \as{While many existing works define a variety of formal models for different aspects of Rust's design (we discuss these in detail in \secref{related}), these do not handle the full set of challenges defined above. We provide a high-level overview of how some related work compares in terms of the objectives of our own work in \tabref{relatedtable}. We note that many existing works have objectives we do not address, such as defining operational semantics for Rust, or defining how type checking itself can/should be done.}

\as{Overall, we have designed our novel model to satisfy the following criteria:}
\begin{enumerate}\label{ourkeycriteria}
\item The model provides a detailed, function-modular account of \emph{how and why} a program type-checks, tracking at each program point in a function's body which places own and borrow which memory, which are currently borrowed-from and how, and when \emph{and how} these borrowed capabilities (to read or write) flow back to their origins. Our model uses a novel representation for sets of borrows, allowing us to connect to the compiler's borrow-checking concepts, while abstracting over the concrete details of any given borrow-checker algorithm or implementation, making it possible to connect to alternative Rust borrow checkers.
\item The model captures precise path-sensitive information for non-recursive control flow, and summarises function calls and loops with abstractions that precisely reflect the shape of constraints imposed by borrow checking (this turns out to be complex for loops that reborrow memory). The former grants consumers the freedom to demand as much information as desired, while the the latter allows supporting general Rust code (e.g.~that following a summarised loop) while providing consumers with an abstraction that could be enriched by further analysis or annotations. In particular, our loop and function summaries are suitable for connection with user specifications in a deductive verification setting.
\item All \jgout{ingredients}parts of the model have a clear and easily-checkable correspondence with notions in the Rust compiler, \jgout{including}\eg{ the program representation and type-checking features}. Consequently, the model can be used to visualise and understand\jgout{the compiler's} type checking itself. This close connection avoids potential gaps between the de-facto definition and semantics of Rust as defined by the compiler and the representations and information comprising instances of our model.
\item Our model handles a large fragment of real-world Rust code (over 97\% of the functions in the top-500 crates), and is automatically generated from compiler runs, ensuring the practical applicability of tools built upon our model, and demonstrating its generality in practice.
\item The model is suitable for consumption by a wide variety of static analysis and verification tools, providing at least the information they currently require, and potentially extending their reach to a wider class of Rust programs; this makes its definition a significant advance for designing and implementing detailed Rust analysis and verification tools. %In addition, although conceptually closely related to borrow-checker information, the model can be connected to a variety of borrow-checker algorithms and implementations.
\end{enumerate}
\as{As we explain in\zgout{ detail in} \secref{related} and illustrate in \tabref{relatedtable}, existing models of Rust's type checking do not meet multiple criteria from this list. Conversely, many existing verification and analysis tools could\zgout{simply} extract the information they rely on from our model: our work makes building\zgout{new (possibly sophisticated)} static tools for Rust a substantially simpler task.}
In summary, this paper's key contributions are as follows:
\begin{enumerate}
\item We define and present \emph{Place Capability Graphs} (PCGs), as a detailed and general-purpose model of Rust's type checking satisfying the full criteria above; in particular, the model directly builds upon the Rust compiler's programmatic representations and analysis results.
\item We provide an implementation of our model, written as a standalone reusable artefact (crate) in Rust; this implementation is open-sourced~\cite{PcgGithub} and the artefact is available online~\cite{PcgArtefact}.

\item We evaluate our model's generality by applying it to the entire codebases of the top-500 Rust crates at time of writing\zgout{(along with a collection of challenging hand-crafted examples)}; we demonstrate support for over 97\% of the published real-world code, and successfully generate models for all program points in these functions.

\item We employ a form of mutation testing to generate Rust programs that the compiler should \emph{not} accept according to our model, and verify that the restrictions captured by our model match the compiler's results; \markus{this provides} confidence that our model accurately captures the \emph{restrictions} imposed by Rust's type system in terms of memory accesses it disallows.

\item We develop new prototype versions of two existing Rust analysis tools as consumers of our model: the information-flow analysis Flowistry~\cite{Flowistry} and the deductive verifier Prusti~\cite{prusti}. Our PCGs\zgout{successfully} replace Flowistry's alias analysis and Prusti's proof annotation algorithm. Integrating our PCGs dramatically simplified aspects of Prusti's implementation; all required annotations to drive its formal proofs are deterministically read off from our\zgout{novel} model.
\end{enumerate}

Our paper\footnote{A version of our paper as a technical report is also available on arXiv~\cite{PcgArxiv}.}
is organised as follows. \secref{motivation} presents specific challenges that arise when reasoning about\zgout{the full complexity of} Rust ownership and borrowing, also presenting\zgout{ the reader with} key background on Rust itself. \secref{approach} introduces our PCG model and defines the rules governing its construction based on information\zgout{extracted} from the Rust compiler. Our evaluation in \secref{eval} demonstrates the accuracy and wide applicability of our model. In \secref{related}, we compare our model to related work, and hypothesise how\zgout{information provided by} our model could benefit existing analysis and verification tools. We conclude and discuss future work in \secref{conclusion}.

%
%
%Furthermore, although the Rust compiler checks that
%
%
%
%analysis tools must also model these sets and integrating this with precise reasoning about individual locations and borrows mutated. Existing tools use the most diverse
%
% be concerned with a potentially-unbounded number of borrows,
%
%, in order to correctly reflect both when which capabilities are
%

%   \input{sections/intro.tex}

% \section{Background}\label{sec:bg}
%   \input{sections/bg.tex}

\section{Capturing Rust}\label{sec:motivation}
In this section, we provide an overview of Rust and present the challenges of modelling Rust's type checking\footnote{We will use ``type checking'' to refer to both classical type checking concerns and Rust-specific notions of borrow checking.} to be both useful to downstream consumers and accurately reflect the compiler.

% describe the challenges that arise when
% We illustrate these challenges

% - this is hard because borrowing is complex, especially when it interacts with control flow, loops, and complex data structures

% - in some sense the goal of the model is to explain the borrow checker
% - source rust includes things like lifetimes, but these are a simplified abstraction provided to the user, and can be omitted and inferred by the compiler instead
% - it is non-trivial to explain HOW and WHY a rust program borrow-checks

% - we defined some criteria which we claim our model satisfies, here is why it is challenging to do so

% - TODO explain why capabilities are a natural way to talk about the Rust type system, cite prior work, say that we will use the term capabilities in the paper to talk about what exactly is transferred

\subsection{Ownership}\label{sec:ownership}

As explained in the introduction, ownership is a core concept in Rust's type system.
\markus{In the simplest case, \emph{move} assignments transfer ownership of the source's memory into the destination. }
\markus{Consider the program \code{replace\_x\_owned} in \figref{motivation:move}, whose assignments are all \emph{move}. }
\markus{Initially, \code{pos} owns an instance of the \code{Pos2D} struct, and \code{new\_x} owns a \code{T}}\footnote{\markus{Unlike in languages like Java/C++, function arguments in Rust always refer to owned, initialised, and distinct memory.}}.
\markus{For many Rust types (including \code{Pos2D}) the ownership of a struct equates to owning each field, which can be moved into and out of individually. }
\markus{The move in \lineref{motivation:partialmove} transfers ownership \emph{only} from \code{pos.x} to \code{old\_x}, therefore, attempting to \jg{read from} \code{pos.x} or \code{pos} immediately after this line will cause a type-checking error. }

\begin{figure}[t]
\begin{minipage}[t]{0.48\textwidth}
\begin{subfigure}{\textwidth}
    \begin{rust}
struct Pos2D<T> { x: T, y: T }

fn replace_x_own<T>(mut pos: Pos2D<T>,
                    new_x: T) -> Pos2D<T> {
    let old_x = pos.x;~\label{line:motivation:partialmove}~
    pos.x = new_x;~\label{line:motivation:unpartialmove}~
    return pos;~\label{line:motivation:partialmovereturn}~
}
\end{rust}
\vspace{-0.3cm}
\caption{}%PA simple function to illustrate Rust's move assignments and ownership of composite types. \jgnote{added an explicit return statement for clarity}}
\vspace{-0.3cm}
\label{fig:motivation:move}
\end{subfigure}
\end{minipage}
\;
\begin{minipage}[t]{0.48\textwidth}
\begin{subfigure}{\textwidth}
    \begin{rust}
fn replace_x<'a, T>(pos: &'a mut Pos2D<T>,
                    new_x: T) {
    let x_ref = &mut (*pos).x;~\label{line:motivation:reborrow}~
    *x_ref = new_x;~\label{line:motivation:derefborrow}~
}
fn caller(mut original: Pos2D<i32>) {
    let pos = &mut original;~\label{line:motivation:mutborrow}~
    replace_x(pos, 0);
}
\end{rust}
\vspace{-0.3cm}
\caption{}%A variation of the example to illustrate mutable borrows and in-place mutation. \jgnote{Added explicit lifetime parameter 'a to fit the prose}}
\vspace{-0.3cm}
\label{fig:motivation:borrowexample}
\end{subfigure}
\end{minipage}
\caption{Examples to showcase Rust move assignments and in-place mutations.}
\end{figure}
%
% as illustrated by the function \texttt{replace\_x\_owned} in \figref{motivation:move}.
% \footnote{In this and future examples, we use \texttt{String} for concreteness where any non-copy Rust type would work just as well.}.

% \footnote{For several built-in primitive types such as integers, Rust treats assignments as \emph{copy assignments}: both source and target end up owning distinct copies of the value assigned. Move assignments are the default and more-common case.}
% ; it  despite its declared type, \code{pos} no longer has this capability according to Rust's rules.

\lineref{motivation:unpartialmove} also transfers ownership, this time from \code{new\_x} to \code{pos.x}.
In combination with the never-removed ownership to \code{pos.y} \markus{this move reconstitutes} a complete \code{Pos2D} instance in \code{pos}, \markus{which is then free to be returned to the caller on \lineref{motivation:partialmovereturn}. }
If \lineref{motivation:unpartialmove} were omitted, this \code{return} would fail to type check \markus{for the same reason we couldn't read from \code{pos.x} immediately following \lineref{motivation:partialmove}. }

%
%
%Initially, \texttt{pair} has \emph{full ownership} of a \texttt{Pair} struct and transitively, ownership of the \texttt{String}s held in \texttt{pair.fst} and \texttt{pair.snd}.
%The assignment in \lineref{motivation:partialmove} then transfers ownership of the \texttt{String} held in \texttt{pair.fst} to \texttt{old\_fst}.
%This move assignment is a \emph{partial move}; \texttt{} retains partial ownership of the \texttt{Pair}, specifically full access to the \texttt{String} held in \texttt{pair.snd}, but owns nothing through \texttt{pair.fst}.
%Critically, Rust now prohibits the function from moving \texttt{pair}.
%This ability is restored along with full ownership in \lineref{motivation:unpartialmove} when \texttt{new\_fst}'s value is moved into \texttt{pair.fst}, and the function uses this to return \texttt{pair} to the caller.
%
%Thus, for a model of Rust's type-checking to explain \emph{why} a program obeys its ownership discipline, the model must keep pace with changes in ownership across program points, tracking ownership in a fine-grained manner for composite data.
%Specifically, we should be able to observe both the decomposition of ownership of a data structure into ownership of its parts as well as the restoration of full ownership from ownership of its parts.

% Takeaways:
% - this is flow-sensitive
% - capabilities are intro'd later

% - ``while you are the owner of memory, you can read and write it freely'' (by default)

\subsection{Borrowing, Borrow Extents, and Borrow Checkers}
\label{sec:motivation:borrows}
% Dependencies:
% - Ownership

% TODO how does this change the story
% TODO we are not giving out ownership
% - when something is borrowed, the owner can't do everything they did before
% - the borrower can do something for a period of time
% - something is being handed to and fro, the thing that is handed around is a capability
% - we can restate the rules in these terms
% - a borrow is handing a capability off, borrower can use until compiler says the borrow is over
% - TODO mention that in practice, the compiler picks when borrows end

\markus{Rust} permits \markus{a controlled form of aliasing} through \emph{borrowing}, with which \markus{the exclusive ability to read or write to a piece of} memory can be \markus{loaned to a reference aliasing it}.
\markus{The function \code{replace\_x} in \figref{motivation:borrowexample} implements an in-place variant of \code{replace\_x\_owned} using \emph{mutable borrows}. }
\markus{Recall that in \code{replace\_x\_owned} we were able to mutate the field \code{x} because the function owned its \code{pos} argument. }
\markus{In \figref{motivation:borrowexample} \code{pos} does not own its target (whose owner is on the call-stack), but still has the exclusive ability to read and write to it. }
% \markus{Instead, because \code{pos} is mutable borrow, it comes referent which is owned by some other variable. }
\markus{This allows \code{replace\_x} to mutate a \code{Pos2D} struct in-place}.
% enabling in-place mutation as illustrated by the variant of the previous example shown in . Like the example given in \figref{motivation:move}, \texttt{replace\_x} replaces the first element of a \code{Pos2D}. Rather than moving values in and out, the function is given a \emph{mutable borrow} which (for the duration of the borrow)  Such borrows can be created as on \lineref{motivation:mutborrow}; the line afterwards passes this borrow as a function argument.

\markus{The mutation in \code{replace\_x} occurs under a \emph{reborrow}.}
\markus{On \lineref{motivation:reborrow} a new borrow \code{x\_ref} temporarily obtains the exclusive ability to write to a field of the target of \code{pos}, which it does on \lineref{motivation:derefborrow}.}\footnote{Rust allows many dereference operators to be elided, but in this section we make them explicit to aid understanding.}
\markus{Like the previous example, \code{x\_ref} only borrows from the \code{x} field of the underlying struct,} and neither \code{(*pos).x} nor \code{*pos} can \markus{be read from right after \lineref{motivation:reborrow} (though \code{(*pos).y} can). }
\markus{Note that it is not ownership of \code{*pos} being lent, but instead the exclusive ability to read and write to it. }
%\markus{In this example, the error is due to the temporary transfer of read and write access to the underlying place \code{original}}, which retains ownership throughout the execution of this function.
Following prior literature on related type systems \cite{Boyland01,concurrency-control,pony,odersky-capabilities,safe-parallelism,AEminium} \markus{in this paper} we refer to these rights as \emph{capabilities}: the (possibly unique) ability to read or write to the underlying borrowed memory. \markusfootnote{Removed sentence about shared borrows: ``Rust's notion of \emph{shared borrows} allows the flexibility of borrowing only read capability; while such borrows are live the borrowed-from places also retain only read capability.
'', could be put back in but it is out of place.  }

\paragraph{Borrow Extents and Borrow Checking.}
The Rust compiler assigns each borrow a \markus{contiguous portion of the program (in the sense of control flow) for which the borrow is allowed to use or loan out the capabilities to its target.}
\markus{We refer to these as \emph{borrow extents}. }
Some borrow extents are annotated in Rust code by \emph{named lifetimes} such as \ticktt{a} in our example above, though in general a Rust program contains vastly more borrow extents than it does named lifetimes.\footnote{Additional borrow extents can arise for a variety of reasons: (1) most lifetimes are elided in source programs, although elaborated types (with lifetimes) are explicit in Rust's Mid-level Intermediate Representation (MIR), (2) even after elaboration, some relevant lifetimes do not appear in the declared types of any variable, but are associated only with temporary sub-expressions, (3) Rust's newer borrow-checking algorithms, in particular Polonius \cite{Polonius,Polonius-update}, work in terms of finer-grained notions of borrow extents than named lifetimes, distinguishing extents for separate sets of borrows that correspond to the same lifetime, \eg~when the same variable is assigned many times. The Polonius term for its notion of borrow extent is an \emph{origin}; we use our terminology to remain generic over different borrow checkers' implementation choices. }

Borrow extents can be constrained \markus{relative to each other}.
For example, a new reborrow (like \code{x\_ref} in our example) \markus{cannot have a longer extent than the borrow it was created from.}
\markus{In addition to these \emph{outlives constraints}, usages of a borrow's target will constrain borrow extents to include a set of program points.}
\markus{\emph{Borrow checking} is the process of ensuring that there exists a solution to a system of borrow constraints.}
\markus{There are multiple borrow checking algorithms for Rust which provide varying degrees of detail about the system's solution, and in fact can describe different solutions altogether! }
Precisely modelling borrow extents (in particular, when borrows end) is critical for downstream analyses that reason precisely about how memory is accessible, aliased and updated via chains of borrows.
For tools which build their analysis around how borrowed capabilities flow back to their lenders (e.g.~Prusti~\cite{prusti} and Aeneas~\cite{aeneas}), and any other analyses demanding precise aliasing information, it is essential to model \markus{how chains of borrows are related to each other, and how the capabilities from those borrows transfer when those borrows expire. }

\subsection{Path-Sensitivity}
\label{sec:motivation:pathsensitivity}
% Dependencies:
% - Borrowing
% - Ownership

% - for owned things that are moved out in a path sensitive manner, we don't really track path sensitive data
% - borrows/moves(?) may be created in one branch and not the others - this can lead to interesting constraints in the borrow checker that our model needs to be able to explain
% - what if borrows are expired in a different order in each branch? (example) - we need to explain this
% - a dataflow analysis like flowistry needs to know about the dependencies created in each branch
% TODO

\begin{figure}[t]
\begin{rust}
fn dec_max(mut pos: Pos2D<i32>, mut z: i32) {
    let max2 = if pos.x > pos.y { // c0/c1~\label{line:motivation:first-if}~
        &mut pos.x                // bb1 (c0)
    } else {
        &mut pos.y                // bb2 (c1)
    };                            // bb3~\label{line:motivation:flowjoin}~
    let max3 = if z > *max2 {     // c2/c3
        pos.x = 0;                // bb4 (c2) max2 borrow must end here!~\label{line:motivation:max2-borrow-end}~
        &mut z~\label{line:motivation:chained-borrow-1}~
    } else {
        &mut *max2                // bb5 (c3)~\label{line:motivation:chained-borrow-2}~
    };
    *max3 -= 1;                   // bb6
}
\end{rust}
  \caption{An example designed to illustrate path-sensitivity with respect to (re)borrowing. \jg{The code is annotated with basic block and branch identifiers for use in \figref{pathsensitive-example}.}}
  \label{fig:motivation:flowsensitive}
\end{figure}

Although Rust's type checking is flow insensitive in terms of \emph{outcomes} (\eg~whether a place currently has capability to read or write its underlying memory does not depend on the control flow taken), the borrows at a given program point can depend on control flow.
Rust analyses concerned with \emph{side-effects} (\eg~verifiers such as Prusti~\cite{prusti} and Creusot~\cite{creusot}) require a model \as{capturing} flow-sensitive borrow information on top of the compiler's more conservative, flow-insensitive checks.

The function \code{dec\_max} given in \figref{motivation:flowsensitive} is designed to illustrate the differences between a flow-sensitive and flow-insensitive model of borrowing.
% (in realistic code the control flow would be more complex and the branching often idiomatically due to pattern-matching).
%
%%
%%
%%
%%We just observed that a lifetime might encompass one of multiple possible borrowed references.
%%However, a lifetime
%%just seen that although the compiler uses lifetimes as a bookkeeping mechanism for borrows, borrows and lifetimes are not necessarily in one-to-one correspondence.
%%
%%there is not a one-to-one correspondence between lifetimes and the extents of borrows
%
%\texttt{f} declares \texttt{c}, but leaves its initialization until the branches of the following \texttt{if} expression.
At \lineref{motivation:first-if}, \code{max2} borrows from a different field of \code{pos} depending on which branch of the \code{if} expression is taken.
Following the join point at \lineref{motivation:flowjoin} \code{max2} is guaranteed to be initialised \markus{no matter which path was taken. Since} its target \code{*max2} could borrow capabilities from either \code{pos.x} or \code{pos.y} the compiler will consider both \code{pos.x} and \code{pos.y} to be borrowed-from.
In reality, \code{max2} borrows only from one of \code{pos.x} or \code{pos.y}, and an analysis tool such as a verifier may need to distinguish \emph{which} and \emph{when}.
Borrows created \emph{before} a control-flow branch may be forced to end during the branches themselves; \markus{it is even possible that this occurs in one branch and not the others. }
\markus{For example,} at \lineref{motivation:max2-borrow-end} the borrow stored in \code{max2} is forced to end in order to access \code{pos.x}, while the same borrow must be live in order to be reborrowed at \lineref{motivation:chained-borrow-2}.

Since code within branches affects constraints between borrows outside the branches, one might assume a precise path-sensitive analysis must analyse each path through a function separately.%\footnote{Indeed, this concern is used to motivate the less-precise join definition in the Aeneas model for Rust \cite{Ho'24}.}
\as{We will show that by careful design of our model, this is not the case}.
By \as{exploiting} the fact that the validity of all memory accesses depends only on the compiler's path-insensitive analysis, we are able to efficiently combine our analysis state at each join point while retaining the ability to produce path-sensitive models of the aliasing relationships present in virtually all real-world Rust code.

\subsection{Function Calls}
\label{sec:motivation:functioncalls}

\begin{figure}[h]
\begin{rust}
fn max<'r, 'a: 'r, 'b: 'r>(rx: &'a mut i32, ry: &'b mut i32) -> &'r mut i32 {
    if *rx > *ry { &mut *rx } else { &mut *ry }
}

fn dec_max_alt<'a>(pos: &'a mut Pos2D<i32>) {
    let rx = &mut pos.x;
    let ry = &mut pos.y;
    let res = max(rx, ry);~\label{line:motivation:call-max}~
    *res -= 1;~\label{line:motivation:res-dies}~
}
\end{rust}
  \caption{A function to illustrate modular type checking of calls concerning reborrowing. \jgnote{Added explicit outlives constraints to fit the existing prose}}
  \label{fig:motivation:functioncall}
\end{figure}

\markus{To enable in-place mutation without sacrificing modularity, Rust allows functions to be parameterised by lifetimes which span their entire call}.
\markus{Based on the function signature, Rust computes constraints between the extents of borrows which are passed into the function and the extents of borrows which will be returned}.
\markus{From a caller's perspective, this summary describes every relationship that can be known about the borrows returned by the function call. }
\markus{Conversely, the function body must ensure that it does not impose any additional constraints between its argument and return borrows. }
\markus{A model of Rust's type checking should obey these same principles. }

As an example, consider the program given in \figref{motivation:functioncall}. \markus{The} function \code{max} is parameterised by three \markus{named lifetimes} (\ticktt{a}, \ticktt{b} and \ticktt{r}, for the extents of the two input parameters and output parameter, respectively). The syntax \code{\ticktt{a}:\ticktt{r}} declares an outlives constraint: the extent of the borrow stored in \code{rx} \markus{must} last at least as long as the extent of the borrow returned by \code{max}. These constraints are sufficient to allow \code{rx} or \code{ry} to be reborrowed in the output of the function.

At \code{max}'s \emph{call-site} on \lineref{motivation:call-max}, the compiler sees from the outlives constraints in the function signature that \markus{the body was free to reborrow the output from either argument}, so treats the returned reference as if it reborrowed from both.
The returned borrow therefore gets an extent which ends after \lineref{motivation:res-dies}.
At this point, \markus{both} intermediate borrows \code{rx} and \code{ry} become usable again\footnote{Technically, both arguments to \verb+max+ are moved in and \verb+rx+/\verb+ry+ should not be usable after the call. The compiler avoids this by moving in temporary reborrows of these (\verb+&mut *rx+/\verb+&mut *ry+) instead. We elide this here for simplicity, but our implementation handles this correctly.}.

\markus{Interactions between different Rust features make function calls involving lifetimes challenging to model. }
\markus{In general,} (1)~Rust types can use lifetimes as generic parameters: \eg~the type \code{\&\ticktt{a} mut Pos2D<\&\ticktt{b} mut i32>} denotes a mutable reference to a struct instance that itself stores references (whose extents may differ), (2)~outlives constraints are often implicit (\eg~usage of a type \code{\&\ticktt{a} mut Pos2D<\&\ticktt{b} mut i32>} implicitly imposes \code{\ticktt{b}:\ticktt{a}} to prevent live references from reaching expired borrows), (3)~outlives constraints interact with \emph{variance rules} for Rust types \markus{which themselves have lifetime parameters}.
\markus{We remark that} modular summaries of calls must still connect with fine-grained reasoning about the borrows explicitly manipulated in the calling context.

\subsection{Loops}

\markus{Accounting for the borrowing that occurs across a loop raises further challenges}.
Consider the program given in \figref{motivation:loop}.
\markus{The} function \code{penultimate\_mut} returns a mutable reference to the second-to-last element of a list via the loop at \lineref{motivation:loopbegin}, at each iteration borrowing \code{(*current).head} into \code{prev} and advancing \code{current} to point to the next node.
Like the example in \figref{motivation:flowsensitive}, the borrow stored in \code{prev} at \lineref{motivation:loopbegin} (if any) depends on which branch \markus{was taken}.
Since the number of loop iterations depends on the length of the list, the exact target of \code{prev} cannot be determined statically.
% \markus{Since} the \markus{number of times the} loop body will be entered \markus{depends on} on the length of \code{list}, the exact target of \code{prev} at the join point cannot be determined statically.

As with function calls, it is not possible to maintain precise borrowing information across zero or more iterations of a loop.
Just as \markus{function signatures summarise the borrowing relationships across the call}, modelling type checking across loops
requires accurately summarizing the effect of arbitrarily many loop iterations: a loop invariant, for some suitable notion
of the term.

% If we lost \emph{all} such information, however, it would not be possible to justify why the body of the loop at \lineref{motivation:loopbody} type-checks.
% Examining only the information given in the function signature, we see that \tick{\texttt{a}} outlives \tick{\texttt{b}}.
% The conservative assumption would be that our mutable borrow with lifetime \tick{\texttt{b}} blocks the borrow with lifetime \tick{\texttt{a}}, but at the loop head in \lineref{motivation:loopbegin}, the function destructs \texttt{current.tail}, reborrowing its contents into \texttt{tail}.
% Subsequently, the function writes to \texttt{acc}, whose lifetime, conservatively, we would have assumed to be expired.
% Instead, we can observe the loop invariant that \texttt{acc} borrows from \texttt{current.head}.
% As a result, \texttt{acc} does not block \texttt{current.tail}, and the operations performed within the loop are legal.
% The borrow checker constructs a similar argument by (TODO something about how the borrow checker has some notion of computing a loop invariant?).
% Thus, not only must we preserve borrowing structure across loops, but our analysis must also compute a loop invariant structure that is at least as precise as the borrow checker.

\begin{figure}[t]
\begin{rust}
struct List<T> { head: T, tail: Node<T> }
type Node<T> = Option<Box<List<T>>>;

fn penultimate_mut<'a>(list: &'a mut List<i32>) -> Option<&'a mut i32> {
    let mut current = &mut *list;
    let mut prev = None;
    while let Some(next) = &mut (*current).tail {~\label{line:motivation:loopbegin}~
        prev = Some(&mut (*current).head);~\label{line:motivation:loopbody}~
        current = &mut *next;
    }~\label{line:motivation:loopjoin}~
    prev
}
\end{rust}
\caption{A function that takes a mutable reference to a list and returns a \emph{reborrow} to its second-to-last element}
\label{fig:motivation:loop}
\end{figure}

\subsection{Nested Lifetimes}
% Dependencies:
% - Borrowing
% - Ownership
% - Function Calls
% - Loops

% TODO figure out where to put this
% - structs themselves can contain unbounded numbers of borrows
% - this further complicates the previous challenges

% when reasoning about borrows that we manipulate directly we want full precision
% once we get to a point where there is unboundedness in types, control flow, modularity
% - we are not precise, but as precise as the compiler is
% - we need to be able to connect these two parts
% - this is challenging because the part of the compiler that deals with these abstractions doesn't deal in places

% - easiest possible example is the function that returns the nth element of a list
% - this is a separate example from the ``sets can change'' example
% - might be worth adding a footnote somewhere that you need std::mem::swap to actually change the sets

\begin{figure}[t]
    \begin{rust}
fn replace_x_rf<'a, 'b>(rf: &'a mut Pos2D<&'b mut i32>, new_x: &'b mut i32) {
    (*rf).x = new_x;~\label{line:motivation:replace-x}~
}

fn cl<'a>(r1: &'a mut i32, r2: &'a mut i32, r3: &'a mut i32, r4: &'a mut i32) {
    let f = Pos2D { x: r1, y: r2 }; // f: Pos2D<&'a0 mut i32>
    let rf = &mut f; // rf: &'a2 mut Pos2D<&'a1 mut i32>~\label{line:motivation:borrow-f}~
    (*rf).y = r3;~\label{line:motivation:replace-y}~
    replace_x_ref(rf, r4);
    *f.x = 0;~\label{line:motivation:call-swap}~
}
\end{rust}
  \caption{Rust functions demonstrating nested borrows. The call to \texttt{replace\_x\_rf()} changes the target of \texttt{f.x}.}
  \label{fig:motivation:nestedborrows}
\end{figure}
\markus{Modelling} borrows of types which in turn contain borrows \markus{raises many previously stated challenges}.
Consider the example given in \figref{motivation:nestedborrows}. The first argument
to \texttt{cl} is a borrow of the type
\texttt{Pos2D<\tick{\texttt{a}} mut i32>}, a type that itself contains a
\emph{nested} borrow with lifetime \tick{\texttt{a}}.
An analysis or verification tool might need to know the set of memory locations which the write on \lineref{motivation:call-swap} could have affected.
Prior to the call, it
is straightforward to identify that \texttt{(*rf).x} transitively borrows
from the target of \texttt{r1} (the memory located at \texttt{*r1}).
However, because \texttt{replace\_x\_ref} is allowed to change the targets of \texttt{*rf}, the target of \texttt{f.x}
cannot be known precisely after the call \markus{without sacrificing modularity}.
A similar situation arises when types with nested lifetimes appear across loops.

This situation only arises for nested lifetimes: \eg~it is not possible for \texttt{replace\_x\_ref} to change the target of \texttt{rf} itself (which memory \texttt{rf} points to at the \texttt{cl} client side). Thus, our model must
soundly over-approximate the set of capabilities borrowed from by types with nested lifetimes,
while retaining more precise information for types without.

% TODO explain that the model needs to adjust the sets of borrows reachable for parameters with compatible nested lifetimes, but in a precise enough way that we can still observe disjointness for other parameters.

% - this might also occur with loops

\section{Place Capability Graphs}\label{sec:approach}
\as{In this section, we introduce Place Capability Graphs (PCGs), whose main components are shown in \figref{approach:overview}, and demonstrate
how they address the problems presented in \secref{motivation}.
A PCG is a directed hypergraph over two kinds of nodes. \emph{Place nodes} associate a Rust \emph{place} (expression denoting a memory location \cite{place-expressions}) with \emph{capabilities}}
%
%
% that defines the organisation of capabilities to access memory, current borrows and the relationships between borrow extents relevant for understanding subsequent code.
%\figref{approach:overview} provides an overview of its different components, which we will introduce throughout this section. \emph{Place nodes} associate Rust places with \emph{capabilities}
(\figref{approach:capabilities}), describing the actions allowed on these places%\footnote{In addition to the
%capabilities listed in \figref{approach:capabilities}, our model includes a capability
%ShallowExclusive (\capshallowex), which is used to track the initialisation
%state of \code{Box} types (representing a \code{Box} that owns currently
%uninitialised memory).}
.
A place node with \emph{exclusive} (\capexclusive)~capability can be freely read from, written to,
borrowed and moved. Read (\capread{}) capability allows
reading or immutable borrowing only;
write (\capwrite{}) capability allows assignment only. The \capnone{} capability allows no actions (typically due to an outgoing PCG edge).

The second type of PCG nodes are \emph{lifetime projection} nodes. Lifetime projections $\lproj{p}{\tick{a}}$ (in which \tick{a} is a lifetime from the type of $p$) are a novel construct used to represent a set of borrows currently \as{\emph{stored in} the} place $p$ \as{and whose borrow extents are defined by the borrow checker's current interpretation of lifetime \tick{a} for $p$.}
%whose borrow extents have not yet ended, and cannot end before (a) have borrow extents that , and, (b) have a borrow extent that has not yet ended.
For example, if a variable \code{x} has type \code{\&\ticktt{a} mut String}, then $\lprojtt{x}{a}$ denotes the (singleton) set of borrows stored in $x$. Lifetime projection nodes will allow us to mediate between precise reasoning about
individual places \as{and borrows}, and abstractions over (potentially unbounded) sets of borrows as considered by Rust's borrow checkers.
\as{Rust's flow sensitive type checking means that the values and sets of borrows considered stored in places depend on the program point: in general we allow our graph nodes to refer \eg{} to places as they \emph{were} at a previous program point by labelling them with a program point label.}
\begin{definition}[PCG Preliminaries]% (\tabref{capabilities})
We assume a labelling of each (MIR) program point with a distinct \emph{program point label}, ranged over by $\ppoint$. We use $p$ to range over \emph{places}.

A \emph{generalised place} $\pcgplace$ is defined by the cases: $\pcgplace :{:}= p \mid p\at\ppoint \mid \remote{i}$; in the second case we call $\pcgplace$ a \emph{labelled place}; in the third ($i$ is a non-negative integer) the node is called a \emph{remote place}\footnote{Remote places are used to represent the borrows stored in original arguments of the function under analysis.}.

\emph{Capability types} $\capability$ are defined by: $\capability :{:}= \capexclusive \mid \capwrite \mid \capread \mid \capnone$ (called \emph{exclusive}, \emph{write}, \emph{read} and \emph{none}).

A \emph{lifetime projection} $\lifetimeproj$ is defined by the cases: $\lifetimeproj :{:}= \lproj{\pcgplace}{\maybelabelledregion} \mid \lproj{\pcgplace}{\maybelabelledregion}\at\ppoint \mid \lproj{\pcgplace}{\maybelabelledregion}\at\future$% \mid \lproj{\const}{\lifetime}$\asfootnote{maybe we don't need the third case in the main version of the paper?}
%$\asfootnote{we could consider making adding "at label" onto something a macro}
%\end{definition}
%
%\begin{itemize}
%\item \textbf{Exclusive} ($\capexclusive$): Both writes and reads are permitted
%\item \textbf{ShallowExclusive} ($\capshallowex$) Some parts of the place have not yet been
%        initialised, we have permission to this place, but not to its subplaces.
%\item \textbf{Write} ($\capwrite$): Permits writes to the location, but not reads.
%\item \textbf{Read} ($\capread$): Permits reads from the location, but not
%writes or drops. This permission is only used for places that are immutably
%borrowed\footnote{The capabilities do not differentiate between owned places
%created with \code{let mut} and \code{let}.}.
%\end{itemize}

%\begin{definition}[Node Identifiers]
We assume an infinite set \NodeIDs{} of \emph{node identifiers (node IDs)} ranged over by $\alpha, \beta, \gamma,$ \etc{}, and a (partitioning)
predicate \placeNode\@ on node IDs, which is true (resp.~false) for an infinite subset of
\NodeIDs{}. When \placeNode{n} is true, $n$ is a \emph{place node ID}; otherwise, $n$ is a
\emph{lifetime projection node ID}.% We use Greek letters $\alpha, \beta,$ etc. to denote node identifiers in our rules.
%
%\emph{Place node contents} $\maybelabelled{:}\capability$ are pairs of a generalised place $\maybelabelled$ and a capability $\capability$.
%
%\emph{Lifetime project node contents}, written
\end{definition}
\as{Our PCG edges represent a variety of key notions arising in Rust's design. Edges from place nodes reflect a current flow of capabilities (that removing the edge would undo) from parent(s) to child(ren): the \emph{leaf} place nodes of our PCGs are typically those with capabilities. From lifetime projection nodes, edges reflect instead where (sets of) borrows might flow to and be used.}
\begin{definition}[PCG Edges]
A \emph{PCG edge} is a tuple $(\edgeType,\nodeSet_0,\nodeSet_1)$, in which $\nodeSet_0$, $\nodeSet_1$ are non-empty disjoint sets of node IDs, and  $\edgeType$ is an \emph{edge type}: a label from one of the following cases (names in parentheses): $ \edgeType\;{:}{:}{=}\;\borrowType\textit{ (borrow)}\;\mid\;\unpackType\textit{ (unpack)}\;\mid\;\borrowFlowType\textit{ (borrow-flow)}\;\mid\;\abstractionType\textit{ (abstraction)}\;\mid\;\aliasType\textit{ (alias)}\;\mid\;\derefType\textit{ (deref)}$.

The first set $\nodeSet_0$ are the \emph{sources} of the edge; the second $\nodeSet_1$ are its \emph{targets}; in general, PCG edges are \emph{hyper-edges} (many-to-many). However, each node type comes with specific requirements on the arities and contents of the two node ID sets (place node IDs vs.~lifetime project node IDs), as summarised by the following edge signatures (where $P$ represents that a place node ID is expected; $L$ that a lifetime projection node ID is expected; $\Vec{.}$ denotes zero or more of the element):
$(\borrowType,\{P\},\{L\})$, $(\unpackType,\{P\},\{\Vec{P}\})$, $(\borrowFlowType,\{L\},\{L\})$, $(\abstractionType,\{\Vec{L}\},\{\Vec{L}\})$, $(\aliasType,\{P\},\{P\})$, $(\derefType,\{P,L\},\{P\})$.\asfootnote{TODO: rearrange this definition into some kind of table-like arrangement so that we show labels, names and signatures together, somehow.}
\end{definition}
\as{We define our graphs over a notion of node identifiers via several indirections, to simplify the presentation of operations which change node content but not graph structure.}
\begin{definition}[Place Capability Graphs]
A \emph{Place Capability Graph (PCG)} is defined by the \as{six} elements $(\nodes{G},\edges{G},\places{G},\capabilities{G},\lifetimeprojs{G},\as{\branchChoices{G}})$, where:
\begin{itemize}
\item $\nodes{G}$ is a finite set of node identifiers;
\item $\edges{G}$ is a finite set of PCG edges, whose sources and targets are all in $\nodes{G}$;%, and which form a (hyper-)DAG;
\item $\places{G}$ is an injective map from the place node IDs in $\nodes{G}$ to generalised places;
\item $\capabilities{G}$ is a map from the image of $\places{G}$ to capabilities;
\item $\lifetimeprojs{G}$ is an injective map from the lifetime projection node IDs in $\nodes{G}$ to lifetime projections; and
\item $\branchChoices{G}$ is a map from $\edges{G}$ to a set of \emph{branch-choice identifiers}; (\cf{}\secref{approach:paths}).
\end{itemize}
For node IDs $m$, $n$, we say \emph{$m$ is parent of $n$ in $G$} iff, for some edge $(\edgeType,\nodeSet_0,\nodeSet_1)\in\edges{G}$, $m\in\nodeSet_0$ and $n\in\nodeSet_1$. %For a set of nodes $S$, we write $\parents{S}{G}$ for the set of all parents of nodes in $S$.
We derive the corresponding notions of \emph{child} nodes, \emph{reachability} (via children), and (hyper-)DAG accordingly. The \emph{roots of $G$}, $\roots{G} \subseteq \nodes{G}$ are its nodes without parents; dually the \emph{leaves of $G$}, written $\leaves{G}$ have no child nodes.
\end{definition}

 \as{Our PCGs are acyclic (hyper-DAGs)\footnote{Technically acyclicity can be violated due to our superimposed representations of path-sensitive graphs explained in \secref{approach:paths}; even so, acyclicity remains true when considering the edges corresponding to any single path through the program.}; their edges represent the flows of capabilities (to places/borrows) and borrows (into sets of borrows).} The majority of our rules for constructing PCGs operate on a small handful of leaf nodes and their parents, leaving the rest of the graph unchanged. To simplify expressing these rules, we introduce a notion of \emph{PCG context}, which can be thought of as a graph with some dangling downward (child) edges, to which a further graph can be connected:

\begin{definition}[PCG Contexts]
We define \emph{(PCG) Contexts} $C$ analogously to PCGs except for:
\begin{itemize}
\item An additional element $\holes{C}$ defines a subset of $\leaves{C}$.
\item The domains of $\places{C}$ and $\lifetimeprojs{C}$ are restricted to only the nodes \emph{not} in $\holes{C}$.
\end{itemize}

For a context $C$ and PCG $G$, the \emph{composition of $C$ with $G$} is a partial operation defining a PCG, written $\comp{C}{G}$; it is defined only when:
\begin{itemize}
\item $(\nodes{C}\setminus\holes{C})\intersect\nodes{G} = \emptyset$ (no non-hole node in common).
\item $\holes{C}\subseteq\roots{G}$
\item The codomains of $\places{C}$ (resp. $\lifetimeprojs{C}$) and $\places{G}$ (resp. $\lifetimeprojs{G}$) are disjoint (and so these pairs of injective maps can be combined to an injective map).
\end{itemize}
When these conditions are met, $\comp{C}{G}$ is the PCG defined by taking the union of all sets and maps constituting $C$ and $G$, ignoring the $\holes{C}$ set.

For a pair of contexts $C$ and $C'$, the \emph{context composition of $C$ with $C'$}, is a partial operation defining a PCG \emph{context}, written $\comp{C}{C'}$, defined entirely analogously to the definition for $\comp{C}{G}$ above. When defined, the resulting context has the same holes as $C'$, \ie{} $\holes{\comp{C}{C'}} = \holes{C'}$.
%\end{definition}
%
%\begin{definition}[Substitution] We find it useful to define some rules, in part, by applying a substitution to the contents of nodes in a PCG or context.
A \emph{node content substitution} $\sigma$ is a pair of maps: one from generalised places to generalised places, and one from lifetime projections to lifetime projections; we write \eg{} $\sigma(\pcgplace)$ for the application of the appropriate map. We lift this definition pointwise to a graph or a context in the natural way, defining $\sigma(G)$ to have the changed maps %\begin{itemize}
%  \item A map from generalised places to generalised places.
%  \item A map from lifetime projections to lifetime projections.
%\end{itemize}
%We also refer to these constituent maps using $\sigma$; whether we are referring to either part or the whole is indicated by the type of object to which $\sigma$ is applied.
%The application of a substitution $\sigma$ to a PCG $G$ is denoted $\sigma(G)$ and is defined as follows:
%\begin{itemize}
%  \item
  $\places{\sigma(G)}(\nodeID) = \sigma(\places(G)(\nodeID))$
  and $\lifetimeprojs{\sigma(G)}(\nodeID) = \sigma(\lifetimeprojs(G)(\nodeID))$.
%\end{itemize}
%Applying $\sigma$ to a context $C$ works in the same way.
The operation $\sigma(G)$ is defined only if $\places{\sigma(G)}$ and $\lifetimeprojs(\sigma(G))$ remain injective.
\end{definition}
We use the following compact notations to describe the rules of our PCG analysis.

\begin{definition}[Graph and Rule Notations]
\as{In a given graph $G$, we write $\node{\pcgplace:\capability}{\nodeID}$ to represent a place node: a place node ID $\nodeID\in\nodes{G}$, for which $\places{G}(\nodeID) = \pcgplace$ and $\capabilities{G}(\pcgplace) = \capability$. Similarly, we write $\LPnode{\lifetimeproj}{\nodeID}$ to represent a lifetime projection node: expressing $\nodeID\in\nodes{G}$ and $\lifetimeprojs{G} = \lifetimeproj$. We will omit node IDs $\nodeID$ when it aids clarity.}

We write $\edge{\edgeType}{S_0}{S_1}$ for an edge $(\edgeType,S_0,S_1)$, comma-separating this notation to describe multiple edges. \as{We will present the nodes in $S_0$, $S_1$ either with the notation above or only with their node IDs (and may also elide explicit set parentheses around node sets), as aids presentation. For example, $\edge{\edgeType}{\{\node{\pcgplace_1:\capability_1}\}}{\{\node{\pcgplace_2:\capability_2}{\nodeID_2}\}}, \edge{\edgeType}{\{\nodeID_2\}}{\{\node{\pcgplace_3:\capability_3}\}}$ denotes a linked-list-like graph with three nodes and two edges; the explicit node ID $\nodeID_2$ captures the same node occurring in both edges.}

% and write \eg{} $\edge{\edgeType'}{\edge{\edgeType}{S_0}{S_1}}{S_2}$ as a (DOT-like) shorthand for the edges $\edge{\edgeType}{S_0}{S_1}, \edge{\edgeType'}{S_1}{S_2}$.

Given a Rust (MIR) statement $s$, we denote a \emph{PCG statement rule for $s$} with the syntax $\triple{G_0}{\ppoint}{s}{G_1}$ in which $\ppoint$ is the program point of the statement (often omitted for simplicity). We call $G_0$ the \emph{pre-state PCG} and $G_1$ the \emph{post-state PCG}. As is standard \eg{}, for Hoare Logic rules, we define PCG rules as rule \emph{schemas}, in which meta-variables (\eg{}, for places, program points or portions of a PCG) can be instantiated to obtain specific rule instances.
We denote a \emph{PCG operation rule} with the analogous syntax $\ghosttriple{G_0}{\ppoint}{\pcgop}{G_1}$, in which $\ppoint$ $\pcgop$ is a description of a PCG operation performed \emph{between} program statements by our analysis; their definition and role will be explained shortly.

To reduce notational clutter further, we also adopt the following conventions for defining rules: when we write a statement rule (and analogously for operation rules) in the form $\triple{\comp{C_0}{G}}{s}{\comp{C_0'}{G'}}$ where $C_0$, $C_0'$ are meta-variables, while $G$, $G'$ constitute explicit notation for nodes and edges, we imply that the nodes and edges of $G$ and $G'$ are exactly those notated. Further, when it is clear which nodes in $G$ and $G'$ correspond to one another, we will omit their IDs; for nodes that are clearly fresh in $G'$ (did not occur correspondingly in $G$) we assume them to have fresh node IDs.

Some of our analysis rules do not apply only at the leaves of a PCG; to specify these rules locally we need to allow for extra graph portions may appear \emph{both} above and below. When we write a rule in the form $\triple{\comp{C_0}{\comp{C}{G_0}}}{s}{\comp{C_0'}{\comp{C'}{G_0'}}}$ where $C_0$, $G_0$, $C_0'$, $G_0'$ are meta-variables, while $C$, $C'$ use explicit notation for nodes and edges, we imply that the nodes and edges of $C$ and $C'$ are exactly those notated; those which are holes of $C$ (resp.~$C'$) are notated with node content (places and capabilities, or lifetime projections) actually determined by the mappings defined in $G_0$ (resp.~$G_0'$).
\end{definition}

\begin{figure}[t]
\begin{minipage}{0.3\textwidth}
\begin{subfigure}{\textwidth}
  \begin{tikzpicture}
    \node[] (ex) at (0,0) {Exclusive (\capexclusive)};
    \node[] (write) at (1.5,-0.8) {Write (\capwrite)};
    \node[] (read) at (-1.5,-0.8) {Read (\capread)};
    \node[] (none) at (0,-1.6) {None (\capnone)};

    \draw[dashed,->] (ex) -- (write);
    \draw[dashed,->] (ex) -- (read);
    \draw[dashed,->] (write) -- (none);
    \draw[dashed,->] (read) -- (none);

  \end{tikzpicture}
  \caption{Capabilities}
  \label{fig:approach:capabilities}
\end{subfigure}
\hfill
\begin{subfigure}{\textwidth}
  \centering
  \begin{tikzpicture}
    \node[pcg] (place) at (0,0) {$p: c$};
    \node[anchor=north] (placetext) at ($(place.south)$) {Place};
    \node[lproj] (lproj) at ($(place) + (3.0, 0)$) {$\lproj{p}{\tick{r}}$};
    \node[anchor=north,align=center] (projtext) at ($(lproj.south)$) {Lifetime Projection\\\footnotesize(\secref{approach:borrow})};
  \end{tikzpicture}
  \caption{Nodes}
\end{subfigure}
\end{minipage}
\hfill
\begin{minipage}{0.6\textwidth}
\begin{subfigure}{\textwidth}
\begin{tikzpicture}

    % Left Column
    % Unpack Edges
    \node[pcg] (unpack-source) at (0,0) {$p: c$};
    \node[pcg] (unpack-field1) at ($(unpack-source) + (2.5,0.5)$) {$p.f_1: c$};
    \node (unpack-dots) at ($(unpack-source) + (2.5, 0)$) {\ldots};
    \node[pcg] (unpack-fieldn) at ($(unpack-field1) - (0,1.0)$) {$p.f_n: c$};
    \node[hyperedge, fit=(unpack-field1)(unpack-fieldn)] (unpack-targets) {};
    \draw[expand] (unpack-source) -- (unpack-targets) node[midway,below,align=center] {Unpack\\\footnotesize(\secref{approach:unpack})};

    % Borrow Edges
    \node[pcg] (borrow-source) at ($(unpack-source) - (0,1.4)$) {$p: c$};
    \node[lproj] (borrow-target) at ($(borrow-source) + (2.5,0)$) {$\lproj{p'}{\tick{r}}$};
    \draw[borrow] (borrow-source) -- (borrow-target) node[midway,below,align=center] {Borrow\\\footnotesize(\secref{approach:borrow})};

    % Borrow Flow
    \node[lproj] (funcabs-source) at ($(borrow-source) - (0,1.3)$) {$\lproj{p}{\tick{r}}$};
    \node[lproj] (funcabs-target) at ($(funcabs-source) + (2.5,0)$) {$\lproj{p'}{\tick{s}}$};
    \draw[outlives] (funcabs-source) -- (funcabs-target) node[midway,below=0.2cm,align=center] {Borrow-Flow\\\footnotesize(\secref{approach:compositeborrows})};

    % Right Column
    % Deref Edges
    \node[pcg] (deref-place) at ($(unpack-field1) + (1.5,0)$) {$p: c$};
    \node[lproj] (deref-lproj) at ($(unpack-fieldn) + (1.5,0)$) {$\lproj{p}{\tick{r}}$};
    \node[pcg] (deref-target) at ($(deref-place) + (2.5,-0.5)$) {$*p: c$};
    \node[hyperedge, fit=(deref-place)(deref-lproj)] (deref-sources) {};
    \draw[expand] (deref-sources) -- (deref-target) node[midway,below,align=center] {Deref\\\footnotesize(\secref{approach:borrow})};

    % Alias Edges
    \node[pcg] (alias-source) at ($(deref-place) - (0,1.9)$) {$p: c$};
    \node[pcg] (alias-target) at ($(alias-source) + (2.5,0)$)  {$p': c$};
    \draw[alias] (alias-source) -- (alias-target) node[midway,below,align=center] {Alias\\\footnotesize(\secref{approach:borrow})};

    % Loop Abstraction (From Place)
    \node[lproj] (absfrom) at ($(alias-source) - (0,1.3)$) {$\lproj{p}\tick{r}$};
    \node[lproj] (absto) at ($(absfrom) + (2.5,0)$) {$\lproj{p'}{\tick{s}}$};
    \draw[abstract] (absfrom) -- (absto) node[midway,below=0.2cm,align=center] {Abstraction\\\footnotesize(\secref{approach:functions})};
    % \node[draw,fit=(place)(lproj4)(he)] {};
\end{tikzpicture}
\caption{Edges}
\end{subfigure}
\end{minipage}
\caption{The PCG capabilities, components of the graph and their
  associated permitted behaviours. Capabilities (a) describe the actions that the
  Rust compiler permits on a value via a particular place. PCG nodes (b) denote
  memory locations: either places, or \emph{lifetime projections} which describe
  a region of memory that the borrow checker permits access to.
  Edges (c) in the PCG describe capability transfers, aliasing, and the flow of borrows.}
\label{fig:approach:overview}
\end{figure}

\subsection{PCG Analysis Workflow}
\as{Our PCG analysis requires as input a MIR-representation of a type-checked Rust program (for which we can access the results of the Rust compiler's analyses, especially borrow checking), in which all program points have been assigned distinct labels $l$, and all branches and corresponding joins, as well as loop-heads and corresponding back-edges have been identified.

Our analysis computes a PCG for every program point by forward analysis; the initial PCG used comprises, for each function argument $p$:
%
%
%\zg{
%A PCG is constructed for each program point as a forward
%analysis\abfootnote{``PCG is constructed ... as a forward analysis'' maybe
%reword?} over the MIR (CFG) representation of a Rust function, driven both by
%the statements of the function body and information from the (already-performed)
%Rust compiler's analyses (especially the borrow checker); the analysis visits
%each basic block in the CFG exactly one time.
%
%An initial PCG for a function is constructed by, for each function argument $p$:
%
(a) a place node $\node{p: \capexclusive}$, and (b) for each lifetime $\tick{a}$ in the type of $p$, a lifetime projection node $\LPnodeTwo{\lproj{p}{\tick{a}}}{}$.
Within a (MIR) CFG basic block, each statement is handled by its corresponding PCG analysis rule. Before doing so, PCG operations are performed (a) to remove edges that concern borrow extents that the borrow checker deems to have ended, and (b) to arrange the graph to satisfy the precondition of the analysis rule in question.

Non-initial basic blocks are analysed starting from a PCG computed as the \emph{join} (\cf{} \secref{approach:paths}) of the PCG graphs found for the final states of its direct ancestor blocks; our analysis follows a reverse postorder traversal (excluding loop back-edges) of the CFG blocks. For loops, the analysis is more-complex (but avoids an iterative fixpoint computation): we describe our algorithm in \secref{approach:loops}.
}
%
%
%Starting from the initial state, the analysis computes a PCG for each program
%point in the initial basic block by visiting each statement in the block.  For
%every statement, we define a corresponding PCG analysis rule consisting of (1)
%preconditions on the pre-existing PCG, and (2) how to calculate the PCG for
%after the statement. Prior to applying the rule, the PCG performs
%\emph{operations} that transform it state: first by removing edges that
%correspond to borrow-extents that the borrow-checker reports to be expired, and
%subsequently by performing the necessary operations to ensure that it is in a
%state that satisfies the precondition of the rule. The PCG state for the
%statement is then updated to reflect the effects of the rule.
%
%The analysis subsequently analyses the remaining basic blocks by performing a
%reverse postorder traversal on the CFG (excluding back edges), such that the
%analysis for each basic block is performed \emph{after} the states for all of
%its ancestors has been computed.
%%
%The analysis for a basic block starts from an initial state that is computed as
%the \emph{join} (\cf{} \secref{approach:paths}) of the PCG states of its direct
%ancestors. For basic blocks that are loop heads, the analysis uses information
%from the compiler and borrow-checker to identify a graph shape corresponding to
%a loop invariant; we describe our approach in \secref{approach:loops}.
%

Overall, our PCG analysis results in a PCG for each program point in the body of the
function. Additionally, it returns a list of \emph{annotations} for each program
point that reflect the operations it performed at that point\as{, including those performed to remove edges when necessary.}%\asfootnote{It would be nice to add something about generating annotations for reborrowing, but there seem to be more important things to do for now.}

\subsection{Fine-Grained Place Accounting with Unpack Edges}\label{sec:approach:unpack}

As motivated in \secref{ownership}, Rust's type checking decomposes
capabilities for composite types into field-specific capabilities (and the
reverse); consumer tools typically need to manage their representations for
these (potentially recursive) types with similar operations. For these reasons,
\as{our model defines explicit \emph{unpack} and \emph{pack} PCG operations, which add
(\resp{} remove) \emph{unpack edges} that represent a decomposition of capabilities.
In this subsection, we present simplified versions of our rules
that are nonetheless equivalent when applied to composite types that do not
contain borrows. In \secref{approach:compositeborrows}, we explain how our rules
are extended to support general types.}

\begin{figure}[t]
%     \begin{subfigure}{0.7\textwidth}
%     \begin{rust}
% struct StringList { hd: String, tl: Option<Box<StringList>> }
% fn take_head(list: StringList) -> String {
%   // Initial: {list: ~\capexclusive~, RETURN: ~\capwrite~}
%   // Prepare: {list.hd: ~\capexclusive~, list.tl: ~\capexclusive~, RETURN: ~\capwrite~}
%   return list.hd; // {list.hd: ~\capwrite~, list.tl: ~\capexclusive~, RETURN: ~\capexclusive~}~\label{line:partialmove}~
% }\end{rust}
% \caption{A function that performs a partial move out of the
% argument~\code{list}}
%     \label{fig:partialmove}
%     \end{subfigure}
    \begin{subfigure}{0.3\textwidth}
        \centering
        \begin{tikzpicture}[scale=0.85]
            \node[pcg] (list) at (0,0) {$\code{pos}$: $\capexclusive$};
            \node[pcg] (old_x) at (1.8,0) {$\code{old\_x}$: $\capwrite$};
            \node (spacer) at (0,-1.6) {};
        \end{tikzpicture}
        \caption{Initial PCG State}
        \label{fig:pcg-init}
    \end{subfigure}
    \begin{subfigure}{0.3\textwidth}
        \centering
        \begin{tikzpicture}[scale=0.85]
            \node[pcg] (list) at (0,0) {$\code{pos}$: $\capnone$};
            \node[pcg] (listhd) at ($(list.south) - (0,1.0)$) {$\code{pos.x}$: $\capexclusive$};
            \node[pcg, anchor=west] (listtl) at ($(listhd.east) + (0.1,0)$) {$\code{pos.y}$: $\capexclusive$};
            \node[pcg] (old_x) at (1.8,0) {$\code{old\_x}$: $\capwrite$};
            \node[hyperedge, fit=(listhd)(listtl)] (targetbox) {};
            %\node (spacer) at ($(targetbox.south) -(0,0.5)$) {};
            \draw[expand] let
            \p1=(targetbox.north),
            \p2=(list.south)
            in (list.south) -- (\x2,\y1);
            % \node[draw,fit=(list)(targetbox)] (rect) {};
        \end{tikzpicture}
        \caption{PCG State after \emph{unpack}}
        \label{fig:pcg-unpack}
    \end{subfigure}
    \begin{subfigure}{0.3\textwidth}
        \centering
        \begin{tikzpicture}[scale=0.85]
            \node[pcg] (list) at (0,0) {$\code{pos}$: $\capnone$};
            \node[pcg] (listhd) at ($(list.south) - (0,1.0)$) {$\code{pos.x}$: $\capwrite$};
            \node[pcg, anchor=west] (listtl) at ($(listhd.east) + (0.1,0)$) {$\code{pos.y}$: $\capexclusive$};
            \node[pcg] (old_x) at (1.8,0) {$\code{old\_x}$: $\capexclusive$};
            \node[hyperedge, fit=(listhd)(listtl)] (targetbox) {};
            %\node (spacer) at ($(targetbox.south) -(0,0.5)$) {};
            \draw[expand] let
            \p1=(targetbox.north),
            \p2=(list.south)
            in (list.south) -- (\x2,\y1);
            % \node[draw,fit=(list)(targetbox)] (rect) {};
        \end{tikzpicture}
        \caption{PCG State after move}
        \label{fig:pcg-final}
    \end{subfigure}
    \caption{\jg{PCGs at three points in \figref{motivation:move}; \figref{pcg-init} is the PCG at the outset of \code{replace\_x\_owned}, \figref{pcg-unpack} is the PCG after \code{pos} is \emph{unpacked}, and \figref{pcg-final} is the PCG after the move assignment on \lineref{motivation:partialmove}.}}
\end{figure}
\figref{motivation:move} (page \pageref{fig:motivation:move}) shows an example in which unpacking is needed.
The source-level assignment statement \code{let old = pos.x} will, in MIR, become a move assignment \code{old = move pos.x}.
The PCG analysis rule for move assignments $p = \code{move}~q$, in the case where $q$ does not contain any borrows, is as follows:
\jg{
$
  \triple{\comp{C}{\node{p: \capwrite}, \node{q: \capexclusive}}}
  {p\code{ = move }q}
  {\comp{C}{\node{p: \capexclusive}, \node{q: \capwrite}}}
$
}

\as{This rule requires a pre-state PCG with leaf place nodes for the moved-in ($p$) and moved-out ($q$) places with capabilities $\capwrite$ and $\capexclusive$, respectively; in the post-state PCG the capabilities are exchanged.
%In the post-state, the capability to $p$ is upgraded to $\capexclusive$ and the capability to $q$ is downgraded to $\capwrite$.
}
\as{In our example, this rule cannot be applied to the initial graph in \figref{pcg-init}, }
%
%initial PCG for \code{replace\_x\_owned}, as shown in \figref{pcg-init}:
%
%
%For the code above, the rule can be applied directly (here points (2) and (3) above do not apply) to the PCG in the initial state, yielding the final PCG state $\{\code{input}:
%\capwrite, \code{RETURN}: \capexclusive\}$.
%
%As explained, the rule for this assignment
as it has no (leaf) place node for $\code{pos.x}:\capexclusive{}$. A PCG operation must first be applied to rearrange the graph: in this case an \emph{unpack}, defined by the rule schema:\label{simpleunpack}
$
  \ghosttriple{\comp{C}{\node{p: \capexclusive} }}
  {\opunpack{p}}
  {\comp{C}{\edge{\unpackType}{\node{p: \capnone}}{\Vec{\node{p.f_i: \capexclusive}}}}}
$
\noindent
in which $p.f_i$ ranges over the immediate projections (\eg{} fields of a struct) of $p$. Instantiated for our example (\cf{} \figref{pcg-init} which is transformed into \figref{pcg-unpack}), the corresponding rule instance is $\ghosttriple{\comp{C}{\node{\code{pos}: \capexclusive} }}
  {\opunpack{\code{pos}}}
  {\comp{C}{\edge{\unpackType}{\node{\code{pos}: \capnone}}{\{\node{\code{pos.x}: \capexclusive}, \node{\code{pos.y}: \capexclusive}\} }}}$ in which $C$ is a context with no holes or edges and a single node $\node{\code{old\_x}: \capwrite}$.
%\jg{
%Unlike the rule for move assignments, this rule is not tied to a particular MIR statement.
%Instead, it can be applied between statements to rewrite the PCG in order to satisfy the precondition of another rule; we refer to such rules as \emph{PCG Operations}.
%When instantiated for \code{pos}, this rule
%\begin{enumerate}
%  \item Updates the node $\node{\code{pos}: \capexclusive{}}$ to $\node{\code{pos}: \capnone{}}$, and adds nodes $\node{\code{pos.x}: \capexclusive}$ and
%        \\$\node{\code{pos.y}: \capexclusive}$
%  \item Inserts an \emph{unpack} hyperedge between these nodes:
%        \\$\edge{\unpackType}{\node{\mathtt{pos}: \capnone}}{\{\node{\mathtt{pos.x}: \capexclusive}, \node{\mathtt{pos.y}: \capexclusive}\}}$
%\end{enumerate}
%}
%
% If \code{list} stored borrows, this rule would create lifetime projection nodes for each field of \code{list} according to its type ($\textit{rp}(p)$ gives the set of lifetimes in the type of place $p$) and link these to the lifetime projections of \code{list}, but since \code{list} stores no borrows this is a no-op.
%
% \begin{enumerate}
%   \item Updates the node \code{list}: \capexclusive{} to \code{list}: \capnone{}, and adds nodes for
%          $\code{list.hd}: \capexclusive$ and $\code{list.tl}:
%          \capexclusive$.
%   \item Inserts an \emph{unpack} hyperedge between these nodes: $\mathtt{list} \rightarrow \{\mathtt{list.hd}, \mathtt{list.tl}\}$
% \end{enumerate}
%
After this operation, the resulting PCG (shown in \figref{pcg-unpack}) is suitable as the precondition of
the move assignment rule.
\jg{
Following the move assignment, the PCG is as shown in \figref{pcg-final} (our analysis proceeds with the next statement).
}

The inverse \emph{pack} PCG operation is given by $
  \ghosttriple{\comp{C}{\edge{\unpackType}{\node{p: \capnone}}{\Vec{\node{p.f_i: \capexclusive}}}}}
  {\oppack{p}}
  {\comp{C}{\node{p: \capexclusive} }}
$
\noindent
in which $p.f_i$ again ranges over the immediate projections of $p$. Note that the PCG in \figref{pcg-final} is not one in which this PCG operation could be applied; the exclusive capability for \code{pos.x} is missing. This reflects that Rust would not allow \code{list} to be moved, borrowed or dropped in this state. Our PCG analysis applies pack or unpack operations as needed between statements, favouring packing whenever possible (keeping the graphs smaller). We record which PCG operations are applied where, so that downstream consumers can drive their own analyses with analogous steps.
%
%
%The unpack hyperedge is added to record the capability
%transfers that would be necessary to regain capability to list: specifically,
%the meaning of the edge is that \code{list.hd} and \code{list.tl}
%constitute the memory of \code{list}. Therefore, capability to \code{list}
%could be obtained by performing the corresponding \emph{pack}, which exchanges
%the exclusive capabilities from \code{list.tl} and \code{list.hd} for
%the exclusive capability for \code{list}, and removes the unpack edge.
%
%To ensure that the PCG for a program state is in its most simplified form, the
%PCG is \emph{repacked} after each statement by applying all \emph{pack}
%operations that are immediately applicable.  However in this case, because
%\code{list.hd} and \code{list.tl} have different capabilities,
%\code{list} remains in its unpacked form.

\subsection{Tracking Borrows with Borrow Edges}\label{sec:approach:borrow}

\begin{figure}[t]
\begin{subfigure}[b]{0.33\textwidth}
    \centering
    \def\boxpadding{0.15}
    \def\vspacing{1.0}
    \def\rectleft{-2.1}
    \def\rectright{2.3}
    \def\recttop{0.5}
    \def\rectbottom{-1.5}
    \begin{tikzpicture}[scale=0.95]
    \node[pcg] (x) at (-1,0) {$\code{(*pos).x}$: $\capnone$};
    \withproj{y}{\code{x\_ref}}{\capexclusive}{a}{(0.8,0)}
    \draw[borrow] (x) -- (yproj.west);
    % \draw (\rectleft,\rectbottom) rectangle (\rectright,\recttop);
    \end{tikzpicture}
    \caption{\code{let x\_ref = \&mut (*pos).x;}}
    \label{fig:pcg-simpleborrow-1}
\end{subfigure}
\hfill
\begin{subfigure}[b]{0.3\textwidth}
    \centering
    \def\boxpadding{0.15}
    \def\vspacing{1.0}
    \def\rectleft{-1.8}
    \def\rectright{2.4}
    \def\recttop{0.5}
    \def\rectbottom{-1.5}
    \begin{tikzpicture}[scale=0.95]
    \node[pcg] (x) at (-1.3,0) {$\code{(*pos).x}$: $\capnone$};
    \node[pcg] (y) at (0.8,0) {$\code{x\_{ref}}$: $\capnone$};
    \node[lproj] (yproj) at (1,-0.45){\lprojtt{x\_ref}{a}~\code{at}~\ref{line:motivation:derefborrow}};
    \node[hyperedge,fit=(y)(yproj)] (hyperedge) {};
    \drawpcgnode{dy}{\code{*x\_ref}}{\capexclusive}{(0,-1.2)}
    \draw[borrow] (x) -- (yproj.west);
    \draw[alias, bend right] (x) to (dy);
    \draw[expand] (hyperedge.south) |- (dy);
    % \draw (\rectleft,\rectbottom) rectangle (\rectright,\recttop);
    \end{tikzpicture}
    \caption{After \code{*x\_ref = new\_x;}}
    \label{fig:pcg-simpleborrow-2}
\end{subfigure}
\hfill
\begin{subfigure}[b]{0.3\textwidth}
    \centering
    \def\boxpadding{0.15}
    \def\vspacing{1.0}
    \def\rectleft{-1.8}
    \def\rectright{1.8}
    \def\recttop{0.5}
    \def\rectbottom{-1.5}
    \begin{tikzpicture}[scale=0.95]
    \node[pcg] (x) at (-0.8,0) {\code{pos}: $\capexclusive$};
    \node[pcg] (y) at (0.9,0) {\code{x\_ref}: $\capwrite$};
    % \draw (\rectleft,\rectbottom) rectangle (\rectright,\recttop);
    \end{tikzpicture}
    \caption{After \tick{\code{a}} expires}
    \label{fig:pcg-simpleborrow-3}
\end{subfigure}
\caption{\jg{Partial PCGs at three points in \figref{motivation:borrowexample}. (a) the result of the mutable borrow on \lineref{motivation:reborrow}, (b) the result of the assignment on \lineref{motivation:derefborrow}, and (c) the PCG after the mutable borrow expires and \code{pos} is packed up.}}
\label{fig:simpleborrow}
\end{figure}

As explained in \secref{motivation:borrows}, Rust's borrowed references are used to
temporarily transfer capabilities (and later return them), enabling controlled aliasing and in-place
mutation. The example in \figref{motivation:borrowexample} shows the creation of such a borrow at \lineref{motivation:reborrow}. The PCGs for this example are shown in \figref{simpleborrow}, where a \emph{borrow} edge records the borrow of \code{(*pos).x} into the
\emph{lifetime projection} \lprojtt{x\_ref}{b} \footnote{To help match place nodes with corresponding lifetime projection nodes, we typically draw example diagrams with the latter overlaid under the former. Note however, that these are distinct graph nodes, with no edges between them.}

Lifetime projection nodes model a set (in this case a
singleton) of borrows potentially stored in \code{x\_ref} whose extents have not
yet been ended (according to the borrow checker). \as{Our PCGs satisfy the invariant that, whenever a place node $\node{p:\capability}$ occurs such that $\capability$ is either $\capread$ or $\capexclusive$, a corresponding lifetime projection node $\LPnode{\lproj{p}{\tick{a}}}$ will occur for each lifetime $\tick{a}$ in the type of $p$, \ie{} whenever we can read from a place, we also track the sets of borrows potentially stored in it.}

The \emph{borrow edge}
$\edge{\borrowType}{\node{\placecaptt{(*pos).x}{\capnone}}}{\LPnode{\lprojtt{x\_{ref}}{a}}}$ records the outstanding borrow of
$\code{(*pos).x}$ in the set of borrows represented by \lprojtt{x\_ref}{b}: concretely, the borrowed memory is aliased by the place
\code{*x\_ref}. This edge reflects \emph{why} \code{(*pos).x} doesn't (currently) have its default
capabilities.

As in the previous subsection, we present a simplified
version of the rules that applies when borrows are not stored in composite data
types or inside other borrows. We explain the full rules in \secref{approach:compositeborrows}.
Our PCG analysis rule for creation of a mutable borrow via \texttt{p = \&mut q} is defined as:\label{simpborrow}
$
\triple{\comp{C}{
  \nodeTwo{\placecaptt{p}{\capwrite}}{},
  \nodeTwo{\placecaptt{q}{\capexclusive}}{}}}%
  {l}{\texttt{p = \&mut q}}
  {\comp{\sigma(C)}{
  \nodeTwo{\placecaptt{p}{\capexclusive}}{},
  \edge{\ab{\borrowType{}}}{\nodeTwo{\placecaptt{q}{\capnone}}{}}{\LPnodeTwo{\lproj{\texttt{p}}{\tick{a}}}{}}}}
$
and the corresponding rule for shared borrows is:
$
\triple{\comp{C}{
  \nodeTwo{\placecaptt{p}{\capwrite}}{},
  \nodeTwo{\placecaptt{q}{\capread}}{}}}%
  {l}{\texttt{p = \&q}}
  {\comp{\sigma(C)}{
  \nodeTwo{\placecaptt{p}{\capexclusive}}{},
  \edge{\ab{\borrowType{}}}{\nodeTwo{\placecaptt{q}{\capread}}{}}{\LPnodeTwo{\lproj{\texttt{p}}{\tick{a}}}{}}
  }}
$
where $\tick{a}$ is the lifetime from $\texttt{p}$'s type, and $\sigma$ is a substitution defined to replace (unlabelled) places $p'$ with corresponding labelled places $p'~\at~l$ exactly when $\texttt{*p}$ is a prefix of $p'$, and analogously for the places in lifetime projection nodes. This substitution is needed in case $\texttt{*p}$ had previously been \emph{reborrowed}, and nodes referring to its prior value existed elsewhere in the graph.
%
%The substitution $\sigma$ is necessary to account for \emph{reborrows} of
%\texttt{*p}. Assigning to \texttt{p} does not invalidate any reborrows of
%\texttt{*p}, but as it changes the reference stored in \texttt{p}, any
%references to \texttt{*p} in the PCG should be updated to reflect that they
%refer to the version of \texttt{*p} at location $l$.
%

In this simple example, our lifetime projection concept may seem overly complex given the (single) created borrow, the usage of these uniformly for any types storing borrows is what enables the generality of our model, and its handling in particular of general function calls and loops.

The rule for the assignment \code{*x\_ref = new\_x} on \lineref{motivation:derefborrow} requires
\code{*x\_ref} to be present in the PCG with capability \capwrite; this capability is conceptually buried in the $\capexclusive$ capability for \code{x\_ref}. Our PCG operation \opderef{} for dereferencing a borrow-typed place, is similar to our unpack operation, but we design the rule to reflect that dereferences in Rust require \emph{a justification from the borrow checking of the program} that the corresponding borrow extent has not ended. This dependency is captured with a lifetime projection node:
%%
%Specifically, the unpack operation is defined by the rule:
$
\ghosttriple{\comp{C}{
  \node{\placecap{p}{\capexclusive}}{},
  \LPnode{\lproj{p}{\tick{a}}}{}}}%
  {l}{\opderef{p}}
  {\comp{C}{
  \edge{\derefType{}}{
     \{
       \node{\placecap{p}{\capwrite}}{},
       \LPnode{\lproj{p}{\tick{a}~\at~l}}{}
     \}
   }{
     \nodeTwo{\placecap{*p}{\capexclusive}}{}
   }
  }}
$

In the above rule, $\LPnodeTwo{\lproj{\code{p}}{\tick{a}~\at~l}}{}$ is a \emph{labelled} lifetime projection, used to record a set of borrows that \emph{used to} be associated with a place node at a prior program point; we label lifetime projection nodes whenever their place nodes no longer have at least read capability. This idea will be used and extended when we explain our handling of nested references in \secref{approach:nestedborrows}.
%
%The role of labelled lifetime projections is a technical one; while a place node is not a leaf, it is possible that the sets of borrows stored within it may change; for this reason, the lifetime projection node may have an out-of-date meaning as a set. The labelled versions of these nodes denote sets as they \emph{were} at that point in the program. This is mostly a convenience for simplifying the definition of our rules in terms of local changes to the edges in our graphs. In particular, the notion of labelled lifetime projection nodes is important for the handling of nested references, explained in .

\as{As a convenience (useful for some analysis applications), our PCG analysis also introduces \emph{alias} edges whenever a pair of place nodes are available, one of which is borrowed, and one of which must be the dereference of the corresponding borrow. In general, this can only be known either for types that only store single borrows (references to types that don't contain lifetimes), or because the borrow itself was created explicitly in the function under analysis and can still (possibly via path-sensitivity) be tracked precisely. For example, in \figref{simpleborrow}, our analysis adds the alias
edge $\edge{\aliasType} {\nodeTwo{\placecaptt{(*pos).x}{\capnone}}{}}
{\nodeTwo{\placecaptt{*x\_ref}{\capexclusive}}{}}$ to reflect that
\code{*x\_ref} is a definite alias of \code{(*pos).x}.}
Our alias edges are optional: they are never required or conditioned upon in our rule definitions. To simplify our presentation, we will omit them from the presentation from here onwards.
%
%\zg{
%After introducing a dereference edge, the analysis inserts \emph{alias} edges
%between the dereferenced place and each place that the corresponding
%reference-typed place could directly borrow from, based on the borrow edges
%present in the graph\footnote{Borrow edges are associated with metadata indicating whether it corresponds to a direct alias.}.  In \figref{simpleborrow}, the analysis adds the alias
%edge $\edge{\aliasType} {\nodeTwo{\placecaptt{(*pos).x}{\capnone}}{}}
%{\nodeTwo{\placecaptt{*x\_ref}{\capexclusive}}{}}$ to reflect that
%\code{*x\_ref} is an alias of \code{(*pos).x}. Such alias
%edges are removed accordingly when the dereferenced place is removed from the graph.
%
%We note that alias edges serve as a convenience for consumers of the PCG
%but are not necessary to reason about place capabilities or the
%borrowing relations as seen by the compiler and borrow-checker; none of our
%rules require the presence or absense of an alias edge to insert nodes or other
%\jgout{(non-alias)}types of edges in the graph. Therefore, to simplify our
%presentation, we omit operations related to the handling of alias edges in our
%rules. Also note that our current rules for introducing alias edges in our
%graph are incomplete, in the sense that alias edges are not always introduced
%between places that can be seen to alias statically. We plan to improve our
%rules to identify aliasing more robustly.
%}
%

Returning to our example, our analysis next performs a PCG pack operation, specialised to reference types:
$
  \ghosttriple{\comp{C}{
  \edge{\derefType{}}{
     \{
       \nodeTwo{\placecap{p}{\capwrite}}{},
       \LPnodeTwo{\lproj{p}{\tick{a}~\at~l}}{}
     \}
   }{
     \nodeTwo{\placecap{*p}{\capexclusive}}{}
   }}}
  {\oppack{p}}
  {\comp{C}{
  \nodeTwo{\placecap{p}{\capexclusive}}{},
  \LPnodeTwo{\lproj{p}{\tick{a}}}{}}}.%
$

At the last line of our example, the compiler prescribes that the borrow
extent associated with \code{x\_ref} ends: conceptually, all PCG edges reflecting capabilities transferred by these borrows should be removed. Our analysis performs an iterative cleanup of the PCG, finding edges whose children are all leaves, and which are either unpack edges that can be packed, or edges whose children include either lifetime projection nodes for ended borrow extents or place nodes for places whose types correspond to ended borrow extents. The latter is partially handled by a PCG operation for expiring a borrow:
%
% this causes our PCG to be updated,
%removing all edges associated with lifetime projection nodes that concern
%definitely-ended borrow extents; in this example, this is performed alongside repacking rules to make \code{pos}
%a leaf once more, and as a result, its capability is updated to reflect that it regains
%exclusive access (when the borrow edge originating from it expires).
$
\ghosttriple{\comp{C}{
  \nodeTwo{\placecaptt{p}{\capexclusive}}{},
  \edge{\ab{\borrowType{}}}{\nodeTwo{\placecaptt{q}{\capnone}}{}}{\LPnode{\lproj{\code{p}}{\tick{a}}}}
}}
  {\opexpire{\lproj{\code{p}}{\tick{a}}}}
  {\comp{C}{
     \nodeTwo{\placecaptt{p}{\capwrite}}{},
     \nodeTwo{\placecaptt{q}{\capexclusive}}{}
  }}
$
Applying this rule in our example, \code{x\_ref}'s capability will become \capwrite{}, reflecting that it can be reassigned but no longer read from (as the borrow it stored has expired).

\subsection{Borrows Inside Composite Data Types}\label{sec:approach:compositeborrows}\label{sec:approach:nestedborrows}

So far we have ignored the possibility of storing borrows inside composite data
types, as well the possibility of \emph{nested} borrows, \ie{} creating
references to places that themselves contain borrows (\eg{} as in
\figref{motivation:nestedborrows}). We now describe how we systematically extend
our rules (including those already presented) to handle these issues in two stages: first for code that stores
borrows inside structs but does not nest borrows, and then finally the
full rules that also account for nested borrows.

Because a composite datatype such as \code{List<\&mut i32>} can hold an
arbitrary number of borrows, reflecting Rust's rules for tracking borrows requires
\emph{sets} of borrows that cannot (always) be enumerated. Our \emph{lifetime
projections} represent such sets: we explain next how to adapt our PCG rules to also account for the lifetime projections associated with a type containing lifetimes.

Two general principles underly our adaptations. Firstly, whenever a rule causes a place node to go from having a read/exclusive capability to no longer having one, its corresponding lifetime projection nodes are no longer meaningful as \emph{current} representations of sets of borrows (they describe borrows stored in a place that cannot be accessed): we replace them with labelled versions that refer to the program point before this change. Secondly, when the place's capabilities have flowed elsewhere (\eg{} via a move or unpack), we should reflect that the borrows may correspondingly also now be accessed by new/different places. This latter idea is reflected by \emph{borrow flow} edges, which connect lifetime projection nodes to indicate where a set of borrows may have derived its borrows from.
%
%
%we label its lifetime projection nodes (reflecting the fact that they refer to a place from which borrows can no longer be accessed). In the (common) case that the \emph{reason} for this loss of capability is that the capability has been inherited by other place nodes in some way, we add \emph{borrow flow}
%
%In general, we extend our rules to track changes to sets of borrows by
%introducing borrow-flow edges between the labelled versions of the corresponding
%lifetime projections (reflecting the set of borrows stored in the place prior to
%applying the rule) to the lifetime projections that represent where the borrows
%were moved to.
%
For example, when unpacking a struct we adapt the rule presented on page \pageref{simpleunpack} to be:
%
%the lifetime projections of the struct are labelled,
%new lifetime projections are introduced for each unpacked place, and hyperedges
%are inserted between the labelled and fresh lifetime projections, as per the
%following rule:
%
$
  \ghosttriple{\comp{C}{\node{p: \capexclusive}, \Vec{\LPnode{\lproj{p}{\tick{a}_i}}}}}
  {l}{\opunpack{p}}
  {\comp{C}{\edge{\unpackType}{\node{p: \capnone}}{\Vec{\node{p.f_j: \capexclusive}}},  \Vec{\edge{\borrowFlowType}{\LPnode{\lproj{p~\texttt{at}~l}{\tick{a}_i~\texttt{at}~l}}}{\LPnode{\lproj{p.f_j}{\tick{a}_i}}}}}}
$
where $\tick{a}_i$ ranges over the lifetimes in $p$, $p.f_j$ ranges over the immediate projections (fields) of $p$, and borrow flow edges are added for those $p.f_j$ whose types contain a lifetime corresponding to $\tick{a}_i$ (observe that in the case of a struct whose field types contain no lifetimes, we have our simplified rule).
The result of applying this rule to unpack \code{rf} at \lineref{motivation:replace-x} of \figref{motivation:nestedborrows} is given in \figref{approach:unpack-rf}.
Our earlier rule for moves is analogously extended to become:\\
$
  \triple{\comp{C}{\begin{array}{c}\node{p: \capwrite}, \node{q: \capexclusive}, \\\Vec{\LPnode{\lproj{q}{\tick{a^q_i}}}}\end{array}}}
  {l}
  {p\code{ = move }q}
  {\comp{\sigma(C)}{\begin{array}{c}\node{p: \capexclusive}, \node{q: \capwrite},\\ \Vec{\edge{\borrowFlowType}{\LPnode{\lproj{q~\at~l}{\tick{a^q_i}}~\code{at}~l}}{\LPnode{\lproj{p}{\tick{a^p_i}}}}}\end{array}}}
$\\
in which  $\tick{a^p_i}$ denotes the $i$'th lifetime in the type of $p$,  $\tick{a^q_i}$ denotes the
corresponding lifetime from $q$, and $\sigma$ extends the substitution given for the borrow rules (page \pageref{simpborrow})
such that $\sigma(\lproj{q.{*}}{\tick{r}})=\lproj{q.{*}~\code{at}~l}{\tick{r}~\code{at}~l}$
where $q.{*}$ is any projection\jgfootnote{changed from any \emph{immediate} projection} of $q$.

\begin{figure}[t]
  %\hspace{-0.5cm}
\begin{subfigure}[b]{0.29\textwidth}
\centering
\begin{tikzpicture}[scale=0.85]
  % \node[pcg] (x) at (-1.3,0) {$\code{x}$: $\capnone$};
  % \node[pcg] (y) at (1.3,0) {$\code{y}$: $\capnone$};
  % \node[lproj] (rxproj) at (0,-1) {\lprojtt{(*rf).x~at~\ref{line:createrx}}{b0}};
  \node[lproj] (rfproj) at (0,-1.1) {\lprojtt{rf~at~\ref{line:motivation:replace-x}}{b0}};
  \node[lproj] (newxproj) at (-0.8,-0.2) {\lprojtt{new\_x~at~\ref{line:motivation:replace-x}}{b1}};
  % \draw[borrow] (x) -- (rxproj);
  % \draw[borrow] (y) -- (ryproj);
  \node[pcg] (pairproj) at (0,-2) {\code{*rf}: $\capnone$};
  \node[lproj] (p0proj) at (-1.3,-3.0) {\lprojtt{(*rf).x}{b0}};
  \node[lproj] (p1proj) at (1.3,-3.0) {\lprojtt{(*rf).y}{b0}};
  \node[hyperedge,fit=(p0proj)(p1proj)] (he) {};
  \node[pcg] (p0) at ($(p0proj) - (-0.2,0.45)$) {\code{(*rf).x}: \capexclusive};
  \node[pcg] (p1) at ($(p1proj) - (-0.2,0.45)$) {\code{(*rf).y}: \capexclusive};
  \draw[outlives, bend right] (rfproj) to (p0proj);
  \draw[outlives, bend left] (rfproj) to (p1proj);
  \draw[outlives, bend right] (newxproj) to (p0proj);
  \draw[expand] (pairproj) -- (he);
\end{tikzpicture}
\caption{After \code{(*rf).x = new\_x}}
\label{fig:approach:unpack-rf}
\end{subfigure}
\hspace{0.3cm}
\begin{subfigure}[b]{0.265\textwidth}
\centering
\begin{tikzpicture}[scale=0.85]
  % Define spacing variables
  \def\vspace{0.8} % vertical spacing between lifetime projection levels
  \def\hspace{0.8} % horizontal spacing unit
  \def\basewidth{2} % base horizontal spacing between main nodes

  % Place nodes
  % \node[pcg] (a) at (0,-\vspace) {$\code{a}$: $\capnone$};
  \node[pcg] (f) at (0,-3*\vspace) {$\code{f}$: $\capnone$};
  \node[pcg] (rf) at (0,-5*\vspace) {$\code{rf}$: $\capexclusive$};
  % \node[pcg] (b) at (0,-2*\vspace) {$\code{b}$: $\capexclusive$};

  % Lifetime projections
  \node[lproj] (fprojold) at (-1.8,-2*\vspace) {\lprojtt{f}{a0}~\code{at}~\ref{line:motivation:borrow-f}};
  \node[lproj] (rfproj) at (-1.8,-3.5*\vspace) {\lprojtt{rf}{a1}};
  \node[lproj] (yprojb) at (0,-4*\vspace) {\lprojtt{rf}{a2}};

  % Future lifetime projection for x
  \node[lproj] (xfuture) at (-1.8,-5*\vspace) {\lprojtt{f}{a0}~at~\future};

  % Edges
  % \draw[borrow] (a) -- (xproj);
  \draw[borrow] (f) -- (yprojb);
  \draw[outlives] (fprojold) to node[midway, left] {$\triangle$} (rfproj);
  \draw[outlives] (rfproj) -- (xfuture) node[midway, left] {$\square$};
  \draw[outlives, bend right] (fprojold.west) to node[midway, left] {$\star$} (xfuture.west);
\end{tikzpicture}
\caption{After \code{rf = \&mut f}}
\label{fig:approach:borrow-f}
\end{subfigure}
%\hfill
\begin{subfigure}[b]{0.4\textwidth}
\centering
\begin{tikzpicture}[scale=0.85]
  % Define spacing variables
  \def\vspace{1.0} % vertical spacing between lifetime projection levels
  \def\hspace{0.8} % horizontal spacing unit
  \def\basewidth{1.1} % base horizontal spacing between main nodes

  % Place nodes
  % \node[pcg] (a) at (0,-\vspace) {$\code{a}$: $\capnone$};
  % \node[pcg] (x) at (0,-2*\vspace) {$\code{x}$: $\capnone$};
  \node[pcg] (y) at (\basewidth,-\vspace) {$\code{rf}$: $\capwrite$};
  \node[pcg] (b) at (0.2,-3*\vspace) {$\code{*r3}$: $\capnone$};
  \node[pcg] (dy) at (\basewidth,-4*\vspace) {$\code{*rf}$: $\capexclusive$};

  % Lifetime projections
  \node[lproj] (fprojold) at (-1.8,-\vspace) {\lprojtt{f}{a0~at~\ref{line:motivation:borrow-f}}};
  \node[lproj] (rfprojold) at (-1.8,-2*\vspace) {\lprojtt{rf}{a1~at~\ref{line:motivation:replace-y}}};
  \node[lproj] (yprojb) at (\basewidth,-2*\vspace) {\lprojtt{rf}{a2~at~\ref{line:motivation:replace-y}}};
  \node[lproj] (dyprojc) at (-1.6,-3*\vspace) {\lprojtt{*rf}{a1}};
  \node[lproj] (rfprojfut) at (-1.8,-4*\vspace) {\lprojtt{rf}{a1~at~\future}};

  % Future lifetime projection for x
  \node[lproj] (fprojfut) at (-1.8,-5*\vspace) {\lprojtt{f}{a0}~at~\future};

  % Edges
  % \draw[borrow] (a) -- (fprojold);
  % \draw[borrow] (x) -- (yprojb);
  \draw[borrow] (b) -- (dyprojc);
  \draw[outlives] (fprojold) to node[midway, left] {$\triangle$} (rfprojold);
  \draw[outlives,bend right] (rfprojold.200) to node[midway, left] {} (rfprojfut.160);
  \draw[outlives] (rfprojfut) to node[midway, left] {$\square$} (fprojfut);
  \draw[outlives] (rfprojold) -- (dyprojc);
  \draw[outlives] (dyprojc) -- (rfprojfut);
  \draw[outlives, bend right] (fprojold.west) to node[midway, left] {$\star$} (fprojfut.west);

  % Hyperedge for deref
  \node[hyperedge,fit=(y)(yprojb)] (yedge) {};
  \draw[expand] (yedge.south) -- (dy.north);
\end{tikzpicture}
\caption{After \code{(*rf).y = r3}}
\label{fig:nested-after-reassign}
\end{subfigure}
\caption{
\jg{Partial PCGs at three points in \figref{motivation:nestedborrows}.
\figref{approach:unpack-rf} shows the result of unpacking a struct containing borrows, as is done at \lineref{motivation:replace-x}. \figref{approach:borrow-f} shows the result of the assignment on \lineref{motivation:borrow-f};} the PCG analysis
introduces a future lifetime projection $\lprojtt{f}{a0~\at~\future}$
and adds edges to it
from
\code{f}'s labelled lifetime projection $\lprojtt{f}{a0~\at~\ref{line:motivation:borrow-f}}$ ($\square$)
and the current lifetime projection $\lprojtt{rf}{c1}$ ($\star$).
\figref{nested-after-reassign} shows
the state after the assignment $\code{(*rf).y =
r3}$: the target of $\triangle$ and the source of $\square$ are updated to newly introduced labelled and future projections.}
\label{fig:borrow-in-struct}
\end{figure}

We now turn our attention to the final case we have postponed: nested borrows.
Without the ability to nest borrows, the sets of borrows stored in a place $p$
can only be modified by operations on $p$ itself. %Correspondingly, the rules for
%Rust statements as we have presented so far only operate on the lifetime
However, in the presence of nested borrows, the set of borrows stored in a place
$p$ can be modified via \emph{references} to (parts of) $p$. To support nested borrows, we
must update our rules to account for changes in the sets of borrows denoted by our lifetime projection nodes in such cases.

We solve this issue using a special kind of placeholder node containing a \emph{future lifetime projection}, denoted $\lproj{p}{\tick{a}~\at~\future}$. These nodes are used to track the borrow flow edges that \emph{will} need to be incoming to a lifetime projection node $\lproj{p}{\tick{a}}$ by the time it next becomes available (when $p$ itself becomes a place node with read/exclusive capability). We employ such future lifetime projection nodes for each lifetime that occurs nested in the type of a borrow of $p$, for as long as $p$ itself remains borrowed; once it becomes available, its lifetime projection nodes take the place of the placeholders.

To maintain appropriate borrow-flow edges into these future nodes, we systematically
adapt each of our rules to: (1) introduce new future nodes and
initialise their edges, (2) update edges into these future nodes as further
borrows/moves occur that could contribute to their eventual sets, (3) replace
the future nodes with regular lifetime projection nodes when the borrows end.

For point (1), whenever a place $p$ is borrowed, we label its previous lifetime
projection nodes (if any), and add fresh future nodes for each. To reflect the
possible sources of borrows eventually stored in these new nodes, we add two
edges into each: one from the labelled previous node, and one from the
corresponding lifetime projection node in the newly-created borrow.  For
example, in \figref{approach:borrow-f}, when the borrow \code{rf = \&mut f} is
created, the PCG analysis introduces a future lifetime projection
$\lprojtt{f}{a0~\at~\future}$ to represent borrows that will be stored in
\code{f} under lifetime $\ticktt{a0}$ once the borrow in \code{rf} expires, and
connects this node to two newly introduced edges ($\star$, $\square$).

For point (2), whenever a place $q$ is moved or borrowed into place $p$, if
$q$'s lifetime projections previously had outgoing edges to future nodes (say,
for a place $r$), these future nodes get updated edges from $p$'s corresponding
lifetime projections, reflecting that $p$ may now be used to change the
stored borrows. When $q$ is \emph{borrowed} by $p$, these new dependencies are added
\emph{via} the new future nodes created for $q$, as $q$ may
become usable again before we unwind to $r$. This is shown in
\figref{nested-after-reassign}: after $\code{(*rf).y = r3}$, the
target of $\triangle$ and the source of $\square$ are changed from
{\lprojtt{rf}{a1}} to the newly-introduced lifetime projections
{\lprojtt{rf}{a1~at~\ref{line:motivation:replace-y}}} and
{\lprojtt{rf}{a1~at~\future}} respectively.

Finally, for point (3), when a place node newly becomes a leaf (with read/exclusive capability), any future nodes for its lifetime projections are replaced by current (unlabelled) lifetime projections: this is the moment that the new sets of borrows become available via the place once more.

Based on these principles, we extend our rule for creating borrows as follows:
\[
\triple{\comp{\comp{C}{
  \begin{array}{l}
  \nodeTwo{\placecaptt{p}{\capwrite}}{},
  \nodeTwo{\placecaptt{q}{\capexclusive}}{},\\
  \Vec{\LPnodeTwo{\lproj{\code{q}}{\tick{a}_i}}{\alpha_i}},\\
  \Vec{\edge{\borrowFlowType}{\alpha_i}{\delta_{ij}}}\\
  \end{array}
  }}{G}}
  {\code{p = \&mut q}}
  {\comp{\comp{\sigma(C)}{
  \begin{array}{l}
  \nodeTwo{\placecaptt{p}{\capexclusive}}{},
  \edge{\borrowType}{\nodeTwo{\placecaptt{q}{\capnone}}{}}{\LPnodeTwo{\lproj{\code{p}}{\tick{b}}}{}}, \\
  \Vec{\edge{\borrowFlowType}{
    \LPnodeTwo{\lproj{\code{q}}{\tick{a}_i~\at~l}}{\alpha_i}
  }{
    \LPnodeTwo{\lproj{\code{p}}{\tick{b}_{i}}}{\beta_i}
  }},\\
  \Vec{\edge{\borrowFlowType}{
    \alpha_i
  }{
    \LPnode{\lproj{\code{q}}{\tick{a}_i~\at~\future}}{\gamma_i}
  }},\\
  \Vec{\edge{\borrowFlowType}{
    \beta_i
  }{
    \gamma_i
  }},
  \Vec{\edge{\borrowFlowType}{\gamma_i}{\delta_{ij}}}\\
  \end{array}
    }
  }{\sigma(G)}}
\]
where $a_i$ ranges over the lifetimes in $\code{q}$ and $b_i$ is the lifetime
corresponding to $a_i$ in $\code{p}$. $\delta_{ij}$ is the target of $j$'th edge
originating from the $i$'th lifetime projection of $\code{q}$ (such a node is
guaranteed to be a future lifetime projection).  We extend the rules for moves,
packs, and unpacks analogously.

% TODO link above and below?

\subsection{Computing Joins and Tracking Path-Sensitive Borrows}\label{sec:approach:paths}

\as{When the control flow of a program branches, PCGs for each branch are
computed separately. These results are \emph{joined} to a single graph from which our analysis continues as usual. Our join is designed to retain relevant path-sensitive information (in particular concerning differences in borrow flow in the branches) while minimising duplication of both work and graph (sub-)structure.

% can then be further built upon (or expired, as borrow extents end), according to our analysis.

% More subtly, we also manage to share maximal information from the graphs before the join point.

Our join for place nodes is straightforward: only those which are leaves in both branches make up the place node leaves of the joined graph; this reflects the fact that Rust's rules for which places can be accessed are not path-sensitive. Although which places are considered borrowed is also not path-sensitive, \emph{how and why} they are borrowed may depend on the path taken; as motivated in \secref{motivation:pathsensitivity} this information is relevant for reasoning about the borrows and the aliases and side-effects they induce.}
%
%Borrows which survive the join will also be represented by a lifetime projection node for one of these joined leaves, but their connections back into the results of analysing the two branches must be \emph{preserved}, to retain precise modelling of side-effects (and borrow expiries) performed later.
%%
%the state after the branch. The analysis then continues from the joined state.
%In this way, the analysis avoids the path explosion that would otherwise occur
%(i.e. by performing a separate analysis for each path in the program).
%
%Computing the join for place nodes is straightforward
%
% is straightforward and is computed in
%a pointwise manner: conceptually, if a place is e.g. moved-out in one branch, it
%must be effectively considered to be moved-out in both branches, as the Rust's
%ownership model is not path-sensitive.  However, precisely tracking borrows in
%the general case requires considering path-sensitive information: the target of
%a borrow may depend on the previous control flow of the program.
We describe next how our PCGs track this path-sensitive information
precisely and efficiently (\as{by sharing maximal graph structure}), using
\figref{pathsensitive-example} as a demonstrative example.
\begin{figure}[t]
    \def\rectleft{-2.4}
    \def\rectright{1}
    \def\recttop{1.4}
    \def\rectbottom{-0.4}
    \def\bigrectbottom{-2.4}
    \def\xpos{-1.4}
    \def\ypos{-0}
%   \begin{minipage}{0.38\textwidth}
%     \begin{subfigure}[b]{\textwidth}
%         \begin{rust}
% fn f(mut x: i32, mut y: i32,
%                  mut z: i32) {
%   let r = if x > y {  // c0/c1
%       &mut x          // bb1 (c0)
%   } else {
%       &mut y          // bb2 (c1)
%   };                  // bb3
%   let s = if z > 5 {  // c2/c3
%       &mut *r         // bb4 (c2)
%   } else {
%       &mut z          // bb5 (c3)
%   };
%   *s = 5;             // bb6
% }\end{rust}
%         \caption{A function with path-sensitive borrows}
%         \label{fig:pathsensitive}
%     \end{subfigure}
%     \end{minipage}
    \begin{subfigure}[b]{0.32\textwidth}
        \centering
        \begin{tikzpicture}
            \node[pcg] (x) at (\xpos, 1) {$\code{pos.x}$: $\capnone$};
            \node[pcg] (y) at (\ypos, 1) {$\code{pos.y}$: $\capexclusive$};
            \node[pcg] (r) at (\ypos, 0) {$\code{max2}$: $\capexclusive$};
            \node[lproj] (lproj) at (\xpos,0) {$\lprojtt{max2}{a}$};

            \draw[borrow] (x.south) -- node[right] {\footnotesize $c_0$} (lproj.north);

            \draw (\rectleft,\rectbottom) rectangle (\rectright,\recttop);
        \end{tikzpicture}
        \caption{\code{bb1}}
        \label{fig:pathsensitive-pcg-bb1}
    \end{subfigure}
    \hfill
    \begin{subfigure}[b]{0.32\textwidth}
        \centering
        \begin{tikzpicture}
            \node[pcg] (x) at (\xpos, 1) {$\code{pos.x}$: $\capexclusive$};
            \node[pcg] (y) at (\ypos, 1) {$\code{pos.y}$: $\capnone$};
            \node[pcg] (r) at (\ypos, 0) {$\code{max2}$: $\capexclusive$};
            \node[lproj] (lproj) at (\xpos,0) {$\lprojtt{max2}{a}$};

            \draw[borrow] ($(y.south) - (0.3,0)$) -- node[right=4pt] {\footnotesize $c_1$} (lproj.north);

            \draw (\rectleft,\rectbottom) rectangle (\rectright,\recttop);
        \end{tikzpicture}
        \caption{\code{bb2}}
        \label{fig:pathsensitive-pcg-bb2}
    \end{subfigure}
    \hfill
    \begin{subfigure}[b]{0.32\textwidth}
        \centering
        \begin{tikzpicture}
            \node[pcg] (x) at (\xpos, 1) {$\code{pos.x}$: $\capnone$};
            \node[pcg] (y) at (\ypos, 1) {$\code{pos.y}$: $\capnone$};
            \node[pcg] (r) at (\ypos, 0) {$\code{max2}$: $\capexclusive$};
            \node[lproj] (lproj) at (\xpos,0) {$\lprojtt{max2}{a}$};

            \draw[borrow] (x.south) -- node[left] {\footnotesize $c_0$} (lproj.north);
            \draw[borrow] (y.south) -- node[right=4pt] {\footnotesize $c_1$} (lproj.north);

            \draw (\rectleft,\rectbottom) rectangle (\rectright,\recttop);
        \end{tikzpicture}
        \caption{\code{bb3}}
        \label{fig:pathsensitive-pcg-join}
    \end{subfigure}
    \begin{subfigure}[b]{0.32\textwidth}
      \centering
      \begin{tikzpicture}
          \node[pcg] (x) at (\xpos, 1) {$\code{pos.x}$: $\capexclusive$};
          \node[pcg] (y) at (\ypos, 1) {$\code{pos.y}$: $\capexclusive$};
          \node[pcg] (r) at (\ypos, 0) {$\code{max2}$: $\capwrite$};
          \node[pcg] (z) at ($(r) - (0,1)$) {\code{z}: \capnone};
          \node[lproj] (sproj) at ($(z) - (1,1)$) {\lprojtt{max3}{b}};
          \draw[borrow] (z.south) -- node[right] {\footnotesize $c_2$} (sproj.north);

          \draw (\rectleft,\bigrectbottom) rectangle (\rectright,\recttop);
      \end{tikzpicture}
      \caption{\code{bb4}}
      \label{fig:pathsensitive-pcg-bb4}
    \end{subfigure}
    \hfill
    \begin{subfigure}[b]{0.32\textwidth}
      \centering
      \begin{tikzpicture}
          \node[pcg] (x) at (\xpos, 1) {$\code{pos.x}$: $\capnone$};
          \node[pcg] (y) at (\ypos, 1) {$\code{pos.y}$: $\capnone$};
          \node[pcg] (r) at (\ypos, 0) {$\code{max2}$: $\capnone$};
          \node[lproj] (lproj) at (\xpos,0) {$\lprojtt{max2}{a}$};

          \draw[borrow] (x.south) -- node[left] {\footnotesize $c_0$} (lproj.north);
          \draw[borrow] (y.south) -- node[right=4pt] {\footnotesize $c_1$} (lproj.north);

          \node[pcg] (dr) at ($(lproj) - (0,1)$) {\code{*max2}: \capnone};
          \node[hyperedge, fit=(r)(lproj)] (hyperedge) {};
          \draw[expand] let
            \p1=($(hyperedge.south) - (0.5,0)$),
            \p2=(dr.north),
            \p3=(lproj)
            in  (\x3,\y1) -- (\x3,\y2);

          \node[lproj] (sproj) at ($(dr) - (0, 1.0)$) {\lprojtt{max3}{b}};
          \draw[borrow] (dr) -- node[right] {\footnotesize $c_3$} (sproj);

          \draw (\rectleft,\bigrectbottom) rectangle (\rectright,\recttop);
      \end{tikzpicture}
      \caption{\code{bb5}}
      \label{fig:pathsensitive-pcg-bb5}
    \end{subfigure}
    \hfill
    \begin{subfigure}[b]{0.32\textwidth}
      \centering
      \begin{tikzpicture}
          \node[pcg] (x) at (\xpos, 1) {$\code{pos.x}$: $\capnone$};
          \node[pcg] (y) at (\ypos, 1) {$\code{pos.y}$: $\capnone$};
          \node[pcg] (r) at (\ypos, 0) {$\code{max2}$: $\capnone$};
          \node[lproj] (lproj) at (\xpos,0) {$\lprojtt{max2}{a}$};

          \draw[borrow] (x.south) -- node[left] {\footnotesize $c_0$} (lproj.north);
          \draw[borrow] (y.south) -- node[right=4pt] {\footnotesize $c_1$} (lproj.north);

          \node[pcg] (dr) at ($(lproj) - (0,1)$) {\code{*max2}: \capnone};
          \node[hyperedge, fit=(r)(lproj)] (hyperedge) {};
          \draw[expand] let
            \p1=($(hyperedge.south) - (0.5,0)$),
            \p2=(dr.north),
            \p3=(lproj)
            in  (\x3,\y1) -- (\x3,\y2);

          \node[lproj] (sproj) at ($(dr) - (0, 1.0)$) {\lprojtt{max3}{b}};
          \draw[borrow] (dr) -- node[right] {\scriptsize{$c_2$}} (sproj);
          \node[pcg] (z) at ($(r) - (0,1)$) {\code{z}: \capnone};
          \draw[borrow] (z.south) -- node[right] {\scriptsize{$c_3$}} (sproj.north);

          \draw (\rectleft,\bigrectbottom) rectangle (\rectright,\recttop);
      \end{tikzpicture}
      \caption{\code{bb6}}
      \label{fig:pathsensitive-pcg-bb6}
    \end{subfigure}
    \caption{Example demonstrating joins in the PCG for \figref{motivation:flowsensitive}: the state at \code{bb3}
    is computed by joining the states at \code{bb1} and \code{bb2} (where $c_0$ and $c_1$ are the respective branch-choice identifiers), and
    \code{bb6} is the join of the states at \code{bb4} and \code{bb6} (where $c_2$ and $c_3$ are the respective branch-choice identifiers).}
    \label{fig:pathsensitive-example}
  \end{figure}

Our analysis represents path-sensitive information by associating each \as{choice at each} branch in the CFG with a unique \emph{branch-choice identifier}\asout{: a unique label representing the flow of control from one basic block to another}.
Our analysis tracks the sequences of branch-choices taken to reach the current basic block; when a new branch is taken, each \as{pre-existing} edge is labelled with its identifier; when an edge is created, it is labelled with the tracked set.

At a join point, the edges from all joined graphs are included (retaining any labels) and edges shared between joined PCGs have their sets of labels \emph{merged} \as{(unioned); whenever this results in a \emph{complete} set of branch-choice identifiers for any given branch point, these are removed (the edge exists for each choice at this branch point).}
As an example, the graph in \figref{pathsensitive-pcg-join} after the join point in \texttt{bb3} retains edges from both branches (labelled $c_{0}$ and $c_{1}$).
Edges which existed before the branch point and \emph{survive} in either graph will be preserved without duplication: the graph for \code{bb3} is not duplicated in the final result.
\jg{
As a result, a PCG can be seen as the superimposition of a set of DAGs, one for each path through the CFG, \as{with maximal sharing between each}.
Given a set of branch-choice identifiers describing a particular path through the program, one can recover the DAG for that path by considering only edges labelled with superset of the path's identifiers.
}

\subsection{Tracking Reborrowing Across Function Boundaries with Abstraction Edges}\label{sec:approach:functions}
Calling a Rust function can cause borrows to be created, ended and returned (via different places) to a caller (\cf{} \secref{motivation:functioncalls}). To support function-modular Rust analyses, our
PCG analysis uses \emph{abstraction edges} to capture, as precisely as \as{is guaranteed by the function's \emph{signature} (and Rust's type-checking rules)}, the
\as{borrow flows and corresponding aliasing potentially resulting from a function call.
Abstraction edges are conceptually similar to borrow-flow edges, but are used to represent borrow flow that could not be (modularly) precisely-tracked: both for function calls and for loops (\cf \secref{approach:loops}).}
%
%which borrows might be taken from which, but
%
%in
%the PCG have a different meaning. Namely, the PCG uses borrow-flow edges to relate memory locations
%where precise aliasing information can be recovered, \eg{} when a borrow is
%stored in a struct. Abstraction edges are used when the relation between the
%memory locations cannot be known statically (in a modular analysis), \ie{} for
%function calls and loops (which we discuss in \secref{approach:loops}).
%
%}

In the MIR, a function call is a statement of the form $p =
f(\textit{operands})$, where \emph{operands} is a list of places that are either
moved out or copied, and $p$ is the place where the result of the function will
be stored. From the caller's perspective, each place in the function arguments
either is unchanged or becomes inaccessible; as a result, the caller cannot
observe any changes that the function makes to the \emph{owned} parts of these
places. However, the function can make observable effects \as{ via the \emph{borrows}
stored within the operands (a) by returning a value that contains \emph{reborrows} taken from borrows in the operands, and (b) by \as{changing} the sets of \emph{nested} borrows passed in the operands.}

We model the first case by introducing abstraction edges between lifetime projections in the
operands and those in the result place: each lifetime projection in the operands is
connected to the corresponding lifetime projections that it \emph{outlives} in the result place. \as{The (compiler-checked) outlives constraint captures whether borrows \emph{could} be assigned in this way, according to the type system. \figref{call_max} illustrates the abstraction edges which (modularly) model a call to the \code{max()} function:}
based on the types of \code{res}, \code{rx}, and \code{ry}, \code{res}
may borrow from either variable.% The PCG reflects this information by adding
%abstraction edges between the labelled lifetime projections for \code{rx} and
%\code{ry} with those in the result \code{res}. To obtain precise aliasing
%information, it would be necessary to consider the definition of the
%\code{max()} function, therefore making the analysis non-modular.

\as{The second case is closely related to \secref{approach:nestedborrows}: changes to a set of nested borrows may become relevant through a previously-borrowed place: in this case, future lifetime projections will be dependent on these sets before the call. As explained (\secref{approach:nestedborrows}), changes to these sets of borrows made by the call must be reflected onto these future nodes. In particular, changes made during a function call might necessitate additional dependencies on function arguments (\cf{} \figref{motivation:nestedborrows}).}

%The borrows that these sets \emph{could end up} storing might flow from other arguments provided to the call.
%
%The second case is somewhat more subtle. In the absence of nested borrows, a function call could use
%borrows stored in the operands to change the \emph{values} stored in the caller's places, but
%because such places themselves would not contain any borrows, none of the caller's borrows could be
%changed. On the other hand, the presence of nested borrows would allow the function to manipulate the sets of borrows stored in the caller's places, as in \figref{motivation:nestedborrows}.
%
\as{To directly model the potentially \emph{changed} sets of borrows relevant to these concerns, our analysis of function calls introduces}
%We observe that the sets of nested borrows in the function operands therefore
%behave as both inputs \emph{and} outputs of the function. Therefore, we model
%the second case by introducing
lifetime projections to represent the
\emph{post-state} of each lifetime projection in the operands. Each lifetime
projection in the operands is connected with abstraction edges to its
corresponding post-state projection as well as the post-state \emph{nested}
lifetime projections that it outlives \as{(analogously to sets of borrows explicitly returned)}. In addition, any edges originating from
lifetime projections in the arguments of the function to future nodes are
redirected to originate from the corresponding post-state lifetime projection
instead.

In summary, we define the PCG analysis rule for function calls as follows:
\[
\triple{\comp{\comp{C}{
  \begin{array}{l}
  \node{p: \capwrite},
  \Vec{\node{q_i: \capexclusive}},\\
  \Vec{\LPnodeTwo{\lproj{q_i}{\tick{r_j}}}{\alpha_{ij}}},\\
  \Vec{\edge{\borrowFlowType}{\alpha_{ij}}{\delta_{ijm}}}
  \end{array}
}}{G}}
{l: p = f(\Vec{q_i})}
{\comp{\comp{C}{
  \begin{array}{l}
  \node{p: \capexclusive},
  \Vec{\node{q_i: \capwrite}},
  \Vec{\LPnodeTwo{\lproj{q_i~\at~l}{\tick{r_j}}~\at~l}{\alpha'_{ij}}},\\
  \Vec{\LPnodeTwo{\lproj{p}{\tick{s_k}}}{\beta_{k}}},
  \Vec{\LPnodeTwo{\lproj{q_i~\at~l}{\tick{r_{j}}}~\at~\code{POST}_l}{\gamma_{ij}}},\\
  \Vec{\mathit{RES}},
  \Vec{\mathit{POST}},
  \Vec{\edge{\borrowFlowType}{\gamma_{ij}}{\delta_{ijm}}}
  \end{array}
}}{G}}
\]
Here, $j$ indexes the lifetime projections in $q_i$, $k$ indexes the lifetimes
in the type of $p$, and $m$ indexes the target (future lifetime projection) nodes $\delta_{ijm}$ (if any) of edges from
$\alpha_{ij}$.  $\mathit{RES}$ is the set of edges to the lifetime projections of $p$;
$\mathit{RES}$ contains the edge
$\edge{\abstractionType}{\alpha'_{ij}}{\beta_{k}}$ for each $i, j, k$ where the
j'th lifetime in the type of $q_i$ outlives the k'th lifetime in the type of
$p_k$.
$\mathit{POST}$ is the set of edges to the post-state lifetime projections of
the operands, containing the edge
$\edge{\abstractionType}{\alpha'_{ij}}{\gamma_{uv}}$ for all $i,j,u,v$ where
either $i,j$ = $u,v$ or the j'th lifetime of $q_i$ is a nested lifetime that
outlives the v'th lifetime of $q_u$.% The (possibly zero) nodes $\delta_{ijm}$ are the future nodes

\begin{figure}[t]
        \centering
        \begin{tikzpicture}
            \node[pcg](pos) at (-5,1) {\placecaptt{pos}{\capnone}};
            \node[pcg](posx) at (-5,0) {\placecaptt{pos.x}{\capnone}};
            \node[pcg](posy) at (-3,0) {\placecaptt{pos.y}{\capnone}};
            \node[hyperedge, fit=(posx)(posy)] (exp) {};
            \draw[expand] (pos) -- (exp);
            \withproj[0]{rx}{rx~at~\ref{line:motivation:call-max}}{\capnone}{a~\at~\ref{line:motivation:call-max}}{(-1.3,1)}
            \withproj[0]{ry}{ry~at~\ref{line:motivation:call-max}}{\capnone}{b~\at~\ref{line:motivation:call-max}}{(1.3,1)}
            \node[lproj] (resproj) at ($(rxproj)!0.5!(ryproj) + (0,-1.0)$) {$\lprojtt{res}{r}~\code{at}~\ref{line:motivation:res-dies}$};
            \node[pcg,anchor=east] (res) at ($(resproj.west) - (0.1,0)$) {$\code{res}$: $\capnone$};
            \node[pcg] (dres) at (0.2,-1.3) {$\code{*res}$: $\capexclusive$};

            % Edges
            \draw[abstract] (rxproj) -- (resproj.north);
            \draw[abstract] (ryproj) -- (resproj.north);
            \draw[borrow] (posx) .. controls +(0,1) and +(0,1.1) .. ($(rxproj.north) + (0.5,0)$);
            \draw[borrow] (posy) .. controls +(0,2) and +(0,1.1) .. ($(ryproj.north) + (0.5,0)$);
            \drawderef{res}{dres}
        \end{tikzpicture}
    \caption{PCG generated for the last line in \figref{motivation:functioncall}, including the abstraction edges added for the call to \texttt{max}.}
    \label{fig:call_max}
\end{figure}

\subsection{Reborrowing Across Loops}\label{sec:approach:loops}
\as{Loops can create, end and move borrows, and reborrow: }
%
%Unlike function calls, Rust loops come with no annotations prescribing the
%
%In the presence of loops, it is not possible to precisely track borrowing: in
%particular,
the extent of borrows created \emph{inside} the loop may extend
beyond the loop. \as{These behaviours cannot be tracked precisely; our analysis constructs a \emph{loop invariant} graph structure to summarise them. The parts of a PCG changed by the loop are replaced by our summarised loop invariant graph. This must have leaf nodes for the places accessible at the loop head (and after the loop), and be connected above to the closest borrowed-from nodes which are \emph{not} touched by the loop, which we call the \emph{loop borrow roots}. Its structure must capture intermediate (re)borrowed places that can become leaves when other borrows end: these can be more-general DAGs than suffice to summarise function calls.}

\as{A key design choice for our lifetime projection nodes is to enable abstracting our detailed graphs to capture
%Our key insight is that the structure of PCGs reflects
the constraints
enforced by the borrow checker on its corresponding notions of tracked borrow
extents: these constraints correspond to reachability between the corresponding lifetime projections.
This property is maintained by careful design of our analysis rules and overall workflow, including queries to the compiler itself for some rules, as explained throughout this section.}

% Although our lifetime projection nodes provide more granularity than any current borrow checker for Rust, for any two such nodes which \emph{do} represent borrow extents tracked by the borrow checker
%
%, their reachability
%
%because whenever we introduce lifetime
%projection nodes that might (according to the borrow checker) correspond to
%borrows on which the borrow checker imposes constraints, we query the borrow
%checker and add the corresponding edges.
%
\as{Exploiting this design, we define an algorithm to automatically find an appropriate loop invariant graph by combining our existing analysis rules with the compiler's borrow-checking results.
}
%%Based on this insight, our analysis inserts the loop invariant into the PCG
%state for the loop head by computing the shape based on information from the
%compiler and borrow checker, and replacing the corresponding concrete subgraph
%in the pre-loop state with the computed shape.
Algorithm~\ref{alg:construct-loop-invariant} provides a high-level overview of the algorithm, which we now describe in detail.

\begin{algorithm}[t]
\caption{ConstructLoopInvariant - Main Algorithm}
\label{alg:construct-loop-invariant}
\begin{algorithmic}[1]
\Procedure{ConstructLoopInvariant}{$\mathcal{G}_{\text{pre}}, P_{\text{blocked}}, P_{\text{blockers}}, l_h$}
    \State $P_{\text{loop}} \gets P_{\text{blocked}} \cup P_{\text{blockers}}$ \Comment{1. Identify relevant loop places}
    \State $P_{\text{blocked}} \gets P_{\text{blocked}} \cup (\bigcup_{p \in P_{\text{loop}}} \Call{FindRoots}{p, \mathcal{G}_{\text{pre}}})$ \Comment{2. Add loop borrow roots}

    \State $\mathcal{A} \gets \Call{MakeInvariantGraph}{P_{\text{blocked}}, P_{\text{blockers}}, l_h}$
    \Comment{3. Make invariant for loop head $l_h$}

    \State $\mathcal{G}_{\text{cut}} \gets \Call{ExtractSubgraph}{\mathcal{G}_{\text{pre}}, \mathcal{A}}$
    \Comment{4. Identify subgraph to abstract}

    \State \textbf{return} $(\mathcal{G}_{\text{pre}} \setminus \mathcal{G}_{\text{cut}}) \cup \mathcal{A}$
    \Comment{5. Replace subgraph with abstraction}
\EndProcedure
\end{algorithmic}
\end{algorithm}

% Step 1

\as{Our loop invariant graph concerns those borrows that are live at the loop head and may
differ across iterations of the loop (all other borrows are either
unchanged or only exist temporarily within the loop). It is defined in terms of places that could store these borrows (the
\emph{blocking places}), as well as the targets of the borrows themselves (the
\emph{blocked places}).

We start by defining the \emph{live loop places}: those places that are live at the loop head and used inside the loop. We define the \emph{blocking places} as the subset
of live loop places whose types allow storing borrows. The \emph{blocked places} are the union of (a) the live loop
places that could become blocked during the loop (identified via a
straightforward analysis of the borrow assignments in the loop), and (b) the closest ancestors of the blocking places which are live (but \emph{not} used in the loop body) and borrowed at the loop head; this latter set are the \emph{loop borrow roots}.}
%
%
% the places
%that were borrowed before the loop and could become unblocked due to the
%reassignment of a borrow inside the loop, which we call the .
%We identify the loop borrow roots by querying the PCG state \emph{before} the loop to
%find the targets of the blocking loop places, concretely by
%taking the \as{set of closest ancestors of each of the} blocking loop
%places.

% \begin{algorithm}
% \caption{FindRoots - Identify Loop Borrow Roots}
% \label{alg:find-roots}
% \begin{algorithmic}[1]
% \Procedure{FindRoots}{$p, \mathcal{G}, \mathcal{N}_{\text{loop}}, l_h$}
%     \State $\mathcal{N}_{\text{roots}}^p \gets \emptyset$;
%     $L \gets \{\lproj{n}{r}~|~r~\text{is a lifetime in the type of}~p\}$
%     \While{$L \neq \emptyset$}
%         \State $n \gets \Call{Pop}{L}$
%         \ForAll{node $n'$ blocked by $n$ in $\mathcal{G}$}
%             \If{$n' \in \mathcal{N}_{\text{roots}}^p$ or $n' \in \mathcal{N}_{\text{loop}}$}
%                 \textbf{continue}
%             \ElsIf{$n'$ is live at $l_h$ or $n'$ is a root of $\mathcal{G}$ }
%                 $\mathcal{N}_{\text{roots}}^p \gets \mathcal{N}_{\text{roots}}^p \cup \{n'\}$
%             \Else{}:
%                 $L \gets L \cup \{n'\}$
%             \EndIf
%         \EndFor
%     \EndWhile
%     \State \textbf{return} associated places of $\mathcal{N}_{\text{roots}}^p$
% \EndProcedure
% \end{algorithmic}
% \end{algorithm}

% Step 3
We then construct the loop invariant shape by adding abstraction edges from
blocked places to blocking places, including edges between their
corresponding lifetime projections (Algorithm~\ref{alg:construct-abstraction-graph}).
This shape is constructed by
iterating over each blocked place $p$, and identifying the subset of blocking
places $\overline{p_b}$ that the borrow checker reports as blocking $p$ at the
loop head. For each blocking place $p_b$, a \emph{borrow} edge is added between
$p$ and the lifetime projection corresponding to the borrows in $p_b$ and
\emph{abstraction} edges are added between the lifetime projections of $p$ and
the lifetime projections of $p_b$ (Algorithm~\ref{alg:connect}). \as{The rules for inserting such edges are analogous to those we described for direct borrows in \secref{approach:borrow} and \secref{approach:nestedborrows}}
%essentially the same as the rule for modelling explicit mutable borrows
(including handling of future nodes and edges).

\begin{algorithm}[t]
\caption{MakeInvariantGraph}
\label{alg:construct-abstraction-graph}
\begin{algorithmic}[1]
\Procedure{MakeInvariantGraph}{$P_{\text{blocked}}, P_{\text{blockers}}, l_h$}
    \State $\mathcal{A} \gets$ new empty graph
    \ForAll{$(p_{\text{blocker}}, p_{\text{blocked}}) \in P_{\text{blockers}} \times P_{\text{blocked}}$}
          \If{$p_{\text{blocker}}$ blocks $p_{\text{blocked}}$ at $l_h$}: $\Call{Connect}{\mathcal{A}, p_{\text{blocked}}, p_{\text{blocker}}}$
          \EndIf
    \EndFor
    \State Label all non-leaf lifetime projection nodes in $\mathcal{A}$ with $l_h$; \textbf{return} $\mathcal{A}$
\EndProcedure
\end{algorithmic}
\end{algorithm}
\begin{algorithm}[t]
\caption{Connect - Add Edges to Abstraction Graph}
\label{alg:connect}
\begin{algorithmic}[1]
\Procedure{Connect}{$\mathcal{A}, p_{\text{src}}, p_{\text{tgt}}$}
    \State Insert $\edge{\borrowType}{\nodeTwo{p_{\text{src}}: \capnone}{}}{\LPnodeTwo{p_{\text{tgt}} \downarrow r}{}}$ into $\mathcal{A}$
    \Comment Where $r$ is the lifetime of $p_{\text{tgt}}$
    \ForAll{lifetime $r$ in the type of $p_{\text{src}}$}
        \State $\RP_{\text{mut}} \gets \{\LPnodeTwo{p_{\text{tgt}} \downarrow r'}{}~|~r' \text{is a lifetime of}~{p_{\text{tgt}}}, r~\text{outlives}~r'~\text{and}~r'~\text{outlives}~r\}$
        \If{$\RP_{\text{mut}} \neq \emptyset$};
          \State Add
            $\edge{\abstractionType}{\LPnodeTwo{p_{\text{src}} \downarrow r}{\alpha}}{\LPnodeTwo{p_{\text{src}} \downarrow
            r~\mathtt{at~\future}}{\beta}}$
            \ForAll {$\rp$ in $\RP_{\text{mut}}$}
            \State Add $\edge{\abstractionType}{\alpha}{\rp}$, $\edge{\abstractionType}{\rp}{\beta}$
            \EndFor
        \EndIf
        \State $\RP_{\text{flow}} \gets \{\LPnodeTwo{p_{\text{tgt}} \downarrow r'}{}~|~r' \text{is a lifetime of}~{p_{\text{tgt}}}, r~\text{outlives}~r'\} \setminus \RP_{\text{mut}}$
        \ForAll {$\rp$ in $\RP_{\text{flow}}$}
            \State Add $\edge{\abstractionType}{\LPnode{p_{\text{src}} \downarrow r}}{\rp}$
        \EndFor
    \EndFor
\EndProcedure
\end{algorithmic}
\end{algorithm}

% Step 5
Having computed the loop invariant shape, our analysis identifies the
corresponding subgraph in the PCG state $\mathcal{G}_\text{pre}$ that this
abstraction will replace. The subgraph is constructed by including each edge $e$
in $\mathcal{G}_\text{pre}$ where there exists nodes $m$ and $n$ such that $e$
is along a path from $m$ to $n$ in $\mathcal{G}_\text{pre}$ and $m$ is an
ancestor of $n$ in the abstraction shape. Replacing the subgraph with the loop
invariant yields the PCG for the \as{program point at the} loop head.

\figref{approach:complex} shows the invariant generated for the loop in
\figref{motivation:loop}. The places \texttt{prev} and \texttt{current} are
considered live at the loop head. Because both of their types can contain
borrows, they are considered as candidate blocker nodes in the graph.  Neither
is borrowed at the loop head, so they are not considered blocked nodes. We
identify the place \texttt{*list} as a loop borrow root because it is the closest
live ancestor of the lifetime projection $\LPnodeTwo{\lprojtt{current}{a}}{}$ in
the PCG just prior to the loop (\figref{approach:looppre}). Because the borrow
checker indicates that \texttt{*list} is blocked by both \texttt{current} and
\texttt{prev}, our algorithm adds borrow edges from \texttt{*list} to the
lifetime projections of \texttt{current} and \texttt{prev}.

\begin{figure}[t]
  \begin{subfigure}[b]{0.48\textwidth}
    \centering
    \begin{tikzpicture}
      \node[pcg] (list) at (0,0) {\placecaptt{list}{\capwrite}};
      \node[lproj] (listproj) at (2,0) {\lprojtt{list}{a~\at~2}};
      \node[pcg] (dlist) at (2,-1) {\placecaptt{*list}{\capnone}};
      \node[lproj] (currentproj) at (2, -2) {\lprojtt{current}{a}};
      \node[pcg] (prev) at (4, 0) {\placecaptt{prev}{\capexclusive}};
      \node[lproj] (prevproj) at (4, -1) {\lprojtt{prev}{a}};
      \node[pcg] (current) at (0, -2) {\placecaptt{current}{\capexclusive}};
      \node[hyperedge, fit=(list)(listproj)] (derefsrc) {};
      \draw[expand] (derefsrc) -- (dlist) {};
      \draw[borrow] (dlist) -- (currentproj) {};
    \end{tikzpicture}
    \caption{PCG state before loop head}
    \label{fig:approach:looppre}
  \end{subfigure}
  \hfill
  \begin{subfigure}[b]{0.48\textwidth}
    \centering
    \begin{tikzpicture}
      \node[pcg] (list) at (0,0) {\texttt{*list}: \capnone};
      \node[lproj] (current) at (-1,-1) {\lprojtt{current}{a}};
      \node[lproj] (prev) at (1,-1) {\lprojtt{prev}{a}};

      \draw[borrow] (list) -- (current);
      \draw[borrow] (list) -- (prev);

    \end{tikzpicture}
    \caption{Loop invariant}
  \end{subfigure}
  \caption{The PCG loop invariant constructed for the \texttt{penultimate\_mut}
  function in \figref{motivation:loop}. The loop invariant is constructed at the
  loop head, without analysing the body of the loop.}
  \label{fig:approach:complexshapeinv}
\label{fig:approach:complex}
\end{figure}

%ALEX: we will incorporate formal details into the existing sections
%\section{Formalisation}\label{sec:form}
%\input{sections/form.tex}

% \section{Implementation}\label{sec:impl}
%   \input{sections/impl.tex}

\section{Evaluation and Implementation}\label{sec:eval}
  In this section, we describe our Rust implementation and its evaluation. We
designed our evaluation to answer the following three questions:

\begin{itemize}
\item[\textbf{(Q1)}] Can our implementation handle most real-world Rust code? (\secref{eval:crates})
% \end{itemize}
% We answer this by testing our implementation on the 500 most downloaded Rust
% crates~(\secref{eval:crates}).
% \begin{itemize}
\item[\textbf{(Q2)}] Can our PCGs serve as the basis for other analysis tools? (\secref{eval:flowistry} and \secref{eval:prusti})
% \end{itemize}
% We answer this by replacing the alias analysis in the Flowistry
% tool~(\secref{eval:flowistry}) and the proof annotation analysis in the Prusti
% verifier~(\secref{eval:prusti}) with our PCG implementation.
% \begin{itemize}
\item[\textbf{(Q3)}] Does our implementation accurately reflect Rust's
borrow-checker constraints? (\secref{eval:mutation})
\end{itemize}
% We answer this by using our implementation to identify mutant versions of Rust
% programs that the borrow checker should \emph{reject}, and checking that these
% mutants are indeed rejected by the compiler~(\secref{eval:mutation}).

\subsection{Implementation}

Our implementation is written in Rust and available as a crate that can be
included as a library. It supports both the current Rust NLL
borrow checker as well as the Polonius borrow checker.\zgout{In addition, the
implementation can also produce graphical debug output that produces diagrams
similar to the figures included in this paper.}

Our library provides multiple interfaces for clients to consume the results of
the analysis. Clients can analyse the PCGs themselves to \eg{},~identify
aliasing relationships (as we do in our Flowistry implementation). Clients can
also consume \emph{annotations} describing the operations that transition
between consecutive PCG states; our prototype Prusti implementation consumes
these annotations to generate Viper assertions. Note that both interfaces expose
path-sensitive information: clients can filter graphs and annotations to include
only those relevant for a particular path.

\subsection{Top 500 Crates}\label{sec:eval:crates}

To evaluate the PCG implementation on a wide range of code, we ran our
implementation on functions from the top 500 most downloaded crates on
\texttt{crates.io}, excluding nine crates that did not compile on our test system
\footnote{
The crates
\texttt{winreg}, \texttt{criterion-plot}, \texttt{tiny-keccak},
\texttt{redox\_users}, \texttt{plotters}, \texttt{darling},
\texttt{derive\_more}, \texttt{tokio-native-tls}, and \texttt{clang-sys} did not compile
either because the operating system was not supported,
necessary external libraries were missing, or the compilation required additional configuration via \texttt{cargo}.}. Table~\ref{tab:pcg-stats} presents statistics \peter{of the PCG analysis per function}.
\begin{table}[t]
\caption{Per-Function Results of PCG Analysis. Failures
\texttt{Expansion\-Of\-Alias\-Type}, \texttt{Indexing\-Non\-Indexable\-Type}, and
\texttt{Assign\-Borrow\-To\-Non\-Reference\-Type} occurred due to \peter{limitations in
the Rust compiler:} normalising a type alias erases lifetime information that is
required for the PCG analysis.
}
\centering
% tabular section generated by analyze_errors crate
\footnotesize
\begin{tabular}{llr}
      \toprule
      \textbf{Status} & \textbf{Details} & \textbf{Count} \\
      \midrule
      Success & & 118,483 \\
      \midrule
      \multirow{7}{*}{Unsupported} & \texttt{DerefUnsafePtr} & 1,584 \\
      & \texttt{CallWithUnsafePtrWithNestedLifetime} & 1,039 \\
      & \texttt{InlineAssembly} & 80 \\
      & \texttt{MoveUnsafePtrWithNestedLifetime} & 73 \\
      & \texttt{ExpansionOfAliasType} & 6 \\
      & \texttt{IndexingNonIndexableType} & 5 \\
      & \texttt{AssignBorrowToNonReferenceType} & 1 \\
      \midrule
      Total & & 121,271 \\
      \bottomrule
      \end{tabular}
\label{tab:pcg-stats}
\end{table}
Our analysis generated PCGs at every program point for the vast
majority of functions (97.7\%) in the evaluated crates. The majority of the remaining functions failed because they used features currently unsupported by our model: unsafe pointers and inline assembly. For 12 functions,
PCG construction failed due to a limitation in resolving type aliases in the
Rust compiler: the compiler's interface for normalising a type also erases
lifetime information, which is necessary for our analysis.

Our evaluation completed in 94 minutes on a six-core Intel Xeon E-2236 with 64
gigabytes of RAM, using the default borrow checker and Rust nightly 2024-12-15.
Although our implementation supports the usage of the next-generation Polonius
borrow checker, we used the standard borrow checker for our evaluation as
\peter{Polonius takes more than an hour on some crates in our suite.}

\subsection{Flowistry}\label{sec:eval:flowistry}

Flowistry~\cite{Flowistry} \jgout{is an IDE tool that}performs an information-flow
analysis to identify statements in a Rust program that might affect the value
of a given expression. \jg{Their implementation includes} a Rust alias analysis.

To \peter{demonstrate} that PCGs can form the basis of other analyses, and
to provide evidence for the correctness of our analysis, we \peter{replaced Flowistry's  alias analysis with a new one that uses our PCG implementation. The resulting} version of Flowistry passed all tests in Flowistry's test suite.

\subsection{Prusti}\label{sec:eval:prusti}

Prusti~\cite{prusti} is a verifier that translates Rust programs into Viper~\cite{viper}, a
verification language based on separation logic. In order to check functional
properties, Prusti first verifies that the borrow checker correctly enforced
memory safety by constructing a \emph{core proof} of memory safety.
To automate this, Prusti uses an ad-hoc analysis to infer a
variety of annotations required for the Viper proof.

We developed a prototype version of Prusti that instead uses our PCG
implementation to generate proof annotations. This significantly simplified
the Prusti codebase while also enabling support for some Rust features (such
as references in structs or reborrowing in loops) that the previous analysis
did not support. Our PCGs enable the inference of all annotations required for Prusti's core proof:

\begin{itemize}
  \item Predicate \emph{fold} and \emph{unfold} statements. These are a key feature that enables support for recursive data types. Due to the similarity of separation logic to Rust's ownership system, these have a one-to-one correspondence with our pack and unpack operations.
  \item Loop invariants that specify the loop's memory footprint. These are required for all loops and correspond to the PCG state at the head of the loop.
  \item Magic wand \emph{package} and \emph{apply} statements. These describe how permissions flow from an expired reference to the original owning data structure and are required for reborrowing methods and reborrowing in loops. They correspond to our abstraction edges.
\end{itemize}

\zgout{
Our prototype Prusti queries the PCG for the annotations to be inserted at every MIR location and interleaves these annotations with its own encoding of the operational semantics of the program.
By cleanly separating these two tasks, our approach makes it easier to identify errors in the encoding of the operational semantics (which, in the original Prusti, were often hidden by the complexity of the ad-hoc analysis). Additionally, PCGs facilitate debugging
such errors since memory safety violations can be explained at the level of the MIR, instead of the elaborated Viper program only.
}

% We focused on implementing the features required to show that our PCG model can help building verifiers for Rust, and omitted some orthogonal features, such as support for slices, arrays, closures\ab{, or raw pointers}; calls to trait methods; or polymorphic methods.
Evaluating our prototype against a subset of the Prusti test suite (the should-verify category with overflow checks disabled), encoding \emph{only} the core proof and not the functional correctness specifications, yielded a result of \ab{155 tests passed, 40 failed, and 67 ignored }due to unsupported features.\ab{ Based on manual inspection, the remaining failures are unrelated to the PCG analysis.}
We also successfully verified simplified versions of all the examples presented in this paper.
%, such as programs with reborrowing inside loops, structs with references, or reborrowing functions with multiple pledges.\pmfootnote{Two of the three examples were already given before the bullet list, and readers probably won't know what pledges are.}

\subsection{Mutation Testing}\label{sec:eval:mutation}
To demonstrate that PCGs accurately reflect the restrictions enforced by the Rust compiler, we built a mutation testing tool which uses the PCG to mutate the MIR of well-typed Rust programs into mutants which violate particular ownership or borrowing rules:
\begin{itemize}
  \item
        \texttt{MutablyLendShared} creates a mutable borrow to a place behind a shared borrow.
  \item
        \texttt{WriteToShared} writes to a place behind a shared borrow.
  \item
        \texttt{ReadFromWriteOnly} reads from a place with \textbf{W} capability.
  \item
        \texttt{MoveFromBorrowed} moves out of a place behind a mutable borrow.
  \item
        \texttt{BorrowExpiryOrder} identifies a place $p_{1}$ which blocks another place $p_{2}$ via a mutable borrow and attempts to use $p_{2}$ before $p_{1}$.
  \item
        \texttt{AbstractExpiryOrder} identifies a place $p_{1}$ which blocks another place $p_{2}$ via a mutable borrow and through an abstraction edge and attempts to use $p_{2}$ before $p_{1}$.
\end{itemize}

\begin{table}[t]
\centering
%\small
\footnotesize
\caption{Results of Mutation Testing on top 250 crates.
  \textit{tot} lists the number of mutants produced by the mutation.
  \textit{kill} lists the number of mutants which failed borrow checking, as expected.
  \textit{live} lists the number of mutants which passed borrow checking.
  \textit{\%kill} lists the percentage of mutants which failed borrow checking.
}
\begin{tabular}{lrrrrl}
  \toprule
  \textbf{Mutation} & \textit{tot} & \textit{kill} & \textit{live} & \textit{\%kill} \\
  \midrule
  \texttt{MutablyLendShared}   & 1172766 & 1172421 & 345 & 99.97\% \\
  \texttt{ReadFromWriteOnly}   & 5078080 & 5078080 & 0   & 100.00\% \\
  \texttt{WriteToShared}       & 1172336 & 1171973 & 345 & 99.96\% \\
  \texttt{MoveFromBorrowed}    & 283649  & 283496  & 153 & 99.94\% \\
  \texttt{BorrowExpiryOrder}   & 17119   & 17119   & 0   & 100.00\% \\
  \texttt{AbstractExpiryOrder} & 4817    & 4817    & 0   & 100.00\% \\
  \midrule
  \textit{Total}               & 7728767 & 7727906 & 843 & 99.99\% \\
  \bottomrule
\end{tabular}
\label{tab:fuzzing-stats}
\end{table}

\tabref{fuzzing-stats} reports the results of running our tool on the top 250 most downloaded crates on \texttt{crates.io}.
Note that \texttt{kill} and \texttt{live} do not add up to \texttt{tot}; \texttt{WriteToShared} caused 18 compiler panics not counted towards \texttt{kill} since their cause may be unrelated to type checking.
We also have no data for 9 crates on which our mutation tool crashed.
99.96\% of mutants fail to type check, demonstrating that PCGs reflect Rust's type rules precisely. Of the live mutants we examined, all appeared to violate some rule.
However, we were unable to replicate any live mutants in surface Rust. We suspect this is because the compiler performs some initial borrow checking before generating MIR\footnote{For example, in the \code{mir\_built} query.}. As we mutate downstream of these checks, the compiler may not reject certain mutants.

\section{Related Work and Potential Consumers}\label{sec:related}
  % \begin{figure}[h]
% \begin{rust}
% struct Pair<T> { fst: T, snd: T }
% fn reborrow_pair<T>(pair: &mut Pair<T>) {
%     let (fst, snd) = (&mut pair.fst, &mut pair.snd);~\label{line:create-mut-borrow}~
%     if false { let _ = &(*pair).snd; let _ = &*fst; } // mutation~\label{line:mutant}~
%     let _ = &mut *fst; let _ = &mut *snd;
% }
% \end{rust}
% \caption{Source Rust approximation of a mutant that passes borrow checking}
% \label{fig:pass-example}
% \end{figure}

\peter{We discuss two aspects of related work: we compare existing models for Rust's type checking to PCGs and discuss how users of these models could benefit from the information provided by PCGs.}

\subsubsection*{Prusti}
Prusti's encoding \cite{prusti} was originally presented in terms of \zgout{the idea of a} \emph{place capability summaries}\footnote{The name for our model here is inspired by this notion.}, presented semi-formally as a representation analogous to the place node leaves in our PCGs. Rules for modelling borrows etc.~are formalised only indirectly via the target separation logic. The resulting proofs require detailed annotations, especially for reborrowing (handled by
proving \emph{magic wands}), demanding a step-by-step justification for \emph{how} borrowed capabilities are returned. Prusti previously inferred such annotations through complex algorithms that proved hard to generalise beyond simple reborrowing patterns \cite{prusti-communication}.
The prototype developed in our evaluation \zgout{not only} generates all
necessary annotations directly from our PCG analysis results and can analyse
programs that are not supported in the current version of Prusti, such as those that reborrow inside loops.

\subsubsection*{Oxide}
\as{Oxide \cite{Oxide} is (to our knowledge) the most mature attempt at a formal semantics for a large portion of Rust's type system, including (unlike our work) formalising how type checking can be \emph{performed}: it defines and implements type checking rules, an operational semantics, and proves a type soundness result for a small language including challenging features such as closures.

For Rust \emph{analysis and verification}, however, several key features are absent. Structs and lifetime parameters are not supported: desugaring of structs to tuples is not compatible with general Rust struct definitions which may feature recursive or opaque types. Rust uses lifetime parameters to abstract over \emph{where} in a type borrows may be stored; our lifetime projections capture these abstractions directly and modularly. Oxide's rules also do not capture path-sensitive information, nor show how loop invariants can be computed to summarise loops that interact with (re-)borrowing. The formalism is also for a custom language representation; it's unclear how these results can be connected with practical analysis tools that typically work on Rust's representations (usually MIR).}
%
% Connecting Rust's lifetime parameters with concrete places where borrows are stored and used is a key challenge addressed by our lifetime projections. Oxide also does not represent path-sensitive information at a control-flow join, or provide an algorithm for constructing suitable loop invariants, which for the modern borrow checkers used by the Rust compiler can be surprisingly complex.
%
%Finally, since Oxide's results are provided against a custom language representation, they are not directly reusable for building analysis or verification tools for Rust, almost all of which target a standard Rust representation such as MIR.}

\subsubsection*{Aeneas}
Aeneas \cite{Ho'24} defines formal models of Rust's type checking, and a verification approach mapping (safe) Rust into a functional language for manual proof in proof assistants. Rust moves and borrows are handled by copying values; where reborrows span function calls and loops, additional \emph{backward functions} are generated, defining how to copy corresponding values back again.

Our comparison with Aeneas is similar to that with Oxide: they formalise more in terms of how Rust type checking works, but do not support as large a fragment of Rust (\eg{} no struct lifetime parameters). Reborrowing inside loops is supported by heuristics that need not always succeed~\cite{Ho'24}, and joins do not retain precise path-sensitive information. Connecting with other analyses is (as for Oxide) limited by the use of a custom program representation separate from Rust.

On the other hand, our PCGs could in principle generate the information needed by Aeneas for its verification approach:
%
%Rust lifetime parameters or precise path-sensitive information. Aeneas does include limited support for reborrowing inside loops, but by heuristics which need not succeed~\cite{Ho'24}.
%
%Aeneas' model is fully formalised and (unlike ours) also \emph{defines its own type and borrow checker} (in fact, a class of potential algorithms). However, their model of borrowing has several limitations when compared to ours: (1) it does not represent precise path-sensitive information for (re)borrows, even for loop- and function-free code, (2) there is no support for reborrowing via nested lifetimes in types (for example, borrows stored in a struct type), %which are not mentioned in the formal model or supported by the tool,
%and (3) their technique for computing loop invariants~\cite{Ho'24} Charon\as{relies on heuristics which need not succeed, and} join operations that can introduce less-precise summaries of loops than Rust's borrow checker. Conversely, our modelling of loops captures Rust's borrow-checking constraints precisely, does not require a fixpoint calculation (we reuse the compiler's results), and also applies to types with nested lifetimes.
%
%We are confident that our model could be used to generate the information required by the Aeneas verifier for its \emph{translation}:
managing recursive representations of types is similar, and the generation of backward functions requires steps similar to the proofs of magic wands in Prusti, for which we have demonstrated proof-of-concept support. \peter{Consequently, it may be possible to lift some of the existing limitations of Aeneas by using the information provided by PCGs (as with our Prusti prototype)}, especially with respect to general loops and nested lifetimes.

\subsubsection*{Aquascope}
Aquascope \cite{Crichton'23} is an analysis that visualises the type checking of Rust programs. The tool extracts ownership and borrowing information \zgout{directly }from a customised version of the Rust compiler, even for programs that do \emph{not} type check. While a notion of capabilities (called permissions in their work) per program point is visible in Aquascope\zgout{, as well as whether or not a place is currently borrowed}, information about \emph{why} and \emph{how} a place is borrowed is not available, nor is any information path-sensitive. Consistent with these choices, loops and function calls are not summarised\zgout{, and to our knowledge, nested lifetimes are not directly visualised}. We believe that \zgout{all }the information needed by Aquascope could be\zgout{quite} easily extracted from PCGs, except that our analysis requires programs that type check.

\subsubsection*{RustBelt, RustHornBelt and Creusot}
RustBelt \cite{rustbelt} is an early formal model for Rust, and defines a logical interpretation of Rust's type checking as an instantiation of the Iris framework \cite{IrisFromTheGroundUp}. This interpretation models Rust ownership using separation logic predicates, and borrows via a custom logic of ghost state called the \emph{lifetime logic}; in particular, this setup allows proofs about \emph{unsafe} Rust, where the standard meanings of borrows might be customised for specific library implementations.

RustHornBelt \cite{rusthornbelt} extends RustBelt to enable functional
specifications on verified libraries, and uses \emph{prophecy values} to reflect
the value of borrows as they \emph{will} be eventually modified. Although
prophecies are an alternative means of reasoning about borrows to showing how
capabilities flow back, the \emph{soundness proof} for RustHornBelt requires
showing the existence of an acyclic relation between the borrows\zgout{,
summarising their dependencies}; this is similar to the borrowing information
our model produces.

These models are not directly comparable with ours: they do not connect directly to Rust programs or the compiler's representations; their languages are loop-free \zgout{(automated loop analysis is particularly challenging in Rust)}. However, the models are fully formalised and can reason about unsafe code, while ours \peter{focuses on safe Rust.}
%concerns only the aspects of Rust that the type system governs directly.
Our rules for creating and dereferencing borrows take some inspiration from RustBelt: the dual usage of place nodes and lifetime projection nodes is reminiscent of RustBelt's rules, which require not only access to the borrowed memory but a justification that the corresponding lifetime is currently live.

Creusot \cite{creusot} is a Rust verifier that translates Rust programs into functional programs but in a different manner to Aeneas; it exposes the notion of prophecy values in specifications, so that users can write specifications concerning (re)borrows that span function calls and loops by relating their current and \emph{future} values. While this technique does not require summaries of borrow information to the same extent, we believe that visualising and understanding these summaries (for example for loops) might still help tool users to understand which specifications could be meaningful to express. We are confident that Creusot's translation could also be driven by the information in our model.

\subsubsection*{Stacked Borrows and Tree Borrows}

\peter{
Stacked Borrows~\cite{stacked-borrows} and Tree Borrows~\cite{tree-borrows} define a runtime semantics using graphs (stacks or trees) to represent dynamic borrow information in each execution state, \as{with a focus on defining when memory accesses via
\emph{unsafe Rust code} are allowed. }%\asout{They use this information to define rules that govern memory accesses and designate undefined behavior in unsafe Rust code.}
In contrast, PCGs define a modular static analysis for safe Rust that precisely captures static borrow-checking information. This different goal requires us to address challenges that do not arise for operational semantics such as Stacked or Tree Borrows, where only bounded memory and behaviours are explored, and modularity is not a concern. In particular, our PCGs must represent (possibly unbounded) sets of borrows stored in \eg{}, structs, provide (re)borrowing rules that integrate with these representations, and use them to summarise the effect of function calls and loops. Our PCGs are also designed to efficiently represent path-sensitive borrowing information.
}

\subsubsection*{Other Rust Semantics}
We are aware of several older attempts to formalise aspects of Rust's type system (\eg{},~\cite{Reed15,Kan18,Wang18,Weiss18}); to our knowledge, these do not cover the detailed borrowing concerns addressed by our model, and are typically for idealised definitions of Rust-like languages. Some are operational semantics, which \zgout{by nature} do not require the same complex rules to summarise Rust's implied properties for general path-sensitive information, loops or function calls.

\subsubsection*{Other Rust Analyses}
Many Rust verification and analysis tools could potentially make use of our model. Verus \cite{verus-extended} is a verifier whose translation to SMT requires similar information to Prusti; it does not currently support reborrows that span function calls or loops, but there is long-standing interest in supporting them in the future \cite{verus-reborrowing}, which would require \peter{information similar to that provided by PCGs.} Flux \cite{Flux} is a Rust verifier based on Liquid Types which also requires compiler-generated information for its translation that our model can provide. Rudra \cite{Rudra} is a mature program analysis for Rust, whose analyses require type-system-related information; to our knowledge these analyses are all flow-insensitive, and our model could easily provide (more than) the required basis.

\peter{Pearce~\cite{Pearce21} presents an early formalisation of Rust's type system, but does not support more-intricate features of Rust types (such as structs storing references) nor control flow\zgout{that make capturing the full system most challenging}. His work aims to define borrow checking, whereas ours focuses on the results of the borrow checker.
}
Rust Analyzer~\cite{rust-analyzer} is a library for semantic analysis
of Rust code. Its analyses are primarily intended to support IDE features, e.g.
finding usages of a variable, and in general are
lightweight versions of analyses that the Rust compiler itself performs. In
contrast, our analysis builds on top of the compiler's existing analyses to
provide an elaborated model of its type and borrow checking.

\subsubsection*{Ownership Types}
Ownership type systems were first designed in the context of object-oriented languages.
Clarke et~al.~\cite{ClarkePN98} present a type system that organises objects in
a tree-shaped ownership structure. They enforce that each reference chain from a
root of the ownership graph to an object goes through all of the object's
(transitive) owners; all other references are ruled out by the type system.
Follow-up work proposed various variations of this idea. For instance, Universe
Types~\cite{Mueller02} distinguishes between write and read-only references and
imposes ownership restrictions only on write references. Other
systems~\cite{ClarkeW03,MullerR07} allow ownership to change dynamically, akin
to move assignments in Rust. Some of these early ownership type systems permit
restricted forms of borrowing (\eg{} External Uniqueness~\cite{ClarkeW03}
permits lexically-scoped borrows), but none can express the control-flow
dependent borrow extents tracked by the Rust compiler.

Boyland's alias burying~\cite{Boyland01} supports unique variables, similar to Rust's owning places. In contrast to prior work, which maintained uniqueness through destructive reads (related to Rust's move assignments), references can be borrowed temporarily. A modular static analysis ensures that the borrowed-from unique variable is not used before the borrow goes out of scope.

\section{Conclusions and Future Work}\label{sec:conclusion}
  We have presented Place Capability Graphs, a novel model of Rust's
type and borrow checking. We demonstrated the construction of PCGs from Rust's
internal programmatic representation. In our evaluation, we showed that PCGs
support a large subset of real-world Rust code, accurately reflect the
constraints of the borrow checker, and can be used as the basis for program
analysis tools.

There are several promising avenues of future work. We plan to continue the formal development of our work, and consider its suitability as a specification of type checking itself. We hope the open-source implementation of
Place Capability Graphs can serve as the basis for a wider range of Rust analysis
tools, beyond our prototype integrations into
Prusti and Flowistry.\zgout{Moving forward, we intend to work with the community to
identify other applications of our model and, if needed, adapt the model to
support additional technical use cases.}

\section*{Acknowledgements}
  We thank Rui Ge, Yanze Li, and Youssef Saleh for helpful feedback on this work and its presentation, and Rui Ge, Yanze Li and Randy Zhu for helping to test our artefact.

We acknowledge the support of the Natural Sciences and Engineering Research Council of Canada (NSERC). This work was funded in part by the Stellar Development Foundation.

\newpage
\section*{Data Availability Statement}
We have made available all aspects of our evaluation (including the stand-alone implementation of our model) as an artefact~\cite{PcgArtefact}, including sufficient instructions to reproduce our various evaluations. Our implementation also produces visualisations similar to the figures used in our paper, which can be used to manually inspect instances of our model as generated on chosen Rust code.

\bibliographystyle{ACM-Reference-Format}
\bibliography{paper}


\begin{thebibliography}{42}

%%% the Web address.

\ifx \showCODEN    \undefined \def \showCODEN     #1{\unskip}     \fi
\ifx \showDOI      \undefined \def \showDOI       #1{#1}\fi
\ifx \showISBNx    \undefined \def \showISBNx     #1{\unskip}     \fi
\ifx \showISBNxiii \undefined \def \showISBNxiii  #1{\unskip}     \fi
\ifx \showISSN     \undefined \def \showISSN      #1{\unskip}     \fi
\ifx \showLCCN     \undefined \def \showLCCN      #1{\unskip}     \fi
\ifx \shownote     \undefined \def \shownote      #1{#1}          \fi
\ifx \showarticletitle \undefined \def \showarticletitle #1{#1}   \fi
\ifx \showURL      \undefined \def \showURL       {\relax}        \fi
% invisible to TeX
\providecommand\bibfield[2]{#2}
\providecommand\bibinfo[2]{#2}
\providecommand\natexlab[1]{#1}
\providecommand\showeprint[2][]{arXiv:#2}

\bibitem[Astrauskas et~al\mbox{.}(2019)]%
        {prusti}
\bibfield{author}{\bibinfo{person}{Vytautas Astrauskas}, \bibinfo{person}{Peter M{\"{u}}ller}, \bibinfo{person}{Federico Poli}, {and} \bibinfo{person}{Alexander~J. Summers}.} \bibinfo{year}{2019}\natexlab{}.
\newblock \showarticletitle{Leveraging {R}ust types for modular specification and verification}.
\newblock \bibinfo{journal}{\emph{Proc. {ACM} Program. Lang.}} \bibinfo{volume}{3}, \bibinfo{number}{{OOPSLA}} (\bibinfo{year}{2019}), \bibinfo{pages}{147:1--147:30}.
\newblock
\urldef\tempurl%
\url{https://doi.org/10.1145/3360573}
\showDOI{\tempurl}


\bibitem[Bae et~al\mbox{.}(2021)]%
        {Rudra}
\bibfield{author}{\bibinfo{person}{Yechan Bae}, \bibinfo{person}{Youngsuk Kim}, \bibinfo{person}{Ammar Askar}, \bibinfo{person}{Jungwon Lim}, {and} \bibinfo{person}{Taesoo Kim}.} \bibinfo{year}{2021}\natexlab{}.
\newblock \showarticletitle{Rudra: Finding Memory Safety Bugs in {R}ust at the Ecosystem Scale}. In \bibinfo{booktitle}{\emph{{SOSP} '21: {ACM} {SIGOPS} 28th Symposium on Operating Systems Principles, Virtual Event / Koblenz, Germany, October 26-29, 2021}}, \bibfield{editor}{\bibinfo{person}{Robbert van Renesse} {and} \bibinfo{person}{Nickolai Zeldovich}} (Eds.). \bibinfo{publisher}{{ACM}}, \bibinfo{pages}{84--99}.
\newblock
\urldef\tempurl%
\url{https://doi.org/10.1145/3477132.3483570}
\showDOI{\tempurl}


\bibitem[Boyland(2001)]%
        {Boyland01}
\bibfield{author}{\bibinfo{person}{John Boyland}.} \bibinfo{year}{2001}\natexlab{}.
\newblock \showarticletitle{Alias burying: Unique variables without destructive reads}.
\newblock \bibinfo{journal}{\emph{Softw. Pract. Exp.}} \bibinfo{volume}{31}, \bibinfo{number}{6} (\bibinfo{year}{2001}), \bibinfo{pages}{533--553}.
\newblock
\urldef\tempurl%
\url{https://doi.org/10.1002/SPE.370}
\showDOI{\tempurl}


\bibitem[Castegren and Wrigstad(2016)]%
        {concurrency-control}
\bibfield{author}{\bibinfo{person}{Elias Castegren} {and} \bibinfo{person}{Tobias Wrigstad}.} \bibinfo{year}{2016}\natexlab{}.
\newblock \showarticletitle{{Reference Capabilities for Concurrency Control}}. In \bibinfo{booktitle}{\emph{30th European Conference on Object-Oriented Programming (ECOOP 2016)}} \emph{(\bibinfo{series}{Leibniz International Proceedings in Informatics (LIPIcs)}, Vol.~\bibinfo{volume}{56})}, \bibfield{editor}{\bibinfo{person}{Shriram Krishnamurthi} {and} \bibinfo{person}{Benjamin~S. Lerner}} (Eds.). \bibinfo{publisher}{Schloss Dagstuhl--Leibniz-Zentrum fuer Informatik}, \bibinfo{address}{Dagstuhl, Germany}, \bibinfo{pages}{5:1--5:26}.
\newblock
\showISBNx{978-3-95977-014-9}
\showISSN{1868-8969}
\urldef\tempurl%
\url{https://doi.org/10.4230/LIPIcs.ECOOP.2016.5}
\showDOI{\tempurl}


\bibitem[Clarke and Wrigstad(2003)]%
        {ClarkeW03}
\bibfield{author}{\bibinfo{person}{Dave Clarke} {and} \bibinfo{person}{Tobias Wrigstad}.} \bibinfo{year}{2003}\natexlab{}.
\newblock \showarticletitle{External Uniqueness Is Unique Enough}. In \bibinfo{booktitle}{\emph{{ECOOP} 2003 - Object-Oriented Programming, 17th European Conference, Darmstadt, Germany, July 21-25, 2003, Proceedings}} \emph{(\bibinfo{series}{Lecture Notes in Computer Science}, Vol.~\bibinfo{volume}{2743})}, \bibfield{editor}{\bibinfo{person}{Luca Cardelli}} (Ed.). \bibinfo{publisher}{Springer}, \bibinfo{pages}{176--200}.
\newblock
\urldef\tempurl%
\url{https://doi.org/10.1007/978-3-540-45070-2\_9}
\showDOI{\tempurl}


\bibitem[Clarke et~al\mbox{.}(1998)]%
        {ClarkePN98}
\bibfield{author}{\bibinfo{person}{David~G. Clarke}, \bibinfo{person}{John Potter}, {and} \bibinfo{person}{James Noble}.} \bibinfo{year}{1998}\natexlab{}.
\newblock \showarticletitle{Ownership Types for Flexible Alias Protection}. In \bibinfo{booktitle}{\emph{Proceedings of the 1998 {ACM} {SIGPLAN} Conference on Object-Oriented Programming Systems, Languages {\&} Applications, {OOPSLA} 1998, Vancouver, British Columbia, Canada, October 18-22, 1998}}, \bibfield{editor}{\bibinfo{person}{Bj{\o}rn~N. Freeman{-}Benson} {and} \bibinfo{person}{Craig Chambers}} (Eds.). \bibinfo{publisher}{{ACM}}, \bibinfo{pages}{48--64}.
\newblock
\urldef\tempurl%
\url{https://doi.org/10.1145/286936.286947}
\showDOI{\tempurl}


\bibitem[Clebsch et~al\mbox{.}(2015)]%
        {pony}
\bibfield{author}{\bibinfo{person}{Sylvan Clebsch}, \bibinfo{person}{Sophia Drossopoulou}, \bibinfo{person}{Sebastian Blessing}, {and} \bibinfo{person}{Andy McNeil}.} \bibinfo{year}{2015}\natexlab{}.
\newblock \showarticletitle{Deny Capabilities for Safe, Fast Actors}. In \bibinfo{booktitle}{\emph{International Workshop on Programming Based on Actors, Agents, and Decentralized Control}} \emph{(\bibinfo{series}{AGERE! 2015})}. \bibinfo{publisher}{ACM}, \bibinfo{pages}{1--12}.
\newblock


\bibitem[Community and Team(2025)]%
        {place-expressions}
\bibfield{author}{\bibinfo{person}{Rust Community} {and} \bibinfo{person}{Language Team}.} \bibinfo{year}{2025}\natexlab{}.
\newblock \bibinfo{title}{Place Expressions and Value Expressions (The {R}ust Reference)}.
\newblock
\newblock
\urldef\tempurl%
\url{https://doc.rust-lang.org/reference/expressions.html#place-expressions-and-value-expressions}
\showURL{%
\tempurl}
\newblock
\shownote{Accessed July 2025}.


\bibitem[Crichton et~al\mbox{.}(2023)]%
        {Crichton'23}
\bibfield{author}{\bibinfo{person}{Will Crichton}, \bibinfo{person}{Gavin Gray}, {and} \bibinfo{person}{Shriram Krishnamurthi}.} \bibinfo{year}{2023}\natexlab{}.
\newblock \showarticletitle{A Grounded Conceptual Model for Ownership Types in {R}ust}.
\newblock \bibinfo{journal}{\emph{Proc. ACM Program. Lang.}} \bibinfo{volume}{7}, \bibinfo{number}{OOPSLA2}, Article \bibinfo{articleno}{265} (\bibinfo{date}{Oct.} \bibinfo{year}{2023}), \bibinfo{numpages}{29}~pages.
\newblock
\urldef\tempurl%
\url{https://doi.org/10.1145/3622841}
\showDOI{\tempurl}


\bibitem[Crichton et~al\mbox{.}(2022)]%
        {Flowistry}
\bibfield{author}{\bibinfo{person}{Will Crichton}, \bibinfo{person}{Marco Patrignani}, \bibinfo{person}{Maneesh Agrawala}, {and} \bibinfo{person}{Pat Hanrahan}.} \bibinfo{year}{2022}\natexlab{}.
\newblock \showarticletitle{Modular information flow through ownership}. In \bibinfo{booktitle}{\emph{Proceedings of the 43rd ACM SIGPLAN International Conference on Programming Language Design and Implementation}} (San Diego, CA, USA) \emph{(\bibinfo{series}{PLDI 2022})}. \bibinfo{publisher}{Association for Computing Machinery}, \bibinfo{address}{New York, NY, USA}, \bibinfo{pages}{1–14}.
\newblock
\showISBNx{9781450392655}
\urldef\tempurl%
\url{https://doi.org/10.1145/3519939.3523445}
\showDOI{\tempurl}


\bibitem[Denis et~al\mbox{.}(2022)]%
        {creusot}
\bibfield{author}{\bibinfo{person}{Xavier Denis}, \bibinfo{person}{Jacques{-}Henri Jourdan}, {and} \bibinfo{person}{Claude March{\'{e}}}.} \bibinfo{year}{2022}\natexlab{}.
\newblock \showarticletitle{Creusot: {A} Foundry for the Deductive Verification of {R}ust Programs}. In \bibinfo{booktitle}{\emph{Formal Methods and Software Engineering - 23rd International Conference on Formal Engineering Methods, {ICFEM} 2022, Madrid, Spain, October 24-27, 2022, Proceedings}} \emph{(\bibinfo{series}{Lecture Notes in Computer Science}, Vol.~\bibinfo{volume}{13478})}, \bibfield{editor}{\bibinfo{person}{Adri{\'{a}}n Riesco} {and} \bibinfo{person}{Min Zhang}} (Eds.). \bibinfo{publisher}{Springer}, \bibinfo{pages}{90--105}.
\newblock
\urldef\tempurl%
\url{https://doi.org/10.1007/978-3-031-17244-1\_6}
\showDOI{\tempurl}


\bibitem[Developers(2025c)]%
        {prusti-communication}
\bibfield{author}{\bibinfo{person}{Prusti Developers}.} \bibinfo{year}{2024-2025}\natexlab{c}.
\newblock \bibinfo{title}{Personal Communication}.
\newblock
\newblock


\bibitem[Developers(2025a)]%
        {PcgArtefact}
\bibfield{author}{\bibinfo{person}{Prusti Developers}.} \bibinfo{year}{2025}\natexlab{a}.
\newblock \bibinfo{booktitle}{\emph{Artefact for "Place Capability Graphs: A General- Purpose Model of Rust's Ownership and Borrowing Guarantees"}}.
\newblock
\urldef\tempurl%
\url{https://doi.org/10.5281/zenodo.15759309}
\showDOI{\tempurl}


\bibitem[Developers(2025b)]%
        {PcgGithub}
\bibfield{author}{\bibinfo{person}{Prusti Developers}.} \bibinfo{year}{2025}\natexlab{b}.
\newblock \bibinfo{title}{PCG Repository}.
\newblock
\newblock
\urldef\tempurl%
\url{https://github.com/prusti/pcg}
\showURL{%
\tempurl}
\newblock
\shownote{Accessed August 21, 2025}.


\bibitem[Developers(2022)]%
        {verus-reborrowing}
\bibfield{author}{\bibinfo{person}{Verus Developers}.} \bibinfo{year}{2022}\natexlab{}.
\newblock \bibinfo{title}{Returning mutable references (discussion)}.
\newblock
\newblock
\urldef\tempurl%
\url{https://github.com/verus-lang/verus/discussions/35}
\showURL{%
\tempurl}
\newblock
\shownote{Accessed March, 2025}.


\bibitem[Gordon et~al\mbox{.}(2012)]%
        {safe-parallelism}
\bibfield{author}{\bibinfo{person}{Colin~S. Gordon}, \bibinfo{person}{Matthew~J. Parkinson}, \bibinfo{person}{Jared Parsons}, \bibinfo{person}{Aleks Bromfield}, {and} \bibinfo{person}{Joe Duffy}.} \bibinfo{year}{2012}\natexlab{}.
\newblock \showarticletitle{Uniqueness and Reference Immutability for Safe Parallelism}.
\newblock \bibinfo{journal}{\emph{SIGPLAN Not.}} \bibinfo{volume}{47}, \bibinfo{number}{10} (\bibinfo{date}{Oct.} \bibinfo{year}{2012}), \bibinfo{pages}{21--40}.
\newblock


\bibitem[Grannan et~al\mbox{.}(2025)]%
        {PcgArxiv}
\bibfield{author}{\bibinfo{person}{Zachary Grannan}, \bibinfo{person}{Aurel Bílý}, \bibinfo{person}{Jonáš Fiala}, \bibinfo{person}{Jasper Geer}, \bibinfo{person}{Markus de Medeiros}, \bibinfo{person}{Peter Müller}, {and} \bibinfo{person}{Alexander~J. Summers}.} \bibinfo{year}{2025}\natexlab{}.
\newblock \bibinfo{title}{Place Capability Graphs: A General-Purpose Model of Rust's Ownership and Borrowing Guarantees}.
\newblock
\newblock
\showeprint[arxiv]{2503.21691}~[cs.PL]
\urldef\tempurl%
\url{https://arxiv.org/abs/2503.21691}
\showURL{%
\tempurl}


\bibitem[Haller and Odersky(2010)]%
        {odersky-capabilities}
\bibfield{author}{\bibinfo{person}{Philipp Haller} {and} \bibinfo{person}{Martin Odersky}.} \bibinfo{year}{2010}\natexlab{}.
\newblock \showarticletitle{Capabilities for Uniqueness and Borrowing}. In \bibinfo{booktitle}{\emph{European Conference on Object-Oriented Programming (ECOOP)}} \emph{(\bibinfo{series}{Lecture Notes in Computer Science}, Vol.~\bibinfo{volume}{6183})}, \bibfield{editor}{\bibinfo{person}{Theo D'Hondt}} (Ed.). \bibinfo{publisher}{Springer}, \bibinfo{pages}{354--378}.
\newblock


\bibitem[Ho et~al\mbox{.}(2024)]%
        {Ho'24}
\bibfield{author}{\bibinfo{person}{Son Ho}, \bibinfo{person}{Aymeric Fromherz}, {and} \bibinfo{person}{Jonathan Protzenko}.} \bibinfo{year}{2024}\natexlab{}.
\newblock \showarticletitle{Sound Borrow-Checking for {R}ust via Symbolic Semantics}.
\newblock \bibinfo{journal}{\emph{Proc. ACM Program. Lang.}} \bibinfo{volume}{8}, \bibinfo{number}{ICFP}, Article \bibinfo{articleno}{251} (\bibinfo{date}{Aug.} \bibinfo{year}{2024}), \bibinfo{numpages}{29}~pages.
\newblock
\urldef\tempurl%
\url{https://doi.org/10.1145/3674640}
\showDOI{\tempurl}


\bibitem[Ho and Protzenko(2022)]%
        {aeneas}
\bibfield{author}{\bibinfo{person}{Son Ho} {and} \bibinfo{person}{Jonathan Protzenko}.} \bibinfo{year}{2022}\natexlab{}.
\newblock \showarticletitle{Aeneas: {R}ust verification by functional translation}.
\newblock \bibinfo{journal}{\emph{Proc. {ACM} Program. Lang.}} \bibinfo{volume}{6}, \bibinfo{number}{{ICFP}} (\bibinfo{year}{2022}), \bibinfo{pages}{711--741}.
\newblock
\urldef\tempurl%
\url{https://doi.org/10.1145/3547647}
\showDOI{\tempurl}


\bibitem[Jung et~al\mbox{.}(2020)]%
        {stacked-borrows}
\bibfield{author}{\bibinfo{person}{Ralf Jung}, \bibinfo{person}{Hoang{-}Hai Dang}, \bibinfo{person}{Jeehoon Kang}, {and} \bibinfo{person}{Derek Dreyer}.} \bibinfo{year}{2020}\natexlab{}.
\newblock \showarticletitle{Stacked borrows: an aliasing model for {R}ust}.
\newblock \bibinfo{journal}{\emph{Proc. {ACM} Program. Lang.}} \bibinfo{volume}{4}, \bibinfo{number}{{POPL}} (\bibinfo{year}{2020}), \bibinfo{pages}{41:1--41:32}.
\newblock
\urldef\tempurl%
\url{https://doi.org/10.1145/3371109}
\showDOI{\tempurl}


\bibitem[Jung et~al\mbox{.}(2018a)]%
        {rustbelt}
\bibfield{author}{\bibinfo{person}{Ralf Jung}, \bibinfo{person}{Jacques{-}Henri Jourdan}, \bibinfo{person}{Robbert Krebbers}, {and} \bibinfo{person}{Derek Dreyer}.} \bibinfo{year}{2018}\natexlab{a}.
\newblock \showarticletitle{RustBelt: securing the foundations of the {R}ust programming language}.
\newblock \bibinfo{journal}{\emph{Proc. {ACM} Program. Lang.}} \bibinfo{volume}{2}, \bibinfo{number}{{POPL}} (\bibinfo{year}{2018}), \bibinfo{pages}{66:1--66:34}.
\newblock
\urldef\tempurl%
\url{https://doi.org/10.1145/3158154}
\showDOI{\tempurl}


\bibitem[Jung et~al\mbox{.}(2018b)]%
        {rust-belt}
\bibfield{author}{\bibinfo{person}{Ralf Jung}, \bibinfo{person}{Jacques{-}Henri Jourdan}, \bibinfo{person}{Robbert Krebbers}, {and} \bibinfo{person}{Derek Dreyer}.} \bibinfo{year}{2018}\natexlab{b}.
\newblock \showarticletitle{RustBelt: Securing the Foundations of the {R}ust Programming Language}.
\newblock \bibinfo{journal}{\emph{{PACMPL}}} \bibinfo{volume}{2}, \bibinfo{number}{{POPL}} (\bibinfo{year}{2018}), \bibinfo{pages}{66:1--66:34}.
\newblock


\bibitem[Jung et~al\mbox{.}(2018c)]%
        {IrisFromTheGroundUp}
\bibfield{author}{\bibinfo{person}{Ralf Jung}, \bibinfo{person}{Robbert Krebbers}, \bibinfo{person}{Jacques-Henri Jourdan}, \bibinfo{person}{Aleš Bizjak}, \bibinfo{person}{Lars Birkedal}, {and} \bibinfo{person}{Derek Dreyer}.} \bibinfo{year}{2018}\natexlab{c}.
\newblock \showarticletitle{Iris from the ground up: A modular foundation for higher-order concurrent separation logic}.
\newblock \bibinfo{journal}{\emph{Journal of Functional Programming}}  \bibinfo{volume}{28} (\bibinfo{year}{2018}), \bibinfo{pages}{e20}.
\newblock
\urldef\tempurl%
\url{https://doi.org/10.1017/S0956796818000151}
\showDOI{\tempurl}


\bibitem[Kan et~al\mbox{.}(2018)]%
        {Kan18}
\bibfield{author}{\bibinfo{person}{Shuanglong Kan}, \bibinfo{person}{David San{\'{a}}n}, \bibinfo{person}{Shang{-}Wei Lin}, {and} \bibinfo{person}{Yang Liu}.} \bibinfo{year}{2018}\natexlab{}.
\newblock \showarticletitle{K-Rust: An Executable Formal Semantics for {R}ust}.
\newblock \bibinfo{journal}{\emph{CoRR}}  \bibinfo{volume}{abs/1804.07608} (\bibinfo{year}{2018}).
\newblock
\showeprint[arxiv]{1804.07608}
\urldef\tempurl%
\url{http://arxiv.org/abs/1804.07608}
\showURL{%
\tempurl}


\bibitem[Lattuada et~al\mbox{.}(2023)]%
        {verus-extended}
\bibfield{author}{\bibinfo{person}{Andrea Lattuada}, \bibinfo{person}{Travis Hance}, \bibinfo{person}{Chanhee Cho}, \bibinfo{person}{Matthias Brun}, \bibinfo{person}{Isitha Subasinghe}, \bibinfo{person}{Yi Zhou}, \bibinfo{person}{Jon Howell}, \bibinfo{person}{Bryan Parno}, {and} \bibinfo{person}{Chris Hawblitzel}.} \bibinfo{year}{2023}\natexlab{}.
\newblock \showarticletitle{Verus: Verifying {R}ust Programs using Linear Ghost Types (extended version)}.
\newblock \bibinfo{journal}{\emph{CoRR}}  \bibinfo{volume}{abs/2303.05491} (\bibinfo{year}{2023}).
\newblock
\urldef\tempurl%
\url{https://doi.org/10.48550/arXiv.2303.05491}
\showDOI{\tempurl}
\showeprint[arXiv]{2303.05491}


\bibitem[Lehmann et~al\mbox{.}(2022)]%
        {Flux}
\bibfield{author}{\bibinfo{person}{Nico Lehmann}, \bibinfo{person}{Adam Geller}, \bibinfo{person}{Gilles Barthe}, \bibinfo{person}{Niki Vazou}, {and} \bibinfo{person}{Ranjit Jhala}.} \bibinfo{year}{2022}\natexlab{}.
\newblock \showarticletitle{Flux: Liquid Types for {R}ust}.
\newblock \bibinfo{journal}{\emph{CoRR}}  \bibinfo{volume}{abs/2207.04034} (\bibinfo{year}{2022}).
\newblock
\urldef\tempurl%
\url{https://doi.org/10.48550/arXiv.2207.04034}
\showDOI{\tempurl}
\showeprint[arXiv]{2207.04034}


\bibitem[Matsushita et~al\mbox{.}(2022)]%
        {rusthornbelt}
\bibfield{author}{\bibinfo{person}{Yusuke Matsushita}, \bibinfo{person}{Xavier Denis}, \bibinfo{person}{Jacques{-}Henri Jourdan}, {and} \bibinfo{person}{Derek Dreyer}.} \bibinfo{year}{2022}\natexlab{}.
\newblock \showarticletitle{RustHornBelt: a semantic foundation for functional verification of {R}ust programs with unsafe code}. In \bibinfo{booktitle}{\emph{{PLDI} '22: 43rd {ACM} {SIGPLAN} International Conference on Programming Language Design and Implementation, San Diego, CA, USA, June 13 - 17, 2022}}, \bibfield{editor}{\bibinfo{person}{Ranjit Jhala} {and} \bibinfo{person}{Isil Dillig}} (Eds.). \bibinfo{publisher}{{ACM}}, \bibinfo{pages}{841--856}.
\newblock
\urldef\tempurl%
\url{https://doi.org/10.1145/3519939.3523704}
\showDOI{\tempurl}


\bibitem[M{\"u}ller(2002)]%
        {Mueller02}
\bibfield{author}{\bibinfo{person}{P. M{\"u}ller}.} \bibinfo{year}{2002}\natexlab{}.
\newblock \bibinfo{booktitle}{\emph{Modular Specification and Verification of Object-Oriented Programs}}. \bibinfo{series}{Lecture Notes in Computer Science}, Vol.~\bibinfo{volume}{2262}.
\newblock \bibinfo{publisher}{Springer}.
\newblock


\bibitem[M{\"{u}}ller and Rudich(2007)]%
        {MullerR07}
\bibfield{author}{\bibinfo{person}{Peter M{\"{u}}ller} {and} \bibinfo{person}{Arsenii Rudich}.} \bibinfo{year}{2007}\natexlab{}.
\newblock \showarticletitle{Ownership transfer in universe types}. In \bibinfo{booktitle}{\emph{Proceedings of the 22nd Annual {ACM} {SIGPLAN} Conference on Object-Oriented Programming, Systems, Languages, and Applications, {OOPSLA} 2007, October 21-25, 2007, Montreal, Quebec, Canada}}, \bibfield{editor}{\bibinfo{person}{Richard~P. Gabriel}, \bibinfo{person}{David~F. Bacon}, \bibinfo{person}{Cristina~Videira Lopes}, {and} \bibinfo{person}{Guy L.~Steele Jr.}} (Eds.). \bibinfo{publisher}{{ACM}}, \bibinfo{pages}{461--478}.
\newblock
\urldef\tempurl%
\url{https://doi.org/10.1145/1297027.1297061}
\showDOI{\tempurl}


\bibitem[M{\"{u}}ller et~al\mbox{.}(2016)]%
        {viper}
\bibfield{author}{\bibinfo{person}{Peter M{\"{u}}ller}, \bibinfo{person}{Malte Schwerhoff}, {and} \bibinfo{person}{Alexander~J. Summers}.} \bibinfo{year}{2016}\natexlab{}.
\newblock \showarticletitle{Viper: {A} Verification Infrastructure for Permission-Based Reasoning}. In \bibinfo{booktitle}{\emph{Verification, Model Checking, and Abstract Interpretation - 17th International Conference, {VMCAI} 2016, St. Petersburg, FL, USA, January 17-19, 2016. Proceedings}} \emph{(\bibinfo{series}{Lecture Notes in Computer Science}, Vol.~\bibinfo{volume}{9583})}, \bibfield{editor}{\bibinfo{person}{Barbara Jobstmann} {and} \bibinfo{person}{K.~Rustan~M. Leino}} (Eds.). \bibinfo{publisher}{Springer}, \bibinfo{pages}{41--62}.
\newblock
\urldef\tempurl%
\url{https://doi.org/10.1007/978-3-662-49122-5\_2}
\showDOI{\tempurl}


\bibitem[Pearce(2021)]%
        {Pearce21}
\bibfield{author}{\bibinfo{person}{David~J. Pearce}.} \bibinfo{year}{2021}\natexlab{}.
\newblock \showarticletitle{A Lightweight Formalism for Reference Lifetimes and Borrowing in {R}ust}.
\newblock \bibinfo{journal}{\emph{{ACM} Trans. Program. Lang. Syst.}} \bibinfo{volume}{43}, \bibinfo{number}{1} (\bibinfo{year}{2021}), \bibinfo{pages}{3:1--3:73}.
\newblock
\urldef\tempurl%
\url{https://doi.org/10.1145/3443420}
\showDOI{\tempurl}


\bibitem[Reed(2015)]%
        {Reed15}
\bibfield{author}{\bibinfo{person}{Eric~W. Reed}.} \bibinfo{year}{2015}\natexlab{}.
\newblock \bibinfo{booktitle}{\emph{Patina: A Formalization of the {R}ust Programming Language}}.
\newblock \bibinfo{type}{{T}echnical {R}eport} UW-CSE-15-03-02. \bibinfo{institution}{University of Washington}.
\newblock


\bibitem[{Rust contributors}(2019)]%
        {Polonius}
\bibfield{author}{\bibinfo{person}{{Rust contributors}}.} \bibinfo{year}{2019}\natexlab{}.
\newblock \bibinfo{title}{The Polonius Reference Implementation for the {R}ust Borrow-Checker}.
\newblock
\newblock
\urldef\tempurl%
\url{https://github.com/rust-lang/polonius}
\showURL{%
\tempurl}
\newblock
\shownote{Accessed April 4, 2019}.


\bibitem[{Rémy Rakic and Niko Matsakis on behalf of The Polonius Working Group}(2023)]%
        {Polonius-update}
\bibfield{author}{\bibinfo{person}{{Rémy Rakic and Niko Matsakis on behalf of The Polonius Working Group}}.} \bibinfo{year}{2023}\natexlab{}.
\newblock \bibinfo{title}{Polonius Update}.
\newblock
\newblock
\urldef\tempurl%
\url{https://blog.rust-lang.org/inside-rust/2023/10/06/polonius-update.html}
\showURL{%
\tempurl}
\newblock
\shownote{Accessed March 24, 2025}.


\bibitem[Stork et~al\mbox{.}(2014)]%
        {AEminium}
\bibfield{author}{\bibinfo{person}{Sven Stork}, \bibinfo{person}{Karl Naden}, \bibinfo{person}{Joshua Sunshine}, \bibinfo{person}{Manuel Mohr}, \bibinfo{person}{Alcides Fonseca}, \bibinfo{person}{Paulo Marques}, {and} \bibinfo{person}{Jonathan Aldrich}.} \bibinfo{year}{2014}\natexlab{}.
\newblock \showarticletitle{AEMinium: A Permission-Based Concurrent-by-Default Programming Language Approach}.
\newblock \bibinfo{journal}{\emph{ACM Trans. Program. Lang. Syst.}} \bibinfo{volume}{36}, \bibinfo{number}{1} (\bibinfo{date}{March} \bibinfo{year}{2014}), \bibinfo{pages}{2:1--2:42}.
\newblock


\bibitem[Systems and contributors(2025)]%
        {rust-analyzer}
\bibfield{author}{\bibinfo{person}{Ferrous Systems} {and} \bibinfo{person}{contributors}.} \bibinfo{year}{2025}\natexlab{}.
\newblock \bibinfo{title}{The rust-analyzer Project}.
\newblock
\newblock
\urldef\tempurl%
\url{https://rust-analyzer.github.io/}
\showURL{%
\tempurl}
\newblock
\shownote{Accessed July 30, 2025}.


\bibitem[VanHattum et~al\mbox{.}(2022)]%
        {Kani}
\bibfield{author}{\bibinfo{person}{Alexa VanHattum}, \bibinfo{person}{Daniel Schwartz{-}Narbonne}, \bibinfo{person}{Nathan Chong}, {and} \bibinfo{person}{Adrian Sampson}.} \bibinfo{year}{2022}\natexlab{}.
\newblock \showarticletitle{Verifying Dynamic Trait Objects in {R}ust}. In \bibinfo{booktitle}{\emph{44th {IEEE/ACM} International Conference on Software Engineering: Software Engineering in Practice, {ICSE} {(SEIP)} 2022, Pittsburgh, PA, USA, May 22-24, 2022}}. \bibinfo{publisher}{{IEEE}}, \bibinfo{pages}{321--330}.
\newblock
\urldef\tempurl%
\url{https://doi.org/10.1109/ICSE-SEIP55303.2022.9794041}
\showDOI{\tempurl}


\bibitem[Villani et~al\mbox{.}(2025)]%
        {tree-borrows}
\bibfield{author}{\bibinfo{person}{Neven Villani}, \bibinfo{person}{Johannes Hostert}, \bibinfo{person}{Derek Dreyer}, {and} \bibinfo{person}{Ralf Jung}.} \bibinfo{year}{2025}\natexlab{}.
\newblock \showarticletitle{Tree Borrows}.
\newblock \bibinfo{journal}{\emph{Proc. ACM Program. Lang.}} \bibinfo{volume}{9}, \bibinfo{number}{PLDI}, Article \bibinfo{articleno}{188} (\bibinfo{date}{June} \bibinfo{year}{2025}).
\newblock
\urldef\tempurl%
\url{https://doi.org/10.1145/3735592}
\showDOI{\tempurl}


\bibitem[Wang et~al\mbox{.}(2018)]%
        {Wang18}
\bibfield{author}{\bibinfo{person}{Feng Wang}, \bibinfo{person}{Fu Song}, \bibinfo{person}{Min Zhang}, \bibinfo{person}{Xiaoran Zhu}, {and} \bibinfo{person}{Jun Zhang}.} \bibinfo{year}{2018}\natexlab{}.
\newblock \showarticletitle{KRust: {A} Formal Executable Semantics of {R}ust}.
\newblock \bibinfo{journal}{\emph{CoRR}}  \bibinfo{volume}{abs/1804.10806} (\bibinfo{year}{2018}).
\newblock
\showeprint[arxiv]{1804.10806}
\urldef\tempurl%
\url{http://arxiv.org/abs/1804.10806}
\showURL{%
\tempurl}


\bibitem[Weiss et~al\mbox{.}(2018)]%
        {Weiss18}
\bibfield{author}{\bibinfo{person}{Aaron Weiss}, \bibinfo{person}{Daniel Patterson}, {and} \bibinfo{person}{Amal Ahmed}.} \bibinfo{year}{2018}\natexlab{}.
\newblock \showarticletitle{Rust Distilled: An Expressive Tower of Languages}.
\newblock \bibinfo{journal}{\emph{arXiv preprint arXiv:1806.02693}} (\bibinfo{year}{2018}).
\newblock


\bibitem[Weiss et~al\mbox{.}(2019)]%
        {Oxide}
\bibfield{author}{\bibinfo{person}{Aaron Weiss}, \bibinfo{person}{Daniel Patterson}, \bibinfo{person}{Nicholas~D. Matsakis}, {and} \bibinfo{person}{Amal Ahmed}.} \bibinfo{year}{2019}\natexlab{}.
\newblock \showarticletitle{Oxide: The Essence of {R}ust}.
\newblock \bibinfo{journal}{\emph{CoRR}}  \bibinfo{volume}{abs/1903.00982} (\bibinfo{year}{2019}).
\newblock
\urldef\tempurl%
\url{http://arxiv.org/abs/1903.00982}
\showURL{%
\tempurl}


\end{thebibliography}

\newpage
%\appendix
%\input{sections/appendix.tex}

%\section{Previous Formalisation Draft}\label{sec:form}
%\input{sections/form.tex}
\end{document}